\documentclass[12pt,preprint]{aastex}
\usepackage{graphicx}
\usepackage{paralist}  % for compactitem, compactenum, inparaenum
\usepackage{soul}

\begin{document}
\pagestyle{empty} % omit page # on 1st page

\title{Swift Observations of Gamma-Ray Burst Pulse Shapes:\\
GRB Pulse Spectral Evolution Clarified}

\author{Jon Hakkila\altaffilmark{1}, Amy Lien\altaffilmark{2,}\altaffilmark{3}, 
Takanori Sakamoto\altaffilmark{2,}\altaffilmark{4}, David Morris\altaffilmark{5}, 
\\ James E. Neff\altaffilmark{6}, and Timothy W. Giblin\altaffilmark{7}
%and Demosthenes Kazanas (?)\altaffilmark{5}
}

\affil{$^1$University of Charleston South Carolina at the College of Charleston, SC   29424-0001 \\
$^2$Center for Research and Exploration in Space Science and Technology  (CRESST)  and  NASA  Goddard  SpaceFlight Center, Greenbelt, MD 20771\\
$^3$Dept. Physics,  University  of Maryland,  Baltimore County, Baltimore, MD 21250\\
$^4$Dept. Physics and Mathematics, Aoyama Gakuin University, Japan\\
$^5$Dept. Physics, University of the Virgin Islands, St. Thomas, VI  00802\\
$^6$ Dept. Physics and Astronomy, The College of Charleston, Charleston, SC  29424-0001\\
$^7$Dept. Physics, United States Air Force Academy, CO  80840 \\
}

\email{hakkilaj@cofc.edu}

\section{Abstract}

Isolated Swift gamma-ray burst (GRB) pulses, like their higher-energy BATSE counterparts, emit the bulk of their pulsed emission as a hard-to-soft component that can be fitted by the \cite{nor05} empirical pulse model. This signal is overlaid by a fainter, three-peaked signal that can be modeled by an empirical wave-like function \citep{hak14}: the two fits combine to reproduce GRB pulses with distinctive three-peaked shapes. The precursor peak appears on or before the pulse rise and is often the hardest component, the central peak is the brightest, and the decay peak converts exponentially decaying emission into a long, soft, power-law tail. Accounting for systematic instrumental differences, the general characteristics of the fitted pulses are remarkably similar. Isolated GRB pulses are dominated by hard-to-soft evolution; this is more pronounced for asymmetric pulses than for symmetric ones. Isolated GRB pulses can also exhibit intensity tracking behaviors that, when observed, are tied to the timing of the three peaks: pulses with the largest maximum hardnesses are hardest during the precursor, those with smaller maximum hardnesses are hardest during the central peak, and all pulses can re-harden during the central peak and/or during the decay peak. Since these behaviors are essentially seen in all isolated pulses, the distinction between ``hard-to-soft'' and ``intensity-tracking'' pulses really no longer applies. Additionally, the triple-peaked nature of isolated GRB pulses seems to indicate that energy is injected on three separate occasions during the pulse duration: theoretical pulse models need to account for this.

\section{Introduction}

Gamma-ray burst (GRB) pulse characteristics remain poorly understood even though pulses are the most common structures in GRB light curves: there is little agreement concerning the physical mechanisms responsible for producing them. Pulse characteristics are difficult to extract due to a mixture of strong spectral-temporal evolution, instrumental and sampling biases, confusion caused by overlapping pulses, and low signal-to-noise measurements. Understanding pulse behaviors is important, as it is difficult to adequately model GRB progenitors and outflow without first understanding the mechanism by which pulses liberate energy, and it is difficult to understand pulse mechanisms without first identifying pulse behaviors.

Data driven analysis approaches are providing new insights into GRB pulse characteristics. These approaches have both advantages and disadvantages over standard theoretical modeling techniques: they can identify behaviors (typically clustering and/or correlations) that theoretical models are not trying to explain and thus may not recognize, but by asking questions about pulse behaviors they do not necessarily couch these questions in the form of specific theoretical models. Data driven approaches have allowed the identification of important pulse characteristics even as theoretical models have been unable to explain these behaviors. 

As an example, consider empirical light curve models, which can be used to provide an accurate representation of GRB pulse behaviors in the absence of identified pulse physics. Several empirical models are able to adequately explain the general shapes of GRB pulse light curves (e.g. see the introduction in \cite{hak14}). Among these, the simple model of \cite{nor05} is useful because it has only four free parameters which allow reasonable fits over the wide range of observed pulse characteristics.

Empirical approaches demonstrate that GRB pulse light curves can be generally represented by a {\em hard-to-soft} spectral evolution. Pulses are generally hardest at the first instant they appear, and they continue to soften during the pulse rise, past the pulse peak, and through the pulse decay \citep{hak11,pen12}. Although hard-to-soft evolution is the most common pulse evolution (e.g. \cite{cri99}), {\em intensity tracking} pulse evolution has been identified in a smaller number of pulses (e.g. \cite{for95, lia96, kan06, pen09,lu10,hak11}). It is difficult to estimate what fraction of pulses exhibit intensity tracking behaviors, since there are many cases where overlapping hard-to-soft pulses appear to have been misclassified as intensity tracking \citep{hak11}.

Hard-to-soft spectral evolution is so central to GRB pulse structure that it can be used as a defining characteristic in rigorous statistical modeling techniques \citep{bro14}. The harder a pulse is, the more pronounced the evolution typically is. A small percentage of pulses appear to harden during the pulse rise, but these {\em intensity tracking} pulses are typically among the very softest, suggesting that instrumental biases might be in part responsible for the initial soft emission. Several of the well-known correlated spectral-temporal GRB pulse properties are simply a result of this hard-to-soft spectral decay: rapid hardness evolution produces short duration pulses with short spectral lags while slow hardness evolution produces long duration ones with long lags. 

The relationship between pulse asymmetry and hard-to-soft evolution is not as clear as the one between duration, lag, and evolution. Asymmetry is anti-correlated with hardness and correlated with duration across a large BATSE GRB pulse sample such that asymmetric pulses are generally longer and softer than symmetric pulses \citep{hak11}. This implies that asymmetric pulses in general do not undergo a significant hardness evolution, since the softening is mild and occurs over a long time interval. However, some of the faintest, softest, longest GRB pulses are symmetric and undergo very little hard-to-soft evolution, while some asymmetric pulses are quite hard. To further confuse matters, the bright, short pulses in classical Short GRBs are also symmetric yet very hard; undergoing a rapid hard-to-soft evolution. Without an obvious connection between hard-to-soft evolution and pulse asymmetry, the possibility exists that asymmetry and hard-to-soft evolution are independent or semi-independent pulse characteristics.

GRB pulses (at least those belonging to the Long and Intermediate GRB classes) exhibit additional intensity fluctuations that overlay the hard-to-soft monotonic pulse component \citep{hak14}; these fluctuations can be seen in the residuals obtained by subtracting the \cite{nor05} modeled monotonic pulse shape from the light curve. These fluctuations, which have the appearance of three peaks separated by two valleys, are more easily observed in high signal-to-noise pulses; this makes bright pulses look like they have variable, multi-peaked light curves, in contrast to the smooth monotonic light curves of fainter pulses. When combined with spectral softening, bright pulses appear to be composed of at least three peaks, with the first one (during or immediately preceding the pulse rise) generally being the hardest, the second (and brightest) one being of intermediate hardness, and the last one (during the long decay) being soft. 

Hard-to-soft evolution and phased non-monotonic intensity fluctuations can be used to help determine if a GRB emission episode is composed of a single pulse or of several merged pulses. A single pulse, at least in the energy ranges of BATSE and Fermi, has pulse residual  parameters that correlate with the pulse asymmetry. The residual light curve waves can be fitted by a simple four-parameter empirical model \citep{hak14} 

This empirical model has several interesting features: 1) the basis of the model is a truncated Bessel function of the first kind, $J_0(x)$ that begins at the true pulse peak, 2) the wave can be modeled by a mirror image of itself in time, with the time-reflected part occurring during the pulse rise and the time-forward part occurring during the pulse decay, 3) the pulse decay portion is a time-stretched version of the mirrored part on the pulse rise, and 4) most if not all of the pulse residual fit parameters correlate with the pulse asymmetry. At the present time, no physical mechanism is known that can provide these observed model characteristics.

It is the reproducibility of the light curve residual fluctuations, combined with the overall hard-to-soft spectral evolution, which demonstrates that the fluctuations all combine to form a single evolving pulse with a non-monotonic shape. All pulse light curves seem to undergo multiple intensity increases and decreases. However, the definition of a pulse being composed of several monotonically rising and falling regions may be a semantic one: a single 'pulse' might indeed be composed of three separate, linked, smaller pulses. If we do choose to believe that each fluctuation identifies a separate pulse, then the coupling between these fluctuations and the overall hard-to-soft evolution indicates that multiple pulses always combine to form similar, temporally-linked, complex shapes that need to be described by a single smooth hardness evolution.

The GRBs used to measure pulse residual fluctuations have thus far all been those observed by BATSE and Fermi's GBM experiment. Since these instruments have spectral responses favoring detection of higher energy photons (25 keV to 2 MeV) over lower ones, it is reasonable to wonder whether these residual features are present at lower energies observed by Swift, or if they are primarily a high energy phenomenon. To this end, we examine a sample of single episodic bursts observed by Swift.

\section{Data Analysis Technique}

The Swift GRBs analyzed here were observed between August 1, 2005 and September 1, 2009, and were selected by eye from a sample having measured redshifts as being those having single emission episodes. The initial sample contains 30 GRBs. Similar to BATSE, single-pulsed bursts appear to comprise roughly one-third of all Swift GRBs (one-quarter if the sample is limited to to Long bursts). Light curves are composed of 64 ms, mask-weighted, background-subtracted, 4-channel data available from Swift\footnote{http://gcn.gsfc.nasa.gov/swift\_gnd\_ana.html}. The Swift four-channel data are similar to those used for BATSE and Fermi GBM, but optimized to the Swift energy range (channel 1 $=10-25$ keV, channel 2 $= 25-50$ keV, channel 3 $=50-100$ keV, and channel 4$=100-350$ keV).
 
The technique used for extracting pulses from GRB light curves has been described elsewhere (e.g. \cite{hak14} and references contained therein). Intervals in the light curve where pulses may reside are first identified using the Bayesian Blocks approach of \cite{sca98}, followed by a pulse-fitting technique that fits the empirical model of \cite{nor05} to each interval using a nonlinear least squares routine \citep{mar09}. For BATSE and Fermi GBM data, Poisson statistics are assumed. For Swift data, intensity uncertainties are Gaussian, and this requires a modification of the Bayesian Blocks code (Norris 2014, private communication). The \cite{nor05} pulse model is defined as

\begin{equation}
I(t) = A \lambda e^{[-\tau_1/(t - t_s) - (t - t_s)/\tau_2]},
\end{equation}
where $t$ is time since trigger, and the four Norris pulse fit parameters are the pulse amplitude $A$, the pulse start time $t_s$, the pulse rise parameter $\tau_1$, and the pulse decay parameter $\tau_2$. The normalization constant $\lambda$ is defined in terms of these parameters as $\lambda = \exp{[2 (\tau_1/\tau_2)^{1/2}]}$. Similarly, the pulse peak time occurs at time $\tau_{\rm peak} = t_s + \sqrt{\tau_1 \tau_2}$. In obtaining the pulse fit, the start and end time values of each Bayesian block are used to obtain initial estimates of $t_s$, $\tau_1$, and $\tau_2$, while the maximum amplitude occurring during the block is used {\bf to} estimate $A$.

Observable pulse parameters are defined from the four free pulse fit parameters. We have previously defined the pulse duration $w$ as the time interval between instances at which the intensity is $e^{-3}$ ($4.98\%$) of its maximum value \citep{hak11},
\begin{equation}
w = \tau_2 [9 + 12\mu]^{1/2},
\end{equation}
where $\mu = \sqrt{\tau_1/\tau_2}$.
The pulse {\em asymmetry} $\kappa$ is also defined from the intensities at $e^{-3}$ of the maximum value, and it has a value of
\begin{equation}
\kappa \equiv [1 + 4\mu/3]^{-1/2};
\end{equation}
it ranges from a value of $\kappa=0$ for a symmetric pulse to $\kappa=1$ for an asymmetric pulse with a rapid rise and slow decay. 

The \cite{nor05} model provides a good general fit to BATSE and Fermi GBM pulses, but close examination have identified small yet distinct deviations in the residuals that are phased to the pulse duration \citep{hak14}. These variations have been found to be contained within the fiducial time interval $w_{\rm fid}$ defined by:
\begin{equation}
w_{\rm fid} = \tau_{\rm end} - \tau_{\rm start} = 4.4 \tau_2 (\sqrt{1+\mu/2}+1) + \sqrt{\tau_1 \tau_2},
\end{equation}
with the fiducial end time $t_{\rm end}$ given by 
\begin{equation}
t_{\rm end} = \frac{w}{2} (1+\kappa)+t_s+\tau_{\rm peak}
\end{equation}
and the fiducial start time $t_{\rm start}$ given by
\begin{equation}
t_{\rm start} = t_s - 0.1 [\frac{w}{2} (1+\kappa) - \tau_{\rm peak} ].
\end{equation}

In the fiducial time interval, the residual variations can be fitted with the empirical four parameter function described by \cite{hak14}:

\begin{equation}
\textrm{res($t$)} = \left\{ \begin{array}{ll}
a  J_0( \sqrt{\Omega [t_0 - t - 0.005]}) & \textrm{if $t < t_0 - 0.005$} \\
a & \textrm{if $t_0 - 0.005 \le t \le t_0 + 0.005$} \\
a  J_0(\sqrt{s \Omega [t - t_0 - 0.005]}) & \textrm{if $t > t_0 + 0.005$.}
\end{array} \right. 
\end{equation}
Here, $J_0(x)$ is an integer Bessel function of the first kind, $t_0$ is the central time of the residuals peak, $a$ is the amplitude of the residual peak, $\Omega$ is the variability factor -- the Bessel function's angular frequency -- which defines the timescales of residual wave (a large $\Omega$ corresponds to a rapid rise and fall), and $s$ is a scaling factor that relates the fraction of time which the function before $t_0$ has been compressed relative to its time-inverted form after $t_0$ (or a stretching of the function after the peak time relative to that before it). Using this model, the time during which the pulse intensity is a maximum needs to be a plateau instead of a peak, with a duration of $w_{\rm plateau} \approx 0.010 w_{\rm fid}$. Since there is no evidence that the Bessel function continues beyond the third zero (following the second half-wave), the function is truncated at the third zeros $J_0 (x = \pm 8.654)$.

In case it is not intuitively obvious what roles each of these four free parameters plays in the residual fit, we have included examples in which we vary each parameter individually within the fiducial time interval. The results are shown in Figure 1a (varying $a$), Figure 1b (varying $t_0$), Figure 2a (varying $\Omega$), and Figure 2b (varying $s$). For these residual models we have assumed that the underlying pulse has an asymmetry of $\kappa=0.8$, and that the fiducial timescale is defined to encapsulate a pulse of this asymmetry.

The fiducial values can be converted back to measured values using the following relationships:
\begin{equation}
a_{\rm meas}=a
\end{equation}
\begin{equation}
s_{\rm meas}=s
\end{equation}
\begin{equation}
t_{0; \rm meas}=t_0 (t_{\rm end}-t_{\rm start}) +t_{\rm start}
\end{equation}
and
\begin{equation}
\Omega_{\rm meas}=\Omega/(t_{\rm end}-t_{\rm start}),
\end{equation}
where $t_{\rm start}$ and $t_{\rm end}$ are the real time values corresponding to the start and end of the fiducial duration.

A convenient way to describe the pulse residual amplitudes is to normalize them to the pulse fit amplitudes, because the result is independent of the instrument's signal-to-noise, allowing amplitude comparisons between different instruments. This produces a relative amplitude $R$ defined by
\begin{equation}
R=a/A.
\end{equation} 

\section{Data Analysis}

We apply these pulse-fitting techniques to our sample of 30 Swift GRBs having single emission episodes, with our initial hypothesis being that each emission episode represents a single pulse. We feel justified in making this assumption because the \cite{hak14} results suggest that the residual pattern underlies all isolated pulse structures. This hypothesis is testable; a pulse exhibiting the residual wave structure should have residual wave properties that correlate with the pulse asymmetry in the manner found previously for BATSE pulses. Thus, we fit the Swift 4-channel GRB data using the \cite{nor05} model and the techniques described by \cite{hak11}, but we increase the NCP prior variable (a prior value for the number of change points occurring in the light curve) in the Bayesian Blocks code \citep{sca98} and/or the level of dual timescale peak flux significance \citep{hak11} so that only a single fitted pulse is returned for each GRB in the sample. 

Even though the Swift GRB sample has been selected from what appear to be bright single-pulsed bursts observed during a predetermined time period, many of the light curves are relatively noisy with large uncertainties. This is not surprising, as GRB emission typically decreases at lower energies where background noise increases. As a result, the four-channel pulse fits are generally not as good as those obtained for BATSE and Fermi GBM data, and two bursts (GRB 080129 and GRB 090313) cannot be fitted using the \cite{nor05} pulse model.  The results of the \cite{nor05} fits to the other 28 pulses in the Swift sample are summarized in Table 1.

We attempt to fit the remaining 28 pulses with both the \cite{nor05} model and the \cite{hak14} residual fit model. We use a {\em $\chi^2$ difference test} \citep{ste85} to determine if the fit has been improved with the addition of the \cite{hak14} residual fit. In this test, the difference in the $\chi^2$ fit parameters between the two models $\Delta \chi^2$ is given as
\begin{equation}
\Delta \chi^2=\chi^2_{\rm p} - \chi^2_{\rm p+r}
\end{equation}
where the difference in degrees of freedom $\Delta df$ is
\begin{equation}
\Delta df=df_{\rm p+r}-df_{\rm p}.
\end{equation}
Here `p' refers to the pulse fit and `r' refers to the residual fit. A resulting probability of $p=p(\Delta \chi^2_{\Delta df})$ is obtained for a $\Delta \chi^2$ value having $\Delta df$ degrees of freedom. Pulse plus residual fits having $p \ge 0.001$ do not show significant improvement with the additional model parameters and are excluded, while pulse plus residual fits having $p < 0.001$ are significant enough to be retained.

As shown in Table 2, only eleven pulses in our Swift sample satisfy the $p < 0.001$ criterion: GRB 051111, GRB 060206, GRB 060708, GRB 060912A, GRB 061004, GRB 061021, GRB 070318, GRB 070117, GRB 080413B, GRB 081222, and GRB 091018. The residual fits and combined fits of these pulses are summarized in Table 3 and are shown in Figures 3 through 13 (panels on the left side indicate the fit to the residuals while panels on the right side indicate the overall fit). 

In order to compare the results of the Swift fits with those previously obtained for BATSE and Fermi pulses, we also apply the $\chi^2$ difference test to the pulses listed in \cite{hak14}. Six BATSE pulses (BATSE 0673, BATSE 0711, BATSE 0727, BATSE 1301, BATSE 7567, and BATSE 7775) and one Fermi pulse (GRB 120326A) are not found to exhibit $\Delta \chi^2$ fit improvements and have been subsequently excluded from the analysis.

\subsection{Residual Fit Intensities}

It seems possible that {\em all} GRB pulses might be characterized by a pulse model that includes the same residual wave pattern found in the BATSE data, but that this residual pattern becomes harder to measure for faint pulses. To test this hypothesis on our isolated pulse sample, we check to see if the residual pattern is easier to identify when pulses are bright. We use the signal-to-noise ratio $S/N$ defined as the peak 1 s pulse intensity divided by the peak 1 s pulse intensity uncertainty; these intensities are available for Swift in Table 1 and for BATSE and GBM from published online GRB catalogs\footnote{http://www.batse.msfc.nasa.gov/batse/grb/catalog/4b/4br\_flux.html}\footnote{https://heasarc.gsfc.nasa.gov/W3Browse/fermi/fermigbrst.html}. The $S/N$ values depend on both instrumental sensitivity and instrumental energy response. In Figure 14 we compare our measure of the fit improvement ($-\log[p(\Delta \chi_{\Delta df}^2)]$) to the $S/N$ for each pulse (a large $-\log(p)$ value represents an improved fit). A Spearman rank order test applied to BATSE pulses (crosses) yields a p-value of $4.6 \times 10^{-26}$ that $p$ and $S/N$ are uncorrelated while a similar test applied to Swift pulses (filled circles) yields a p-value of $5.5 \times 10^{-10}$.  The trend also appears to hold for the three Fermi pulses (open diamonds). These results demonstrate that the fits improve significantly as the signal-to-noise increases, which is consistent with our hypothesis. 

The pulses not substantially improved by the residual fits also support the hypothesis that the residual wave is present in all isolated pulses: these unimproved pulses are typically among the faintest pulses in the sample. The mean $S/N$ of the pulses whose fits have been improved ($p < 0.001$) is $\langle S/N \rangle = 25.8$ while the the mean $S/N$ of the unimproved pulses ($p \ge 0.001$) is $\langle S/N \rangle = 17.4.$ A Student's t-test indicates only a $4\%$ likelihood that these distributions have the same mean.

The characteristics of the wave patterns extracted from the improved Swift GRB pulse residuals are found to be similar to those of BATSE GRB pulse residuals in that they exhibit similar correlations with pulse asymmetry $\kappa$. Figure 15 demonstrates the relationship between the variability parameter $\Omega$ with $\kappa$, Figure 16 demonstrates the relationship between the expansion parameter $s$ and $\kappa$, and Figure 17 demonstrates the relationship between the fiducial residual peak time $t_0$ and $\kappa$.

These correlations show a clear linkage between the shape of the episode's bulk emission and the more rapidly-varying component contained within the episode. Since the peak of the residuals $t_0$ occurs close in time to the fitted \cite{nor05} pulse peak $\tau_{\rm peak}$, both $t_0$ and $\tau_{\rm peak}$ correlate with $\kappa$. Since the residual peaks are closely associated with pulse rise and decay, the stretching parameter $s$ correlates with $\kappa$. And, since an asymmetric pulse has a short rise time relative to a longer decay time, it has a large variability and therefore a large $\Omega$ to go with a large $\kappa$.

There is no observed correlation between $R$ and $\kappa$, nor is there any clear correlation between $R$ and $S/N$.  Figure 18 shows that the mean value of $R$ is $\langle R\rangle \approx 0.23$, which indicates that the residual peak intensity $a$ is generally $23\%$ of the fitted peak intensity $A$; this is true for all pulses in which the residuals can be fitted, regardless of the instrument making the measurements. Pulse residuals describing the pulse shape are just as identifiable in symmetric pulses as they are in asymmetric pulses.  This does not mean that symmetric pulses are as easy to observe as asymmetric pulses: the fitted pulse amplitude $A$ has been found to correlate with $\kappa$ for BATSE pulses \citep{hak11}, indicating that asymmetric pulses are generally brighter than symmetric pulses.

The residual pattern is not only easier to detect in bright pulses; it is also easier to detect at higher energies, in pulses with intrinsically larger $R=a/A$ values, and in asymmetric pulses ($\kappa > 0.4)$. Figure 19 contrasts the characteristics of pulses that have not been improved by the residual fits ($p > 0.001$; in blue) with the characteristics of pulses that have been improved by the residual fits ($p \le 0.001$; in black). Improved Swift fits generally require higher $S/N$ ratios than improved BATSE fits. Since BATSE and Swift pulses have similar minimum $S/N$ ratios indicative of comparable instrumental sensitivities, this difference between residual intensities most likely results from the different spectral responses of the instruments, and suggests that the residual structures are more apparent at high energies than at low energies. The left panel of Figure 19 demonstrates an anti-correlation between the normalized residual amplitude $R=a/A$ and the signal-to-noise $S/N$: faint pulses with small $S/N$ ratios tend to have larger values of $R$. Unimproved pulses near the threshold tend to balance this anti-correlation somewhat (this is especially true for BATSE pulses) by having smaller $R$-values. The right panel of Figure 19 demonstrates that symmetric pulses tend to be fainter and rarer than asymmetric pulses; most of the symmetric pulses in the sample have low $S/N$ ratios.

Overall, we find that eleven out of thirty isolated mask-weighted Swift GRB pulses are improved by the addition of the residual model. These pulses exhibit complex shapes consistent with those detected in BATSE and Fermi data, compared to 43 out of 50 BATSE GRB pulses and two out of three Fermi GRB pulses. 
When the potential model improvement resulting from the residual fits is considered as functions of $S/N$ (Figure 14), $R$ (Figure 19, left panel), and $\kappa$ (Figure 19, right panel), it appears that residual curves may be present in all isolated pulses, but are harder to identify in faint pulses. It also appears that the residual features are harder to identify at Swift's lower energies, even though Swift and BATSE have similar instrumental sensitivities: {\em a reasonable interpretation is that the residual features are less pronounced at lower energies.} Finally, residuals are easier to extract when pulses have large normalized residual peaks ($R=a/A$); since symmetric pulses seem to have smaller $R$ values than asymmetric pulses, this suggests that there might be an observational bias favoring asymmetric pulses.

\subsection{Pulse Spectral Evolution}

Figures 20 through 25 summarize the hardness evolutions of the eleven Swift GRB pulses having measurable pulse shapes. The Swift hardness is defined to be the ratio of counts in the 50 to 300 keV energy range relative to the counts in the 15 to 50 keV energy range. In order to explore how hardness changes with pulse shape, each hardness plot is marked with the five maxima and minima in the residual light curve: the precursor peak (downward facing red arrow), the dip between the precursor peak and the central peak (upward facing red arrow), the central plateau or peak (downward facing black arrow), the dip between the central peak and the decay peak (upward facing blue arrow), and the decay peak (downward facing blue arrow). These figures demonstrate the same general hard-to-soft evolution that characterizes BATSE and GBM GRB pulses: the pulse rise phase is spectrally harder than the pulse decay phase. Furthermore, there seems to be a relationship between the times at which pulses harden and the three intensity peaks. For example, two pulses (GRB 061004 and GRB 080413B) and possibly two others (GRB 070318 and GRB 091018) have hardnesses that start soft and increase (reaching maximum hardness at roughly the time of the central peak) before subsequently decreasing in the normal fashion; it is interesting to note that these three or four pulses are among the softest in the sample. Additionally, many of the pulses either undergo a brief re-hardening near the time of the decay peak: this is easily noticed for hard pulses GRB 051111, GRB 060912A, and GRB 061021 yet harder to discern for soft pulses. Table 4 summarizes the visual pulse structure as it relates to the maximum pulse hardness for Swift pulses.

We wonder if similar coupled spectral and temporal behaviors exist for the previously-analyzed GBM and BATSE pulses, and therefore we have expanded the time periods of all plots from \cite{hak14} to correspond to the fiducial duration. Even though this redistribution of counts might result in low signal-to-noise, we recognize that the hardness evolution may be important over the entire life cycle of a GRB pulses, and particularly in the time intervals corresponding to the residual wave features (for this analysis, we note that the hardness ratio used BATSE data is the ratio of counts in the 100 keV to 1 MeV energy range relative to the counts in the 25 to 100 keV energy range). Individual pulse results for forty BATSE pulses having well-measured hardness evolutions have been obtained for the entire fiducial timescale \citep{hak14}.

The hardness-related temporal structures for the two measurable GBM pulses are summarized in Table 5, while the 44 measurable BATSE pulse properties are shown in Table 6. Tables 4, 5, and 6 demonstrate interesting similarities between Swift, GBM, and BATSE GRB pulse evolutions: 1) hard pulses tend to have hard precursor peaks relative to the central peak, while soft pulses do not, and 2) hard pulses typically demonstrate a re-hardening of the decay peak, while soft pulses do not. These results appear to be generally valid even though the energy ranges of the experiments are different. Thus, the complex pulse shapes uncovered for BATSE GRB pulses are apparently ubiquitous, and span a larger energy range than just of that detected by BATSE.

We note the rather remarkable result that the hardness evolution seems to differ for hard pulses and soft pulses. Specifically, hard pulses seem more likely to evolve hard-to-soft while soft pulses are less likely to show this behavior. Furthermore, the times at which pulse evolution changes its decay rate appear linked to the times of the precursor peak, the central peak, and the decay peak. 

To explore in greater detail the hypothesis that pulse spectral evolution is associated with the residual structures, we examine the hardness evolution of BATSE pulses by grouping together those with similar maximum hardness ratios. Since the BATSE pulses in this sample all have different amplitudes, durations, and background count rates, it is difficult to average their time-dependent hardnesses in a reasonable yet unbiased way. The approach we choose is to sum together the 64 ms photon counts found in like parts of the fiducial window for similar pulses, then ratio these in different energy channels to obtain an evolving counts hardness ratio.

Figure 26 indicates the resulting typical hardness evolution of the BATSE GRB pulses, sorted into five groups using individual maximum hardness ratios. Hardnesses are taken by dividing the summed high energy counts (100 keV to 1 MeV) in each partition by the summed low energy counts (25 keV to 100 keV). Hardness groups are HR$_{\rm max} \ge 1.6$ (dark blue), $1.1 \le $HR$_{\rm max} < 1.6$ (light blue), $0.8 \le $HR$_{\rm max} < 1.1$ (dark green), $0.6 \le $HR$_{\rm max} < 0.8$ (light green), and HR$_{\rm max} \le 0.6$ (orange). Similar to the arrows shown in Figures 20 through 25,
the down-facing arrows indicate the averaged fiducial positions of the precursor peak, the central peak, and the decay peak, while the up-facing arrows indicate the averaged fiducial positions of the gaps in the residual function between the peaks. 

{\em This figure demonstrates that all GRB pulse groups evolve from hard to soft in that they are softer at the end of their fiducial time intervals than at the beginning; in this sense all GRB pulses evolve from hard to soft}.  Within this general hard-to-soft evolution, it is also true that GRB pulse hardness is reinvigorated at the times of the three pulse peaks.  In particular, the precursor peak seems to have a profound influence on the subsequent pulse evolution, since the pulses with the largest maximum hardness ratios have the most abrupt hard-to-soft evolutions and are hardest during the precursor peak, while those with the smallest maximum hardness ratios reach these values during the central peak. There is weaker evidence in {Figure 25} for a decrease in the hardness decay during the decay peak than there is for the other two peaks.

\cite{hak11} have shown that hardness evolution is a function of pulse asymmetry: the short rise phases of asymmetric pulses cause them to have shorter lags than their symmetric counterparts, and lags are indicative of the rate of hardness decay. In this study we have found that pulses with larger maximum hardnesses also decay faster, suggesting that maximum hardness should correlate with asymmetry. Figure 27 demonstrates that this correlation indeed exists and is in fact fairly robust: a Spearman Rank Order test yields a p-value of only $p=6.1 \times 10^{-4}$ that these parameters are uncorrelated.

Since asymmetry correlates with peak hardness, we expect to see that BATSE pulses grouped by asymmetry should exhibit hardness evolution that is also related to the timing of the three pulse peaks. Figure 28 demonstrates the effects of this grouping in a manner similar to Figure 26, but using $\kappa$ as the key parameter rather than maximum hardness. Asymmetry groups are $\kappa \ge 0.85$ (dark blue), $0.75 \le \kappa < 0.85$ (light blue), $0.65 < \kappa < 0.75$ (dark green), $0.45 \le \kappa < 0.65$ (light green), and $\kappa \le 0.45$ (orange). The same general hard-to-soft evolution can be seen in the figure, although grouping the pulses by asymmetry may cause the precursor peak to be weaker than it is in Figure 26, but the rapid decay starting from the central peak is more pronounced. 

Maximum hardness is potentially a fragile classification characteristic, as is depends on how the hardnesses are binned in time (e.g. bin size, quantity, and bin center time). To assure ourselves that the correlation of pulse evolution with hardness is robust, we group the pulses by hardness (HR$_S$) as measured from the ratio of the published energy fluences ($S_n$, where $n$ represents some combination of BATSE energy channels):

\begin{equation}
{\rm HR}_{43/21}=\frac{S_4+S_3}{S_2+S_1.}
\end{equation}
The results, with hardness groups ${\rm HR}_{43/21} \ge 5.0$ (dark blue), $2.75 \le {\rm HR}_{43/21} < 5.0$ (light blue), $2.0 \le {\rm HR}_{43/21} < 2.75$ (dark green), $1.0 \le {\rm HR}_{43/21} < 2.0$ (light green), and ${\rm HR}_{43/21} < 1.0$ (orange), are shown in Figure 29. These hardness groupings are entirely consistent with those shown in Figure 26. The fact that they are also consistent with those shown in Figure 28 demonstrates that GRB pulse evolution has a similar functional dependence on both spectral hardness and asymmetry.

\section{Discussion}

Isolated Swift GRB pulses, like their BATSE and GBM counterparts, exhibit three-peaked pulse shapes even though Swift observes the sky at lower energies than BATSE or GBM. It is becoming increasingly clear that GRB pulse structure, particularly at high signal-to-noise, cannot be adequately modeled by simple monotonically increasing and decreasing functions, and spectrally-dependent variations must be treated as an integral part of each pulse. 

Isolated pulses generally evolve from hard to soft, although re-hardening often occurs coincident with each of the three peaks. Hard pulses are more likely to have a hard precursor peak and exhibit hard-to-soft evolution throughout. Soft pulses are more likely to have a soft precursor peak, causing them to reach maximum hardness at the time of the central pulse peak. All isolated pulses are capable of having an additional hardening at the time of the decay peak; this hardening is most often seen as a change in the rate of hardness decay. {\em These results strongly suggest that the commonly-used nomenclature of `hard-to-soft' and `intensity tracking' pulses is not correct; all isolated GRB pulses exhibit both behaviors simultaneously and to different degrees.}

The re-hardening of GRB pulse spectra coincident with the three temporal peaks suggests that a single-component evolving spectral model might be insufficient for producing characteristics of isolated GRB pulses: it is reasonable to suspect that three separate but overlapping evolving components might exist. \cite{pre15} have recently demonstrated through modeling that overlapping hard-to-soft pulses produce a composite hard-to-soft pulse exhibiting intensity tracking behaviors at the peaks. If each of the three pulse components in isolated GRB pulses evolves hard-to-soft, then our simple residual model is inadequate for describing these because it assumes that the first and third peaks are mirror images of one another, whereas the hardness evolution of these events would not be time-symmetric. More detailed, physical models are needed to understand the true nature of the underlying complex GRB pulse structure.

The standard model for GRB prompt spectral emission is kinematic energy injection into a medium via collisionless relativistic shocks; the preferred mechanism by which radiative energy is thought to be released is in the form of a synchrotron spectrum (e.g.~\cite{ree94}). The physical internal shock model of \cite{kin04}, naturally accounts for three separate but temporally overlapping emission regions via a forward shock, a reverse shock, and the region containing the contact discontinuity, and this model may provide a physical basis for the underlying physical structures. The model also predicts that the shock interactions overlap to a considerable extent, which can make them difficult to unambiguously observed separately. Kinematic models such as this one need to be merged with multi-component evolving spectral models, as recent multiple component spectral models (e.g. \cite{bur11}), which thus far have not addressed the possibility that spectral evolution might be associated with specific light curve elements within an individual pulse. \cite{bol14} have considered the radiation mechanisms of kinematically evolving synchrotron shocks, but find that the model better explains observed pulse spectral evolution if (i) the distribution of shock-accelerated electrons is steeper than generally assumed, (ii) the peak energy dependence is reduced by altering microphysics parameters, (iii) the Lorentz factor undergoes more pronounced initial variations than previously thought, and (iv) the relativistic ejection proceeds with a constant mass flux rather than a constant kinetic energy flux. \cite{bol14} indicate that these conditions are hard to justify, but that their ability to test microphysics conditions are limited by existing data. Further modeling is needed to see under what conditions the synchrotron shock model is consistent with observations.

Single-episodic models of GRB pulse shapes, such as those in which the asymmetric pulse shape is produced by curvature in a relativistic outflow (e.g. \cite{fen96, rrf00, qin02, der04, wil10}) predict neither the multi-episodic shapes nor the spectral rehardenings observed here. These models likely need significant modifications in order to explain the observations.

Complex GRB pulses containing many more than three peaks may be simply the result of observing overlapping light curves from several isolated pulses. However, it is also possible that there are types of GRB pulses different than those described here. {\em This latter hypothesis cannot even be adequately explored without first having a working definition of a `gamma-ray burst pulse' such as has been summarized here for isolated GRB pulses.}

The results obtained in this study allow us to make general predications about the spectral evolution of GRB-like pulses at other intrinsic energies. For example, low-energy bursting events have been observed by transient experiments and include XRFs (x-ray flashes) and XRRs (x-ray rich GRBs): these appear to be low-energy counterparts to GRBs (e.g. \cite{sak05}). Similarly, the correlated behaviors of x-ray flares (e.g. \cite{chi10, mar10}) and optical flares  (e.g. \cite{li12}) observed in GRB afterglows suggest that these are simply late, low-energy GRB pulses. {\em The correlated pulse behaviors identified in this manuscript suggest that low-energy pulses should exhibit time histories dominated by intensity tracking spectral behaviors, with a soft precursor peak and potentially a soft or non-existent decay peak.} We are currently exploring these possibilities in follow-up studies involving x-ray data and optical/infrared observations using Etelman Observatory at the University of the Virgin Islands \citep{mor14}.

\section{Conclusions}

In this paper and two prior ones \citep{hak11, hak14} we have re-conceived the observational definition of an isolated gamma-ray burst pulse. We have been able to do this because isolated GRB pulses have complex but well-defined and easily-replicated characteristics. Pulse spectral and temporal evolutions are intertwined in a way suggesting that the underlying physics is orderly and reproducible, even though titanic energies are unleashed through each GRB in a relatively short period of time. 

This study finds that Swift GRB pulses exhibit the same general characteristics as BATSE GRB pulses: individual pulse light curves are characterized by triple-peaked profiles that generally evolve from hard to soft. The three peaks (the precursor peak, the central peak, and the decay peak) are aligned to coincide with the asymmetry of the pulse shape. The precursor and decay peaks can exhibit separate spectral evolution from that of the central peak, causing some pulses to follow intensity tracking evolution overlaying the natural hard-to-soft evolution. The specific form that the evolution takes depends on the maximum spectral hardness of the pulse: hard pulses begin hard, and the overall hardness decays rapidly throughout the pulse. However, during the decay, a re-hardening can occur as the decay peak intensity increases. Pulses of intermediate hardness have precursor peaks having hardnesses similar to those of the central peak which decay at a moderate rate. Soft pulses can begin soft and harden at the central peak, after which the overall hardness decays slowly throughout the pulse. The hardness of any pulse can remain constant or even increase during the time leading up to the decay peak. {\em As a result, `hard-to-soft' and `intensity matching' GRB pulse definitions no longer appear to be valid, since all isolated pulses show both of these behaviors in ways that depend upon asymmetry and/or maximum hardness.}

There is a strong interplay between a GRB pulse's hardness and the symmetry of its temporal evolution: asymmetric pulses are generally hard and show a strong hard-to-soft evolution. Symmetric pulses are more often soft and exhibit intensity tracking spectral evolution. A pulse's duration indicates the timescale over which a system is radiating; this can be due to radiative cooling and/or to the timescale over which kinematic events occur. A pulse's asymmetry and hard-to-soft evolution indicate that the rate of cooling is changing and/or that the kinematic events have more energy to release initially than at later times.  The triple-peaked nature and spectral re-hardening of isolated GRB pulses seems to indicate that energy is injected on as many as three separate occasions during the pulse duration. Theoretical models of GRB prompt emission need to account for these observed pulse spectrotemporal characteristics.

\section{Acknowledgements}
We would like to acknowledge the helpful comments of the anonymous referee, as these greatly helped to improve the manuscript. This work has been supported by NASA EPSCoR grant NNX13AD28A.

\clearpage

\begin{deluxetable}{lrrrrrrrrrrrrrrrr}
\tabletypesize{\scriptsize}
\rotate
\tablecaption{GRB pulses used in this analysis.\label{tbl-1}}
\tablewidth{0pt}
\tablehead{
\colhead{Pulse ID} & \colhead{$w$} & \colhead{$\sigma_w$} & \colhead{$\kappa$} &
\colhead{$\sigma_\kappa$} & \colhead{$p_{\rm 1024}$} & \colhead{$\sigma_{\rm p1024}$} &
%\colhead{$B$} & \colhead{$\sigma_{\rm B}$} & \colhead{$S_B$} & \colhead{$\sigma_{\rm S_B}$} &
\colhead{$t_s$} & \colhead{$\sigma_{t_s}$} & \colhead{$A$} & \colhead{$\sigma_A$} &
\colhead{$\tau_1$} & \colhead{$\sigma_{\tau_1}$} & \colhead{$\tau_2$} & \colhead{$\sigma_{\tau_2}$} &
\colhead{$\chi^2_\nu$} & \colhead{$\nu$}
}
\startdata

GRB 050801 & 5.8 & 4.1 & 0.999 & 0.100 & 1.46 & 0.09 & 0.08 & 0.12 & 0.21 & 1.35 & 2.92E-06 & 1.50E-02 & 1.94E+00 & 2.12E-01 & 1.00 & 1556 \\
GRB 050908 & 19.8 & 56.0 & 0.053 & 0.286 & 0.70 & 0.09 & -77.87 & 134.36 & 0.06 & 0.01 & 8.15E+03 & 4.10E+04 & 7.91E-01 & 1.32E+00 & 1.01 & 6233 \\
GRB 051111 & 43.4 & 1.9 & 0.708 & 0.032 & 2.66 & 0.13 & -7.88 & 0.36 & 0.28 & 0.01 & 5.69E+00 & 1.03E+00 & 1.02E+01 & 4.91E-01 & 1.02 & 6243 \\
GRB 060206 & 8.3 & 1.4 & 0.209 & 0.064 & 2.79 & 0.10 & -7.18 & 1.95 & 0.35 & 0.01 & 1.55E+02 & 9.76E+01 & 5.75E-01 & 1.19E-01 & 1.01 & 6243 \\
GRB 060512 & 18.3 & 111.3 & 0.032 & 0.373 & 0.88 & 0.12 & -140.92 & 1103.85 & 0.06 & 0.01 & 1.06E+05 & 2.44E+06 & 1.94E-01 & 1.50E+00 & 0.95 & 697 \\
GRB 060708 & 9.2 & 1.2 & 0.518 & 0.099 & 1.94 & 0.09 & -1.85 & 0.51 & 0.18 & 0.01 & 6.63E+00 & 3.43E+00 & 1.58E+00 & 2.54E-01 & 1.11 & 774 \\
GRB 060912A & 5.1 & 0.3 & 0.750 & 0.042 & 8.58 & 0.27 & -0.44 & 0.05 & 0.97 & 0.03 & 4.28E-01 & 1.09E-01 & 1.26E+00 & 7.68E-02 & 1.08 & 774 \\
GRB 060926 & 9.7 & 1.5 & 0.840 & 0.094 & 1.09 & 0.09 & -0.25 & 0.13 & 0.13 & 0.01 & 2.63E-01 & 2.00E-01 & 2.71E+00 & 4.55E-01 & 1.06 & 618 \\
GRB 061004 & 7.1 & 2.3 & 0.161 & 0.099 & 2.54 & 0.10 & -7.49 & 4.45 & 0.26 & 0.01 & 3.04E+02 & 3.79E+02 & 3.83E-01 & 1.58E-01 & 1.04 & 6243 \\
GRB 061021 & 11.5 & 0.4 & 0.844 & 0.024 & 6.11 & 0.16 & 1.66 & 0.04 & 0.66 & 0.01 & 2.95E-01 & 5.85E-02 & 3.23E+00 & 1.19E-01 & 1.17 & 3118 \\
GRB 070318 & 51.1 & 2.2 & 0.921 & 0.026 & 1.76 & 0.09 & -0.83 & 0.10 & 0.17 & 0.00 & 2.84E-01 & 1.04E-01 & 1.57E+01 & 7.14E-01 & 1.01 & 4681 \\
GRB 070506 & 9.2 & 10.5 & 0.114 & 0.245 & 0.96 & 0.08 & -13.75 & 28.46 & 0.09 & 0.01 & 1.14E+03 & 4.87E+03 & 3.52E-01 & 5.02E-01 & 1.00 & 3899 \\
GRB 070611 & 18.6 & 327.4 & 0.012 & 0.413 & 0.82 & 0.13 & -377.70 & 8441.82 & 0.06 & 0.01 & 1.90E+06 & 1.26E+08 & 7.60E-02 & 1.69E+00 & 1.04 & 696 \\
GRB 070810A & 13.1 & 2.3 & 0.444 & 0.119 & 1.90 & 0.12 & -5.10 & 1.24 & 0.18 & 0.01 & 1.82E+01 & 1.20E+01 & 1.94E+00 & 4.04E-01 & 1.01 & 9986 \\
GRB 071010B & 18.3 & 0.4 & 0.624 & 0.017 & 7.70 & 0.18 & -2.40 & 0.11 & 0.85 & 0.01 & 5.28E+00 & 4.63E-01 & 3.81E+00 & 9.47E-02 & 1.08 & 4680 \\
GRB 071031 & 44.7 & 12.2 & 0.585 & 0.204 & 0.50 & 0.06 & -10.89 & 3.87 & 0.04 & 0.00 & 1.82E+01 & 1.90E+01 & 8.70E+00 & 2.79E+00 & 1.00 & 3119 \\
GRB 071117 & 5.4 & 0.2 & 0.659 & 0.029 & 11.3 & 0.4 & -0.80 & 0.05 & 1.44 & 0.02 & 1.13E+00 & 1.72E-01 & 1.19E+00 & 4.84E-02 & 1.10 & 462 \\
GRB 080129 &  &  &  &  & 0.50 & 0.12  \\
GRB 080413B & 4.8 & 0.3 & 0.450 & 0.040 & 18.70 & 0.49 & -1.94 & 0.15 & 2.07 & 0.04 & 6.33E+00 & 1.39E+00 & 7.25E-01 & 4.90E-02 & 1.03 & 3274 \\
GRB 080430 & 14.8 & 0.8 & 0.725 & 0.040 & 2.60 & 0.12 & -0.94 & 0.14 & 0.32 & 0.01 & 1.65E+00 & 3.80E-01 & 3.58E+00 & 2.18E-01 & 1.05 & 1087 \\
GRB 080710 & 23.0 & 5.0 & 0.853 & 0.132 & 1.00 & 0.12 & -0.70 & 0.39 & 0.08 & 0.01 & 5.13E-01 & 5.81E-01 & 6.56E+00 & 1.50E+00 & 0.99 & 2337 \\
GRB 080804 & 36.4 & 2.8 & 0.781 & 0.060 & 3.10 & 0.24 & -1.64 & 0.41 & 0.28 & 0.01 & 2.17E+00 & 8.45E-01 & 9.47E+00 & 8.12E-01 & 0.98 & 2337 \\
GRB 080805 & 74.4 & 5.4 & 0.787 & 0.043 & 1.10 & 0.06 & -5.62 & 0.57 & 0.11 & 0.00 & 4.16E+00 & 1.18E+00 & 1.95E+01 & 1.57E+00 & 1.02 & 2337 \\
GRB 081007 & 9.9 & 1.5 & 0.672 & 0.117 & 2.60 & 0.24 & -0.69 & 0.35 & 0.27 & 0.02 & 1.85E+00 & 1.17E+00 & 2.23E+00 & 3.84E-01 & 1.01 & 2337 \\
GRB 081118 & 98.4 & 41.0 & 0.276 & 0.198 & 0.60 & 0.12 & -38.57 & 40.86 & 0.04 & 0.00 & 7.52E+02 & 1.15E+03 & 9.05E+00 & 4.70E+00 & 0.99 & 3274 \\
GRB 081222 & 15.6 & 0.5 & 0.453 & 0.023 & 7.70 & 0.12 & -3.23 & 0.26 & 0.94 & 0.01 & 1.98E+01 & 2.48E+00 & 2.35E+00 & 9.24E-02 & 1.51 & 1087 \\
GRB 090313 &  &  &  &  &  0.80 & 0.18  \\
GRB 090809A & 11.9 & 11.1 & 0.191 & 0.331 & 1.10 & 0.12 & -12.14 & 17.55 & 0.10 & 0.01 & 2.97E+02 & 1.05E+03 & 7.54E-01 & 8.86E-01 & 0.97 & 696 \\
GRB 090927 & 3.0 & 0.5 & 0.920 & 0.131 & 2.00 & 0.12 & 0.01 & 0.02 & 0.31 & 0.04 & 1.71E-02 & 3.18E-02 & 9.35E-01 & 1.57E-01 & 1.08 & 2337 \\
GRB 091018 & 5.4 & 0.3 & 0.399 & 0.034 & 10.30 & 0.24 & -1.55 & 0.18 & 1.14 & 0.02 & 1.11E+01 & 2.24E+00 & 7.13E-01 & 4.52E-02 & 1.03 & 2336 \\
\enddata
%% Text for table notes should follow after the \enddata but before
%% the \end{deluxetable}. Make sure there is at least one \tablenotemark
%% in the table for each \tablenotetext.
\tablecomments
%\caption
{Columns: (1) Burst ID, (2) pulse duration $w$, (3) duration uncertainty $\sigma_w$, (4) asymmetry $\kappa$, (5) asymmetry uncertainty $\sigma_\kappa$,  (6) peak flux $p_{\rm 1024}$ ($cm^{-2} s^{-1}$), (7) peak flux uncertainty $\sigma_{p_{\rm 1024}}$ ($cm^{-2} s^{-1}$), (8) start time $t_s$ (s), (9) start time uncertainty $\sigma_{t_s}$ (s), (10) amplitude $A$, (11) amplitude uncertainty $\sigma_A$ (s), (12) rise parameter $\tau_1$, (13) rise parameter uncertainty $\sigma_{\tau_1}$, (14) decay parameter $\tau_2$, (15) decay parameter uncertainty $\sigma_{\tau_2}$, (16) $\chi^2_\nu$ per degree of freedom $\nu$, and (17) degrees of freedom $\nu$. GRB080129 and GRB 090313 could not be fitted.}
\end{deluxetable}

\clearpage

\begin{table}
\scriptsize
\begin{center}
\caption{The goodness-of-fit ($\chi_\nu^2$) parameters for the \cite{nor05} model (p) and for the combined \cite{nor05} plus \cite{hak14} model (p+r), using the number of degrees of freedom ($\nu$) found for data constrained to the fiducial time interval $t_{\rm start}$ to $t_{\rm end}$, as well as the $\Delta \chi^2$ p-value for $\Delta df$ (here 4) degrees of freedom. Improvement is only considered to be significant if $p < 0.001$. GRB 080129 and GRB 090313 could not be fitted.\label{tbl-2}}
\begin{tabular}{lrrrrrrcc}
\tableline\tableline
GRB pulse & $t_{\rm start}$ & $t_{\rm end}$ & $\nu_{\rm p}$ & $\chi^2_{\nu; \rm p}$ & $\nu_{\rm p+r}$ & $\chi^2_{\nu; \rm p+r}$ & p for $\Delta \chi^2$ & Significant?\\
\tableline
GRB 050801 & -1.501 & 15.890 & 1.00 & 267 & 1.00 & 263 & $1.0$ & \\
GRB 050908 & -80.452 & 28.326& 0.98 & 1695 & 0.97 & 1691 &  $2.7 \times 10^{-2}$ & \\
GRB 051111 & -16.495 & 84.408 & 1.04 & 1572 & 1.01 & 1568 & $2.0 \times 10^{-10}$ & yes\\
GRB 060206 & -8.108 & 11.548 & 1.14 & 303 & 1.07 & 299 & $3.3 \times 10^{-5}$ & yes\\
GRB 060512 & -142.254 & 15.719 & 0.95 & 2504 & 0.95 & 2500 & $9.4 \times 10^{-1}$ & \\
GRB 060708 & -3.384 & 16.702 & 1.34 & 308 & 1.17 & 304& $1.2 \times 10^{-12}$ & yes\\
GRB 060912A & -1.519 & 11.104 & 1.17 & 187 & 0.93 & 183 & $1.2 \times 10^{-9}$ & yes\\
GRB 060926 & -2.495 & 23.078 & 1.12 & 394 & 1.12 & 390 & 1.0 &  \\
GRB 061004 & -8.235 & 10.784 & 1.58 & 235 & 1.21 & 231 & $3.1 \times 10^{-21}$ & yes \\
GRB 061021 & -0.760 & 26.887 & 1.87 & 427 & 1.44 & 423 & $5.2 \times 10^{-40}$ & yes \\
GRB 070318 & -13.591 & 128.84 & 1.08 & 2221 & 1.06 & 2217 & $6.9 \times 10^{-9}$ & yes \\
GRB 070506 & -14.649 & 15.275 & 1.02 & 463 & 1.02 & 459 & $6.4 \times 10^{-1}$ &  \\
GRB 070611 & -379.247 & 17.260 & 1.00 & 4013 & 1.00 & 4009 & 1.0 &   \\
GRB 070810A & -7.116 & 20.954 & 0.99 & 463 & 0.98 & 459 & $3.0 \times 10^{-1}$ &  \\
GRB 071010B  & -5.841 & 36.544 & 1.19 & 621 & 1.18 & 617 & $1.8 \times 10^{-1}$ &  \\
GRB 071031 & -18.935 & 82.172 & 1.00 & 1546 & 1.00 & 1542 & $5.9 \times 10^{-1}$ &  \\
GRB 071117 & -1.857 & 10.901 & 1.36 & 194 & 1.31 & 190 & $4.8 \times 10^{-3}$ & yes  \\
GRB 080129 & & & & & & & &  \\
GRB 080413B  & -2.668 & 7.669 & 1.94 & 156 & 1.45 & 152 & $8.1 \times 10^{-17}$ & yes \\
GRB 080430 & -4.023 & 32.346 & 1.05 & 563 & 1.05 & 559 & $4.9 \times 10^{-1}$ &  \\
GRB 080710 &-6.119 & 55.364 & 1.03 & 956 & 1.03 & 952 & $2.6 \times 10^{-1}$ &  \\
GRB 080804 & -9.643 & 82.956 & 1.01 & 1442 & 1.00 & 1438 & $6.3 \times 10^{-1}$ &  \\ 
GRB 080805 & -22.089 & 168.045 & 1.06 & 2965 & 1.05 & 2961 & $6.6 \times 10^{-1}$ &  \\
GRB 081007 & -2.651 & 20.996 & 1.01 & 365 & 1.00 & 361 & $2.2 \times 10^{-1}$ &  \\
GRB 081118 & -46.324 & 121.500 & 1.00 & 3372 & 1.00 & 3368 & 1.0 &  \\
GRB 081222 & -5.638 & 27.706 & 2.14 & 516 & 1.85 & 512 & $4.7 \times 10^{-33}$ & yes \\
GRB 090313 & & & & & & & &  \\
GRB 090809A & -13.438 & 15.806 & 1.01 & 452 & 1.01 & 450 & 1.0 &   \\
GRB 090927 & -0.750 & 7.742 & 1.10 & 128 & 1.04 & 124 & $1.5 \times 10^{-2}$ &  \\
GRB 091018 & -2.324 & 9.042 & 1.49 & 172 & 1.15 & 168 & $1.2 \times 10^{-12}$ & yes \\
\tableline
\end{tabular}
\end{center}

\normalsize
\end{table}

\clearpage

\begin{table}

\scriptsize

\begin{center}
\caption{GRB pulse residual fit parameters. Results are shown for pulses significantly improved by the combined \cite{nor05} plus \cite{hak14} model (p+r) described in Table 2. Parameters are the pulse start time $t_0$ (column 1) having uncertainty $\sigma_{t_0}$ (column 2), residual amplitude $a$ (column 3) having uncertainty $\sigma_a$ (column 4), pulse variability $\Omega$ (column 5) having uncertainty $\sigma_\Omega$ (column 6), and stretching parameter $s$ (column 7) having uncertainty $\sigma_s$ (column 8). Also included is the ratio of the residual amplitude to the fitted amplitude R (column 9).\label{tbl-3}}
\begin{tabular}{lrrrrrrrrrrrrr}
\tableline
\tableline
GRB pulse & $t_0$ & $\sigma_{t_0}$ & $a$ & $\sigma_a$ & $\Omega$ & $\sigma_\Omega$ & $s$ & $\sigma_s$  & R \\
\tableline
GRB 051111 & -0.01 & 0.56 & 0.053 & 0.003 & 9.5 & 0.9 & 0.18 & 0.02 & 0.19 \\
GRB 060206 & 2.93 & 0.12 & 0.047 & 0.004 & 9.7 & 0.8 & 1.00 & 0.13 & 0.13 \\
GRB 060708 & 1.55 & 0.04 & 0.097 & 0.005 & 22.1 & 1.4 & 0.53 & 0.04 & 0.55 \\
GRB 060912A & 0.33 & 0.02 & 0.189 & 0.010 & 73.0 & 10.3 & 0.15 & 0.02 & 0.20 \\
GRB 061004 & 2.97 & 0.05 & 0.098 & 0.004 & 21.9 & 1.4 & 0.53 & 0.04 & 0.38 \\
GRB 061021 & 2.78 & 0.04 & 0.186 & 0.007 & 32.2 & 2.4 & 0.19 & 0.02 & 0.28 \\
GRB 070318 & 2.52 & 0.18 & 0.031 & 0.002 & 11.9 & 2.0 & 0.11 & 0.02 & 0.19 \\
GRB 071117 & 0.61 & 0.10 & 0.090 & 0.020 & 54.8 & 16.3 & 0.70 & 0.29 & 0.06 \\
GRB 080413B  & 0.40 & 0.02 & 0.423 & 0.020 & 35.8 & 2.5 & 0.37 & 0.03 & 0.20 \\
GRB 081222 & 4.10 & 0.08 & 0.113 & 0.005 & 18.0 & 1.7 & 0.20 & 0.02 & 0.12 \\
GRB 091018 & 1.02 & 0.04 & 0.146 & 0.008 & 35.6 & 3.1 & 0.31 & 0.035 & 0.13 \\
\tableline
\end{tabular}
\end{center}

\normalsize
\end{table}

\clearpage

\begin{table}
\scriptsize
\begin{center}
\caption{Hardness evolution of Swift pulses. The Swift maximum hardness is defined as the maximum ratio of counts in the 20 to 50 keV range relative to the 50 to 300 keV range. The hardness of the precursor peak is taken relative to that of the central peak ('moderate' indicates that the two have similar amplitudes). The hardest pulses tend to exhibit complex spectral evolution characterized by hard precursor peaks and re-hardening preceding their decay peaks. whereas soft pulses tend to exhibit simple, spectral evolution characterized by monotonically increasing and decreasing hardnesses centered on their central peaks.}\label{tbl-4}
\begin{tabular}{lccc}
\tableline\tableline
Swift pulse & max cts hardness & precursor peak & decay peak re-hardening? \\
\tableline
GRB 081222 & $1.64 \pm 0.22$ & hard & yes \\
GRB 071117 & $1.61 \pm 0.48$ & hard & yes \\
GRB 051111 & $1.38 \pm 0.57$ & hard & yes \\
GRB 061021 & $1.00 \pm 0.13$ & moderate & maybe \\
GRB 060912A & $0.94 \pm 0.27$ & hard & yes \\
GRB 060708 & $0.69 \pm 0.09$ & moderate & yes \\
GRB 080413B & $0.66 \pm 0.03$ & soft &  maybe \\
GRB 070318 & $0.66 \pm 0.04$ & moderate & no \\
GRB060206 & $0.66 \pm 0.14$ & hard & maybe \\
GRB 061004 & $0.50 \pm 0.04$ & soft & no \\
GRB 091018 & $0.42 \pm 0.16$ & moderate & maybe \\
\tableline
\end{tabular}
\end{center}

\normalsize
\end{table}

\clearpage

\begin{table}
\scriptsize
\begin{center}
\caption{Hardness evolution of Fermi pulses. The notation is identical to that given in Table 4, although the Fermi maximum hardness is defined as the maximum ratio of counts in the 100 keV to 1 MeV range relative to the 25 to 100 keV range.\label{tbl-5}}
\begin{tabular}{lccc}
\tableline\tableline
Fermi pulse & max cts hardness & precursor peak & decay peak re-hardening? \\
\tableline
GRB 100707A & $3.87 \pm 0.59$ & hard & yes \\
GRB 081224 & $1.86 \pm 0.15$ & soft & maybe \\
\tableline
\end{tabular}
\end{center}

\normalsize
\end{table}

\clearpage

\begin{table}
\scriptsize
\begin{center}
\caption{Hardness evolution of BATSE pulses. The notation is identical to that given in Table 4, although the BATSE maximum hardness is defined as the maximum ratio of counts in the 100 keV to 1 MeV range relative to the 25 to 100 keV range. \label{tbl-6}}
\begin{tabular}{lccc}
\tableline\tableline
BATSE pulse & max cts hardness & precursor peak & decay peak re-hardening? \\
\tableline
BATSE 3040 & $2.30 \pm 1.56 $ & hard & yes \\
BATSE 1883 & $2.14 \pm 0.38$ & hard & yes \\
BATSE 563 & $2.08 \pm 0.20$ & moderate & maybe \\
BATSE 907 & $1.78 \pm 0.06$ & moderate & maybe \\
BATSE 3257 & $1.75 \pm 0.63$ & hard & no \\
BATSE 7711 & $1.73 \pm 0.17$ & hard & maybe \\
BATSE 1406 & $1.71 \pm 0.36$ & hard & maybe \\
BATSE 469 & $1.47 \pm 0.07$ & moderate & yes \\
BATSE 2662 & $1.44 \pm 0.16$ & moderate & maybe \\
BATSE 2665 & $1.43 \pm 0.40$ & hard & yes \\
BATSE 3143 & $1.42 \pm 0.47$ & hard & maybe \\
BATSE 3003 & $1.41\pm 0.12$ & hard & maybe \\
BATSE 658 & $1.22 \pm 0.22$ & hard & yes \\
BATSE 3164 & $1.22 \pm 0.61$ & soft & yes \\
BATSE 1580 & $1.18 \pm 0.12$ & moderate & maybe \\
BATSE 1446 & $1.07 \pm 0.18$ & soft & maybe \\
BATSE 332 & $1.05 \pm 0.16$ & hard & yes \\
BATSE 1467 & $1.03 \pm 0.15$ & hard & maybe \\
BATSE 1806 & $0.98 \pm 0.26$ & hard & yes \\
BATSE 7614 & $0.93 \pm 0.11$ & moderate & no \\
BATSE 501 & $0.9 \pm 0.41$ & moderate & no  \\
BATSE 2958p1 & $0.87 \pm 0.27$ & hard & maybe\\
BATSE 7903 & $0.87 \pm 0.34$ & moderate & yes \\
BATSE 7638 & $0.83 \pm 0.07$ & soft & yes \\
BATSE 7843 & $0.76 \pm 0.07$ & soft & maybe \\
BATSE 7989 & $0.75 \pm 0.03$ & soft & yes  \\
BATSE 3168 & $0.74 \pm 0.08$ & soft & no \\
BATSE 914 & $0.73 \pm 0.04$ & soft &  no \\
BATSE 1145 & $0.73 \pm 0.27$ & moderate & maybe \\
BATSE 1306 & $0.72 \pm 0.32$ & moderate & no \\
BATSE 2958p2 & $0.72 \pm 0.02$ & soft & no \\
BATSE 680 & $0.68 \pm 0.12$ & soft & maybe \\
BATSE 795 & $0.62 \pm 0.04$ & soft & no \\
BATSE 1039 & $0.62 \pm 0.09$ & moderate & no \\
BATSE 540 & $0.57 \pm 0.29$ & moderate & maybe \\
BATSE 2862 & $0.54 \pm 0.07$ & soft & maybe \\
BATSE 3026 & $0.53 \pm 0.05$ & moderate & maybe \\
BATSE 8112 & $0.53 \pm 0.14$ & soft & no \\
BATSE 493 & $0.50 \pm 0.04$ & soft & no \\
BATSE 1432 & $0.48 \pm 0.04$ & moderate & maybe \\
BATSE 1200 & $0.44 \pm 0.03$ & moderate & maybe \\
BATSE 8121 & $0.32 \pm 0.01$ & soft & no \\
BATSE 1319 & $0.31 \pm 0.15$ & soft & no \\
BATSE 1379 & $0.12 \pm 0.05$ & soft & no \\
\tableline
\end{tabular}
\end{center}

\normalsize
\end{table}

\clearpage

\begin{figure}
  \centering
  \begin{tabular}[b]{@{}p{0.45\textwidth}@{}}
    \centering\includegraphics[width=1.\linewidth]{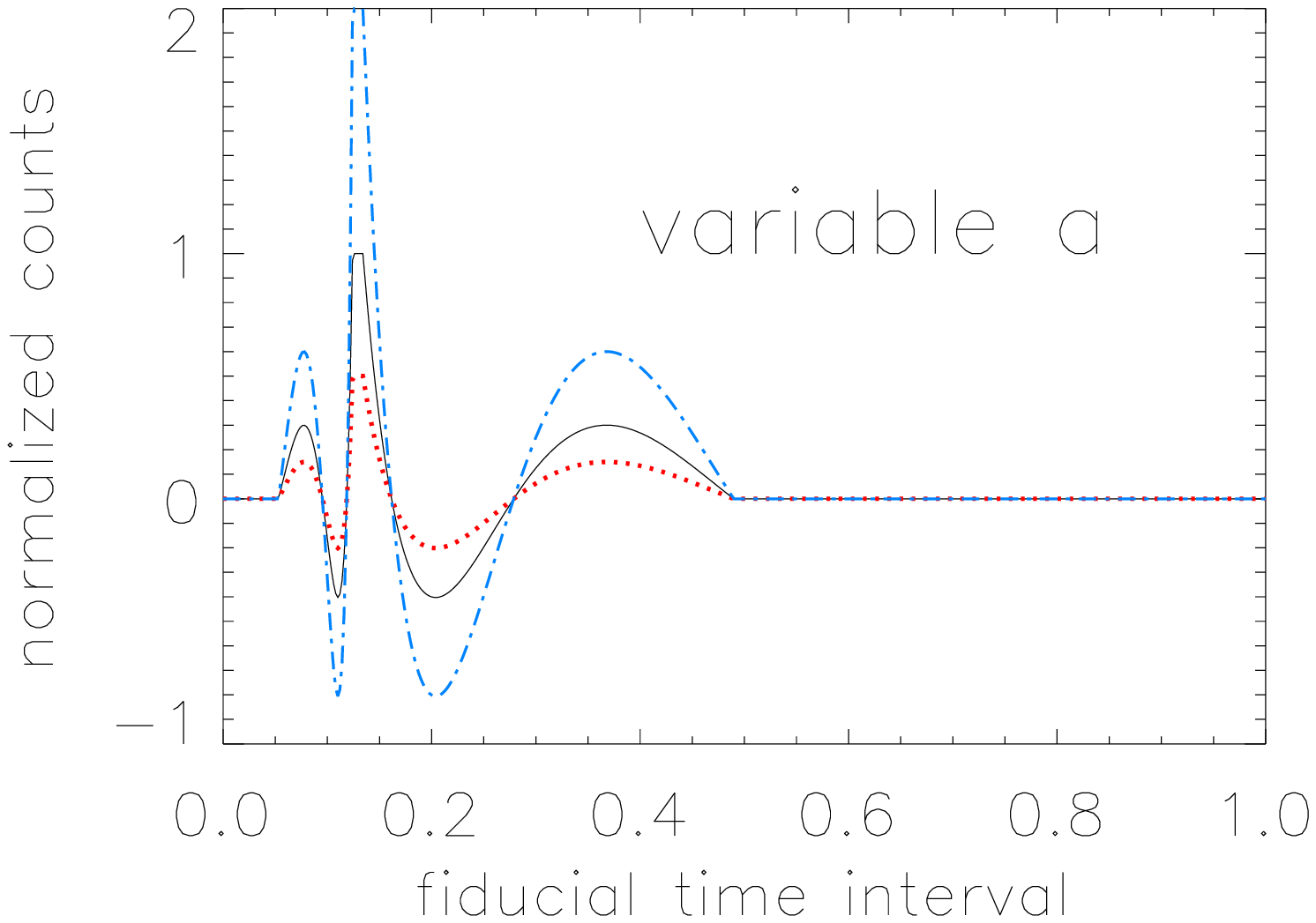} \\
    \centering\small Fig. 1a.
  \end{tabular}%
  \quad
  \begin{tabular}[b]{@{}p{0.45\textwidth}@{}}
    \centering\includegraphics[width=1.\linewidth]{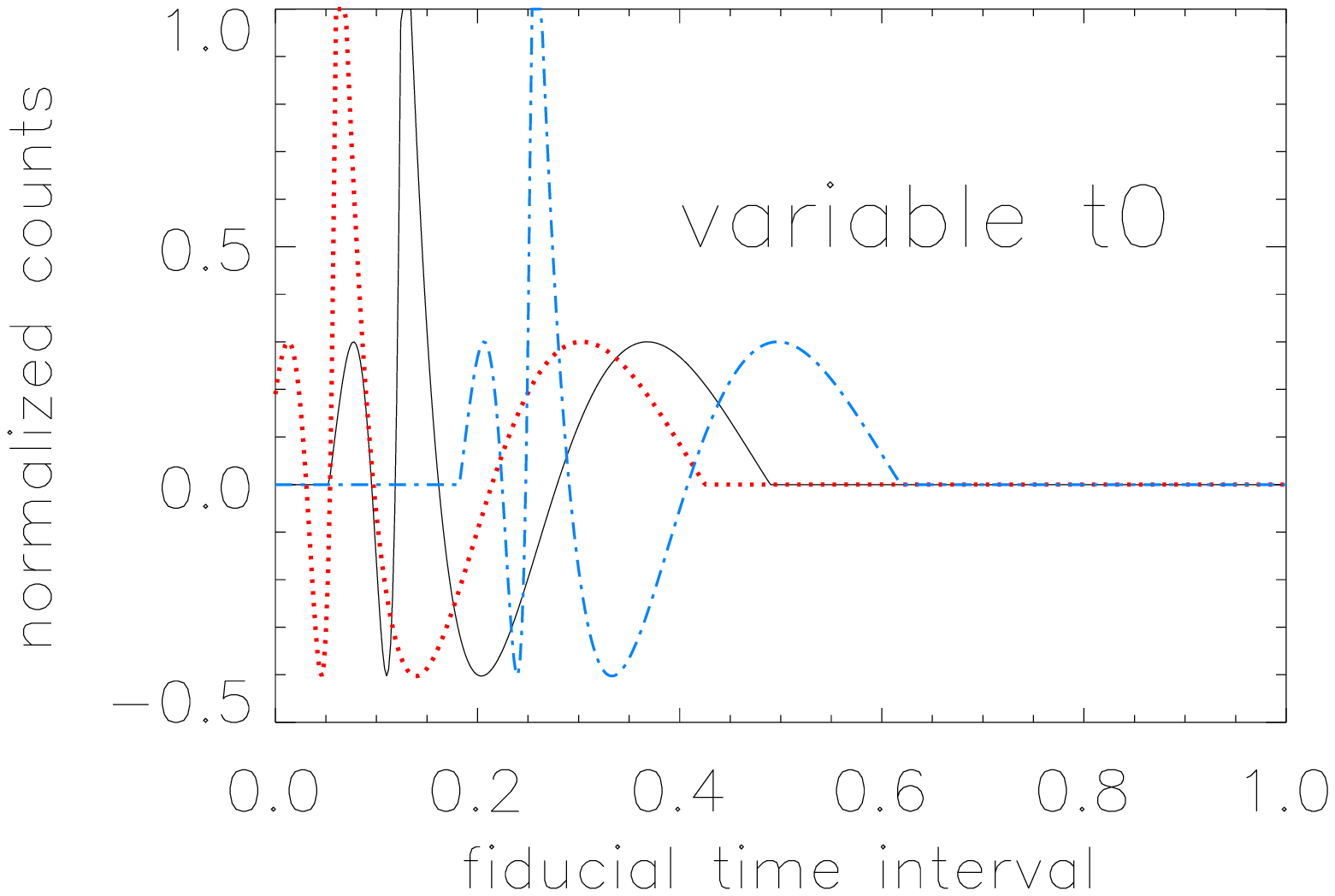} \\
    \centering\small Fig. 1b.
  \end{tabular}
  \caption{The effects of changing individual pulse residual fit parameters (left) peak amplitude $a$ and (right) peak time $t_0$, for a sample pulse with asymmetry $\kappa=0.8$. The expected plot is denoted a solid line (black), while a dot-dashed (blue) line indicates the parameter is doubled from its expected value, and a dashed (red) line indicates that the parameter is halved from its expected value. The plots with the adjusted fit parameters do not necessarily fit within the fiducial time interval ($0 \le t_{\rm fid} \le 1$) because this interval is dependent on $\kappa$. }
  \label{fig:figures}
\end{figure}

\clearpage

%\begin{figure}
%\plottwo{A.eps}{t0.eps}
%\caption{The effects of changing individual pulse residual fit parameters (left) peak amplitude $a$ and (right) peak time $t_0$, for a sample pulse with asymmetry $\kappa=0.8$. The expected plot is denoted a solid line (white), while a dot-dashed (blue) line indicates the parameter is halved from its expected value, and a dashed (red) line indicates that the parameter is doubled from its expected value. The plots with the adjusted fit parameters do not necessarily fit within the fiducial time interval ($0 \le t_{\rm fid} \le 1$) because this interval is dependent on $\kappa$.   \label{fig1}}
%\end{figure}

%\clearpage

\begin{figure}
  \centering
  \begin{tabular}[b]{@{}p{0.45\textwidth}@{}}
    \centering\includegraphics[width=1.\linewidth]{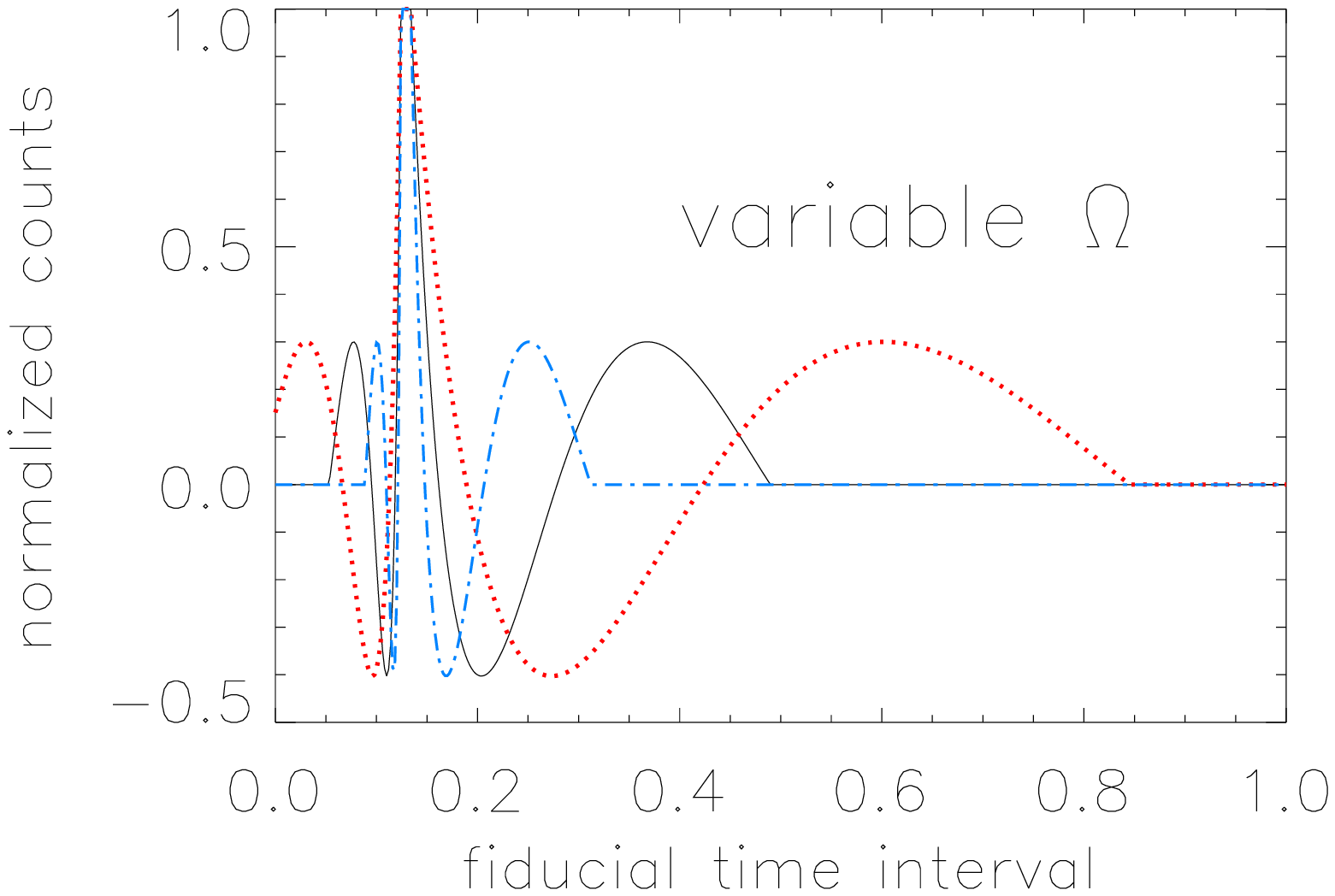} \\
    \centering\small Fig. 2a.
  \end{tabular}%
  \quad
  \begin{tabular}[b]{@{}p{0.45\textwidth}@{}}
    \centering\includegraphics[width=1.\linewidth]{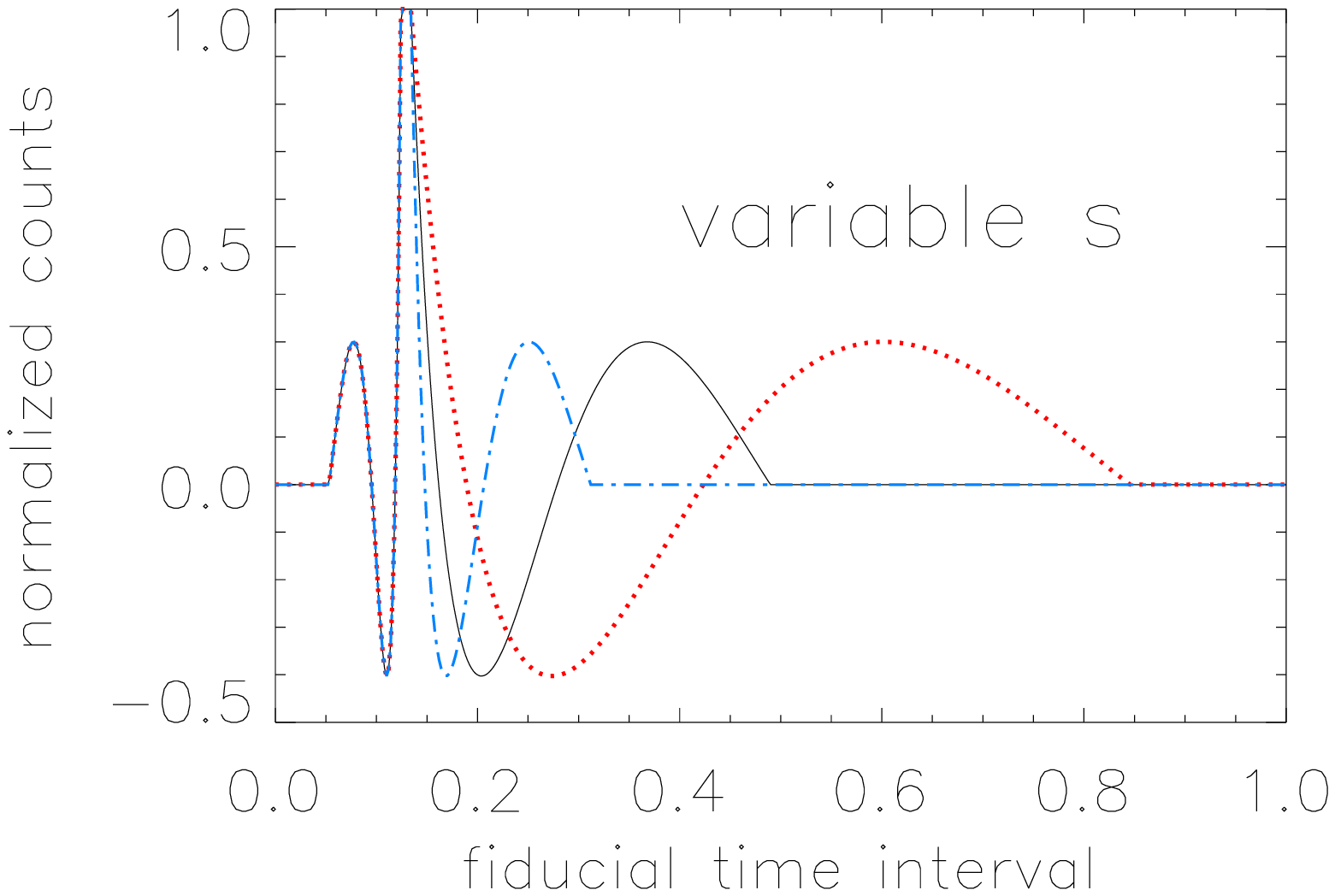} \\
    \centering\small Fig 2b.
  \end{tabular}
  \caption{The effects of changing individual pulse residual fit parameters (left) variability parameter $\Omega$ and (right) stretching parameter $s$, for a sample pulse with asymmetry $\kappa=0.8$. The expected plot is denoted a solid line (black), while a dot-dashed (blue) line indicates the parameter is doubled from its expected value, and a dashed (red) line indicates that the parameter is halved from its expected value. }
  \label{fig:figures}
\end{figure}

%\clearpage

%\begin{figure}
%\plottwo{omega.eps}{s.eps}
%\caption{The effects of changing individual pulse residual fit parameters (left) variability parameter $\Omega$ and (right) expansion parameter $s$, for a sample pulse with asymmetry $\kappa=0.8$. The expected plot is denoted a solid line (white), while a dot-dashed (blue) line indicates the parameter is halved from its expected value, and a dashed (red) line indicates that the parameter is doubled from its expected value.   \label{fig2}}
%\end{figure}

\clearpage

\begin{figure}
\plottwo{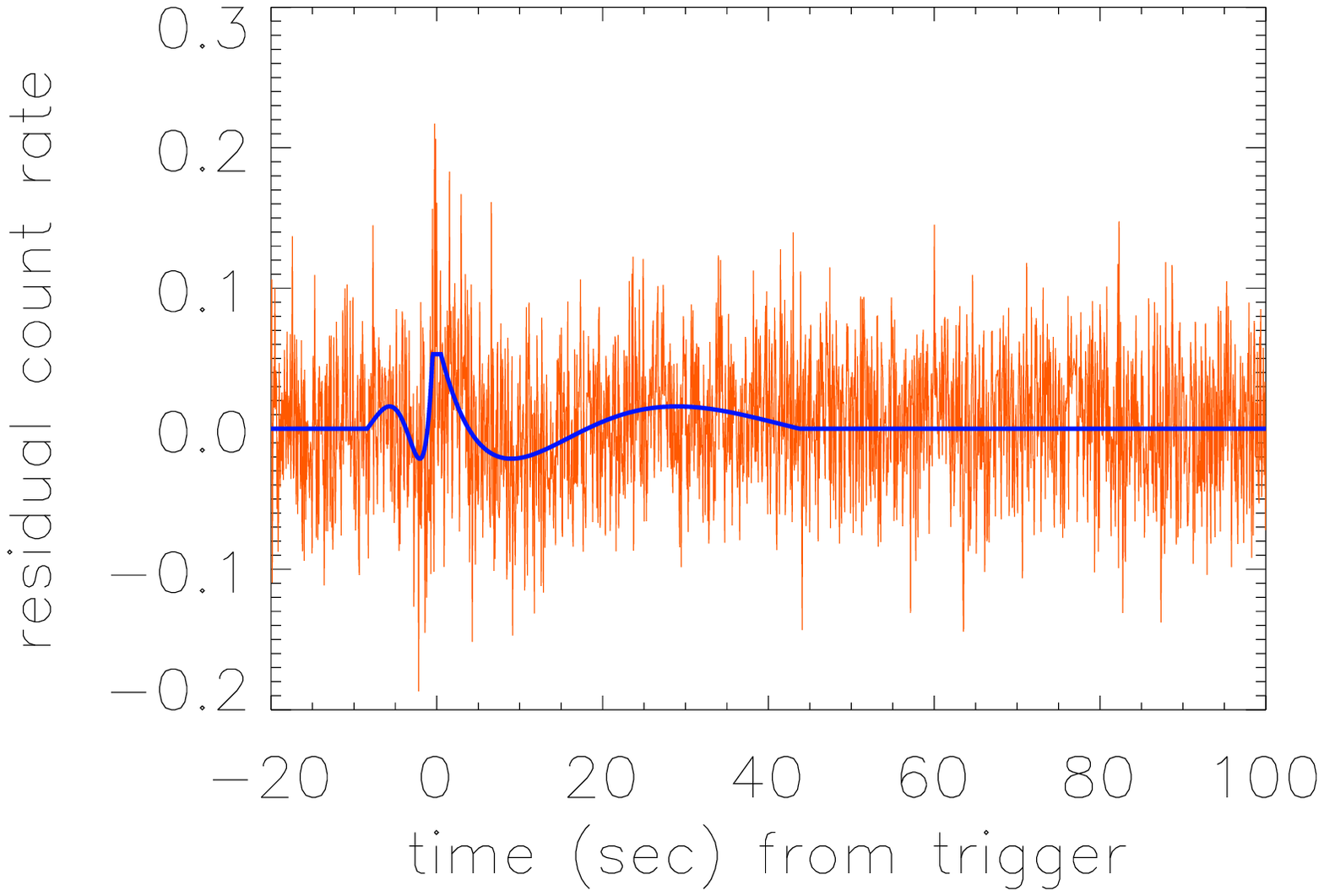}{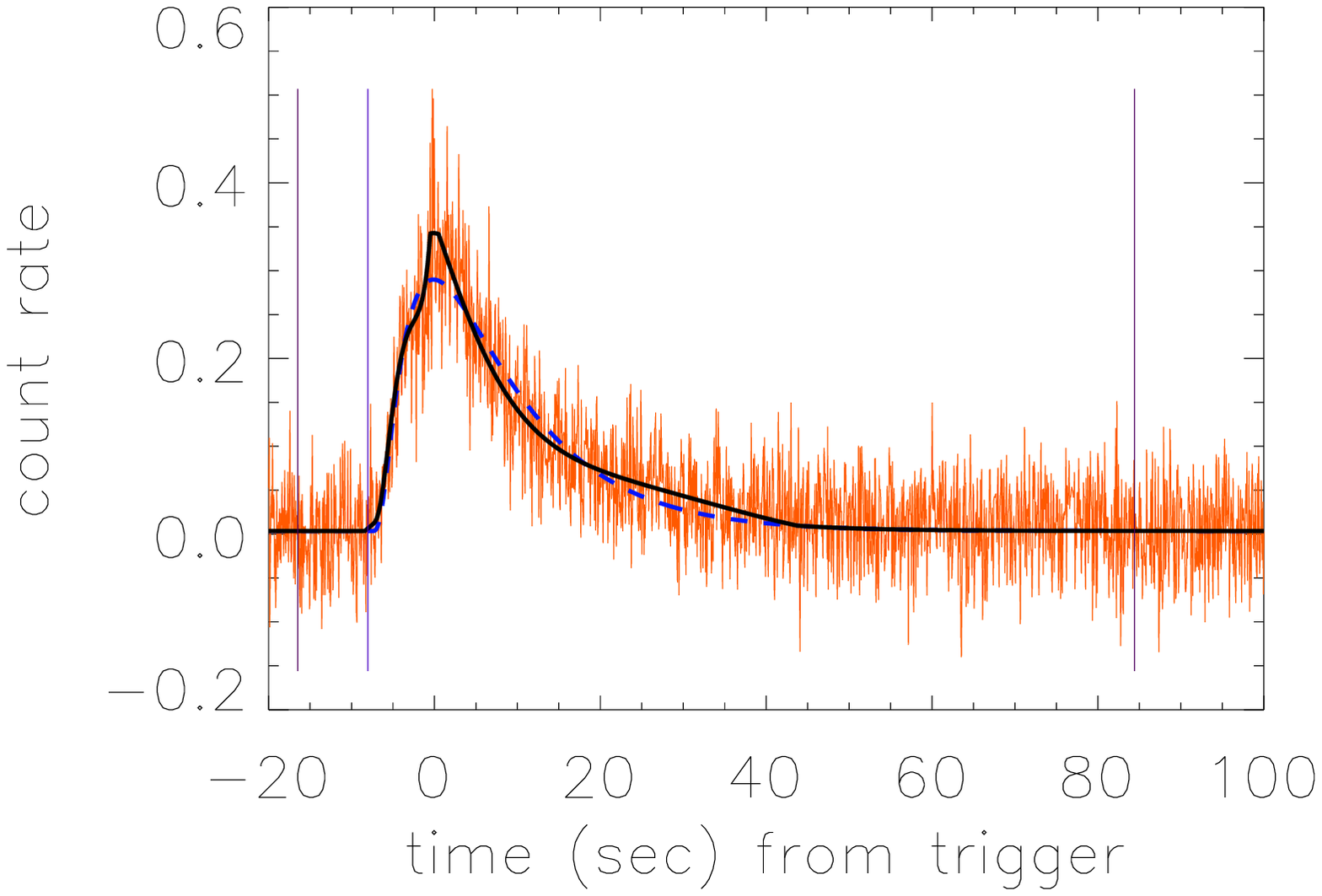}
\caption{GRB 051111: (left) fit to the residuals, (right) fit to the residuals plus pulse model. In the right panels shown in Figures 3 through 13, vertical black lines refer to the $t_{\rm start}$ and $t_{\rm end}$ times which bound the fiducial time interval, and the vertical blue line indicates the pulse start time ($t_s$) fitted from the \cite{nor05} pulse model.} \label{fig3}
\end{figure}

%\clearpage

\begin{figure}
\plottwo{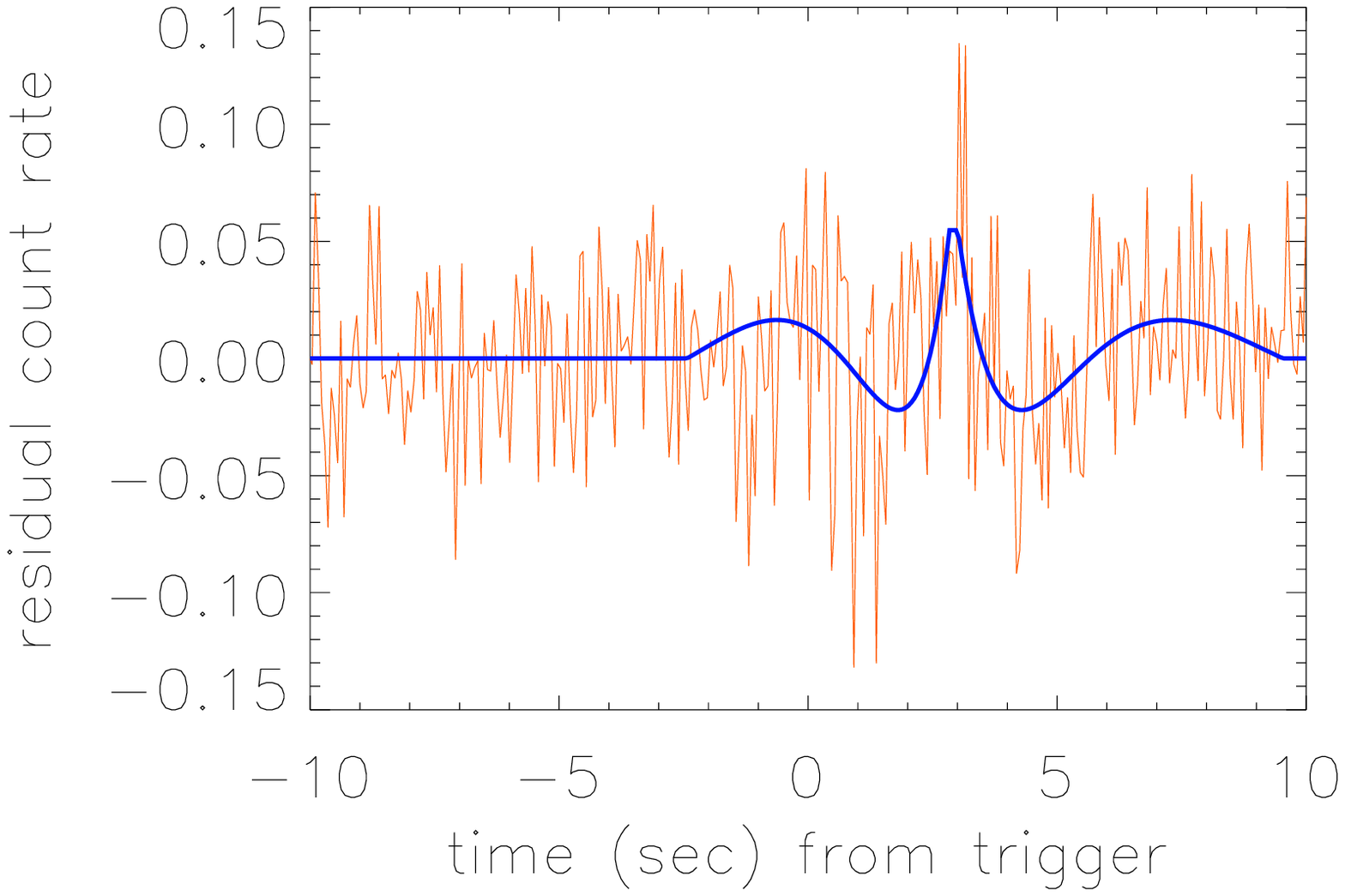}{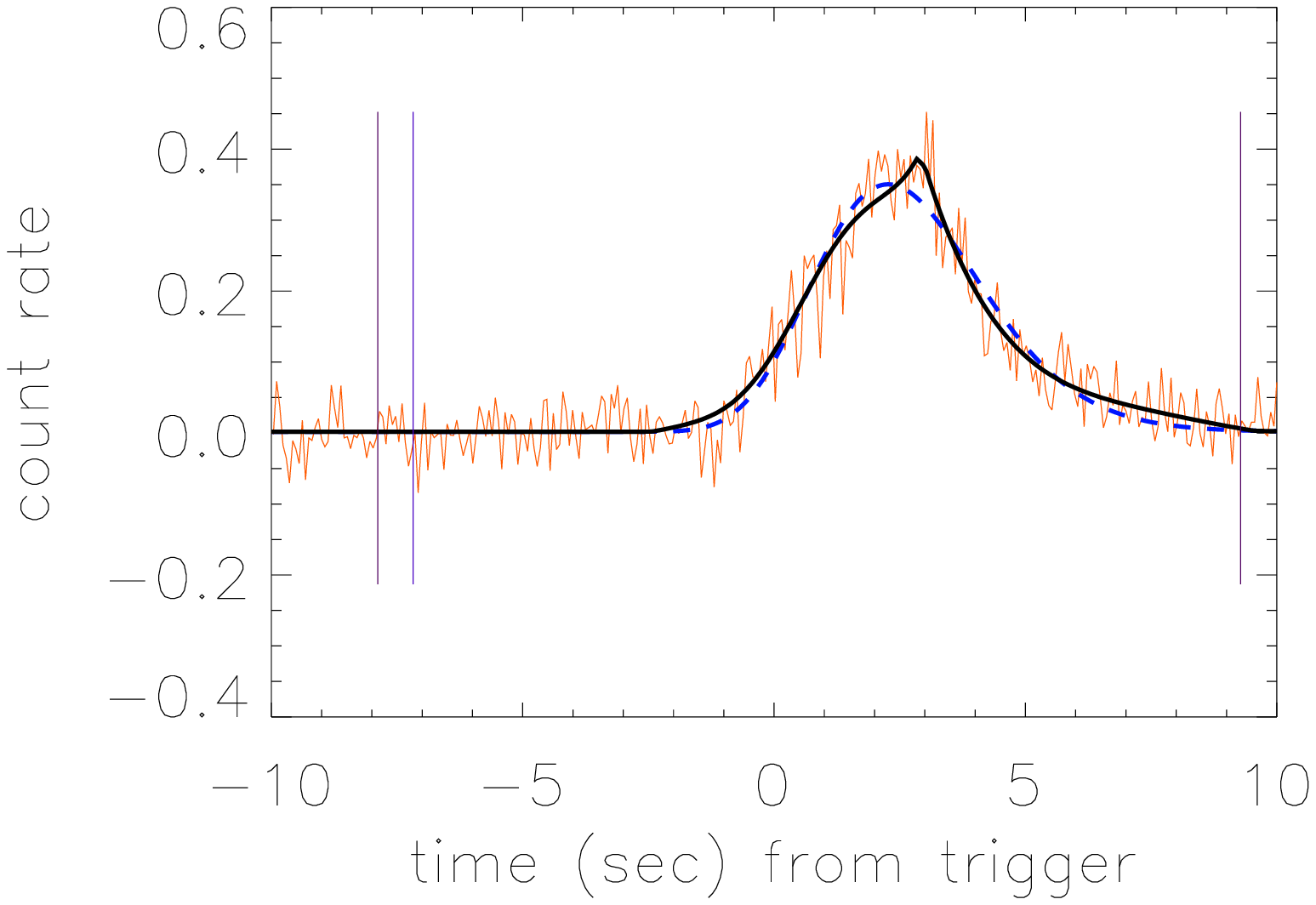}
\caption{GRB 060206: (left) fit to the residuals, (right) fit to the residuals plus pulse model.  \label{fig4}}
\end{figure}

%\clearpage

\begin{figure}
\plottwo{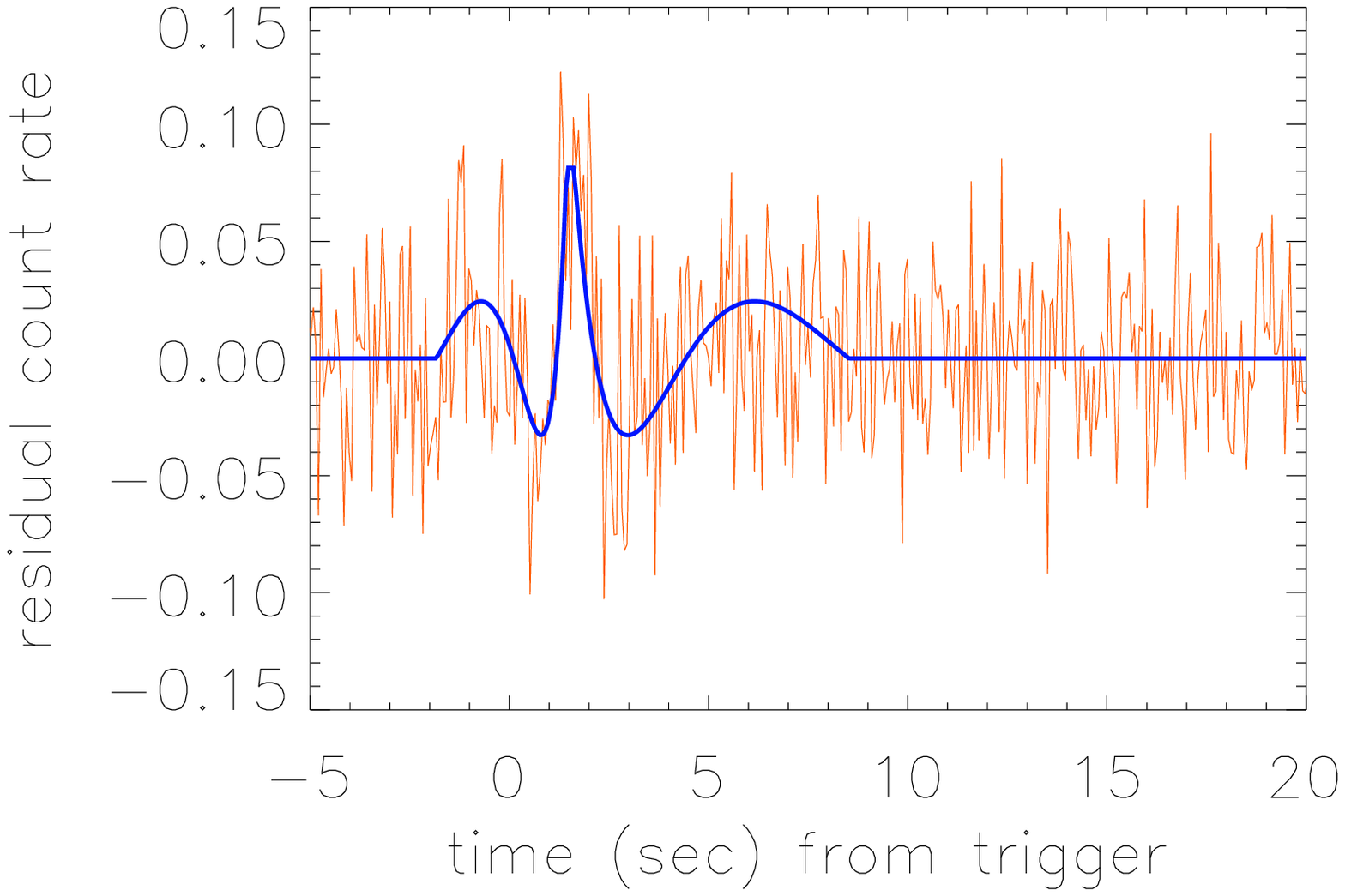}{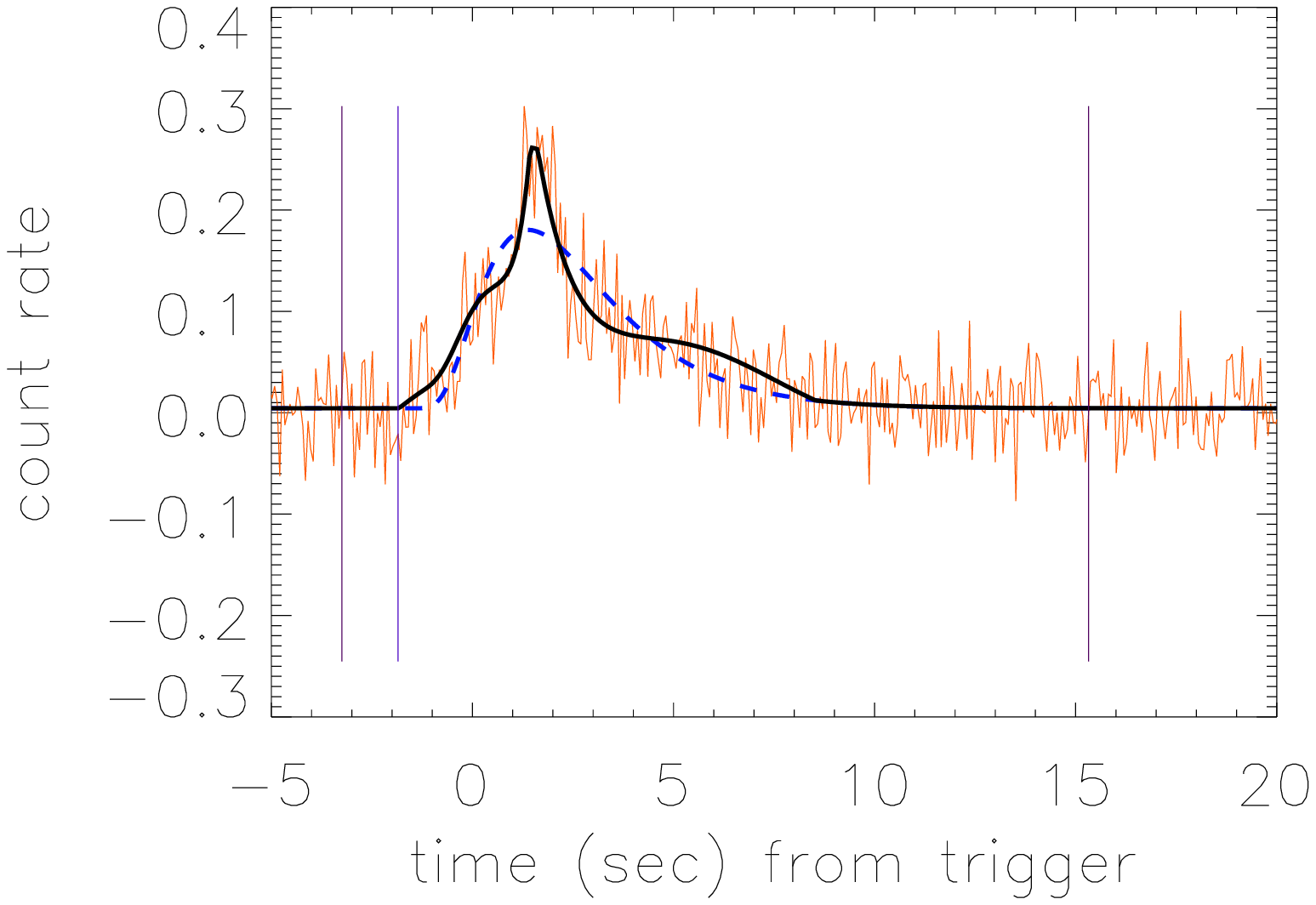}
\caption{GRB 060708: (left) fit to the residuals, (right) fit to the residuals plus pulse model.  \label{fig5}}
\end{figure}

\clearpage

\begin{figure}
\plottwo{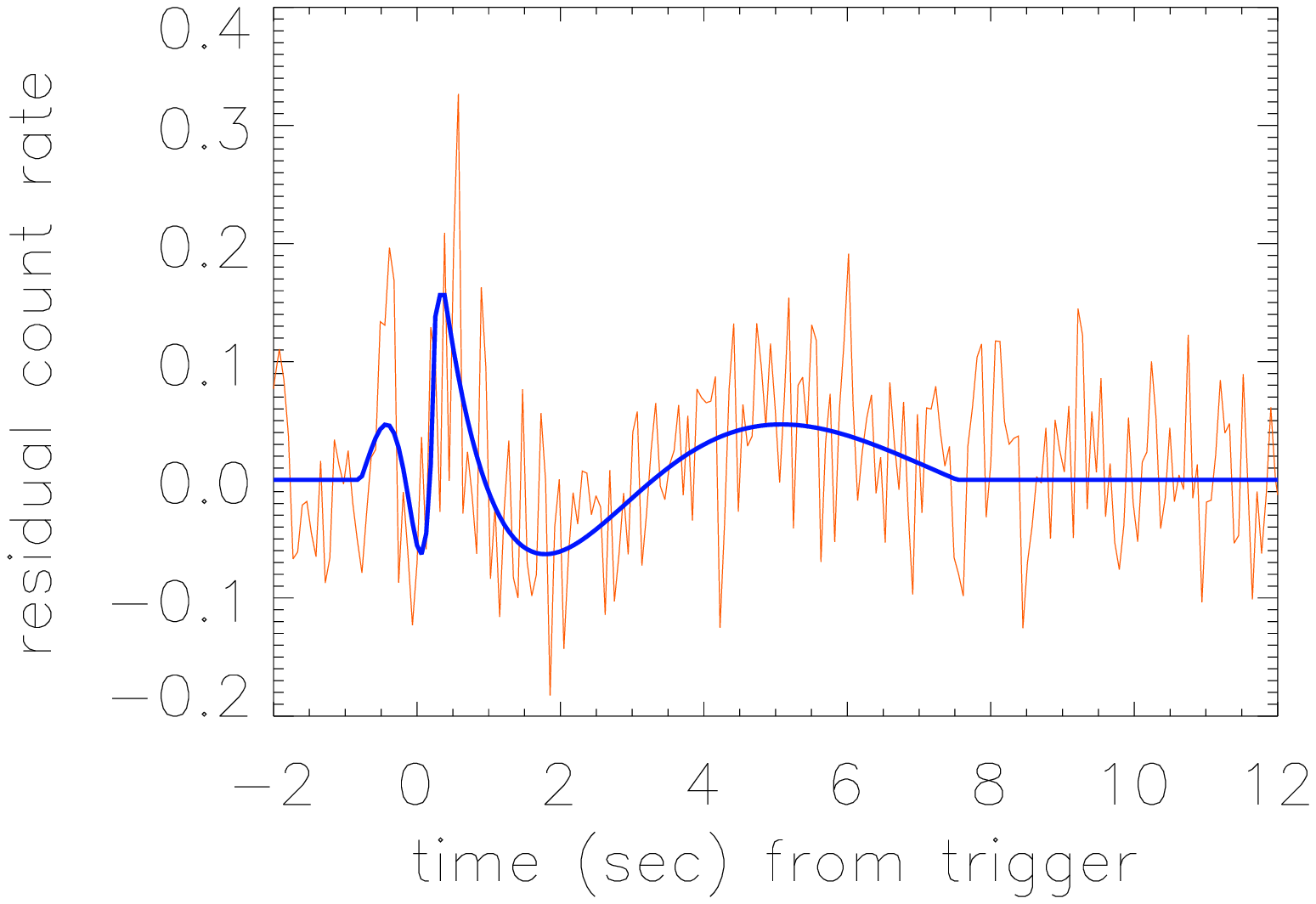}{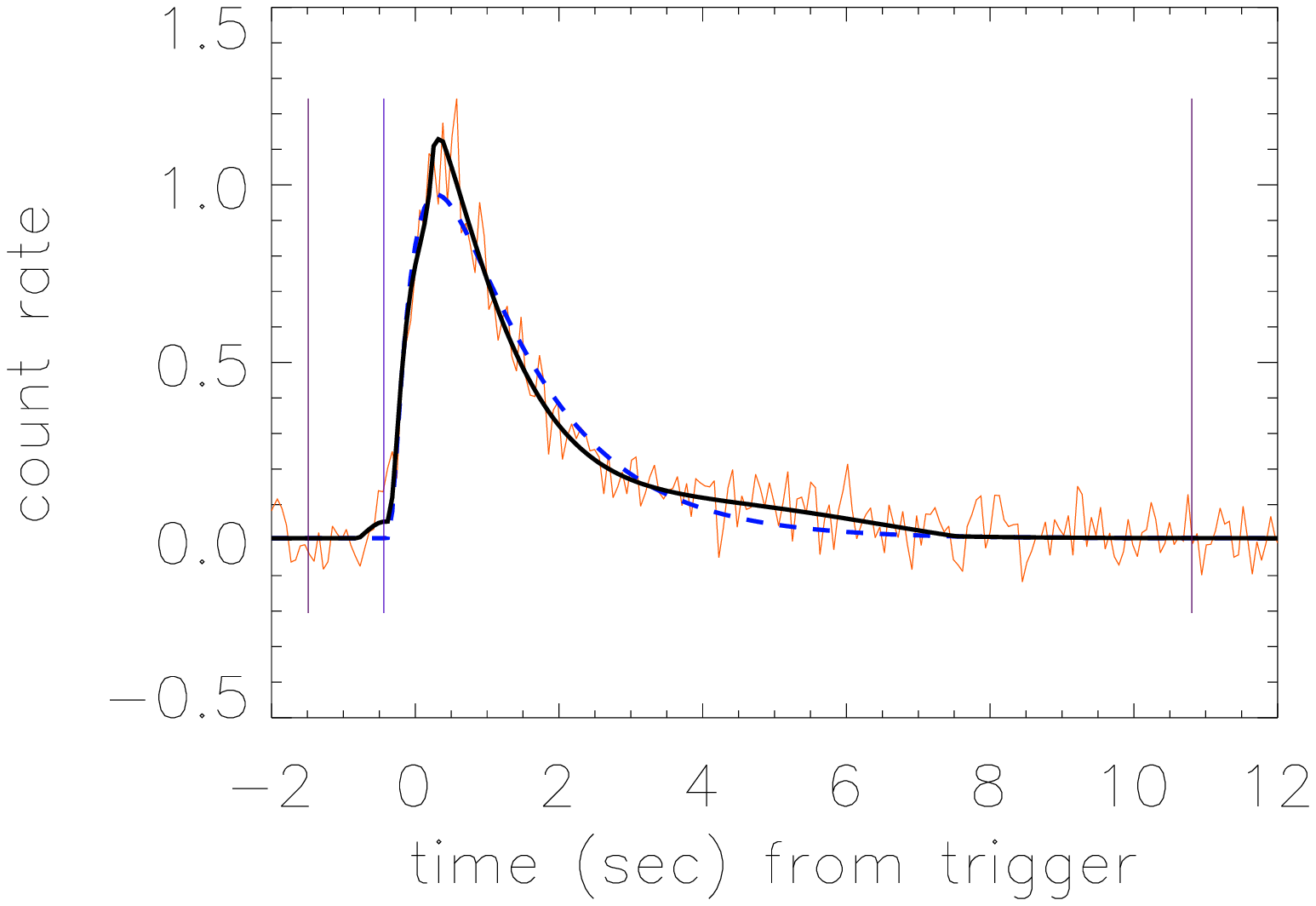}
\caption{GRB 060912A: (left) fit to the residuals, (right) fit to the residuals plus pulse model.  \label{fig6}}
\end{figure}

%\clearpage

\begin{figure}
\plottwo{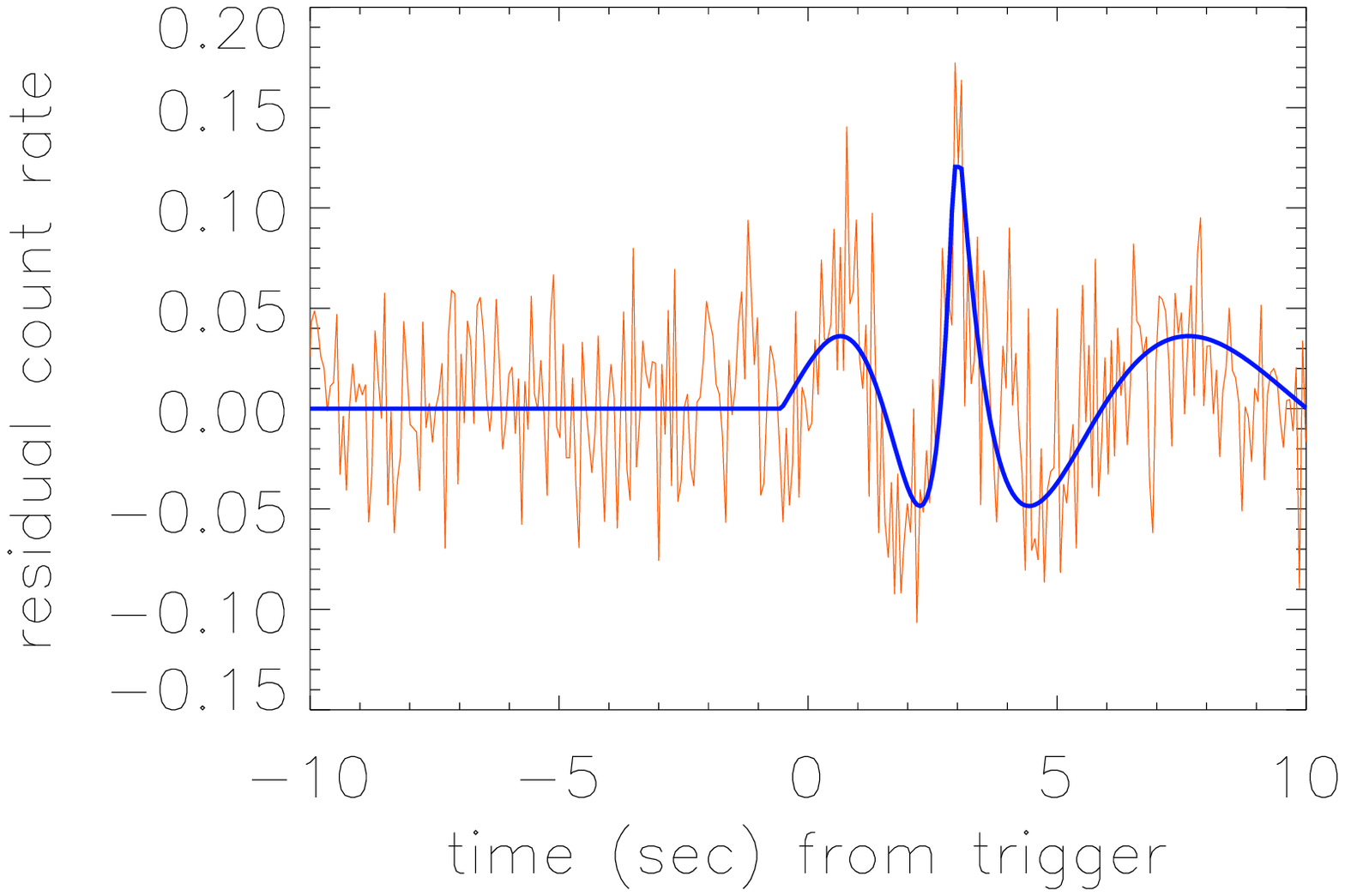}{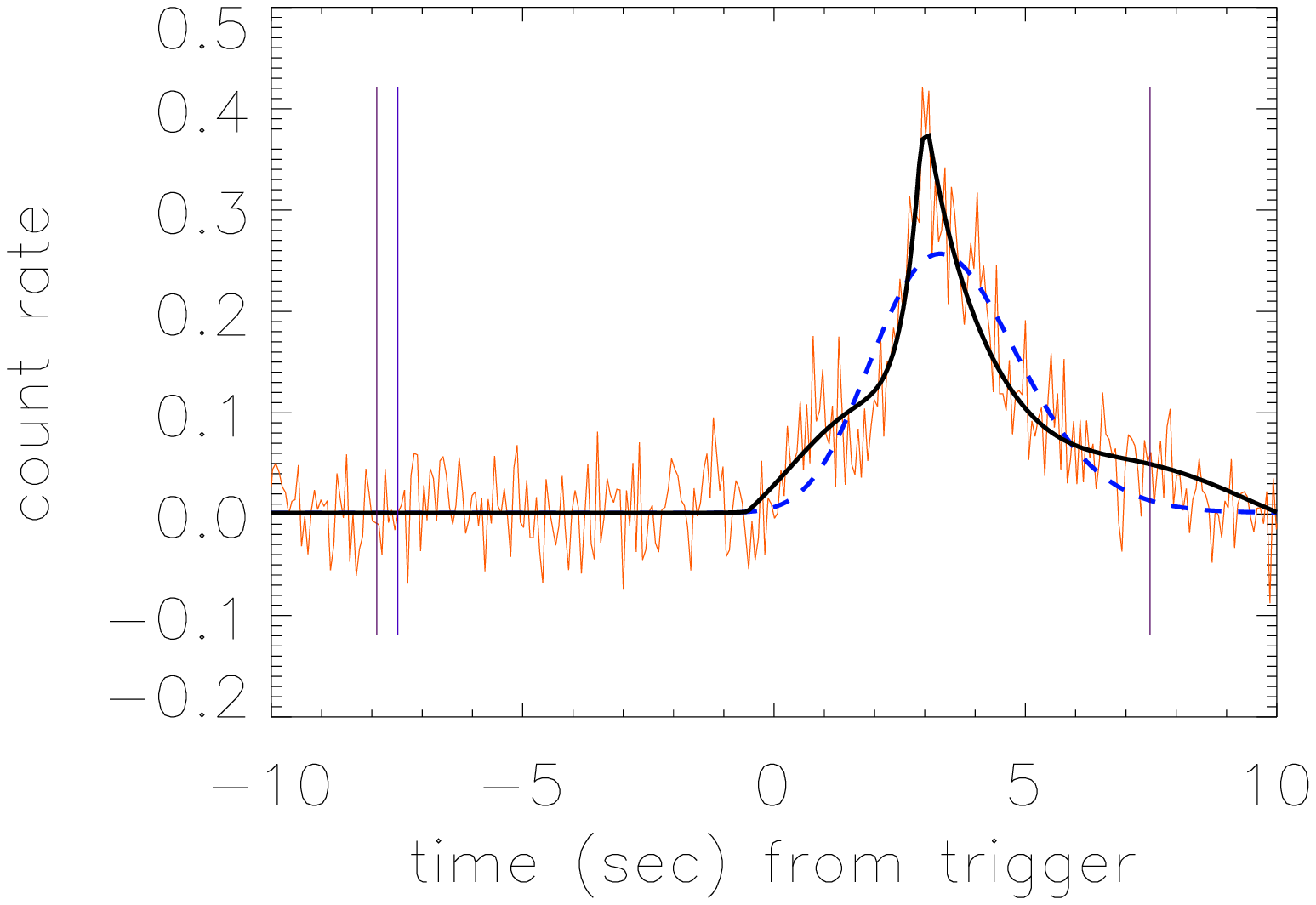}
\caption{GRB 061004: (left) fit to the residuals, (right) fit to the residuals plus pulse model.  \label{fig7}}
\end{figure}

%\clearpage

\begin{figure}
\plottwo{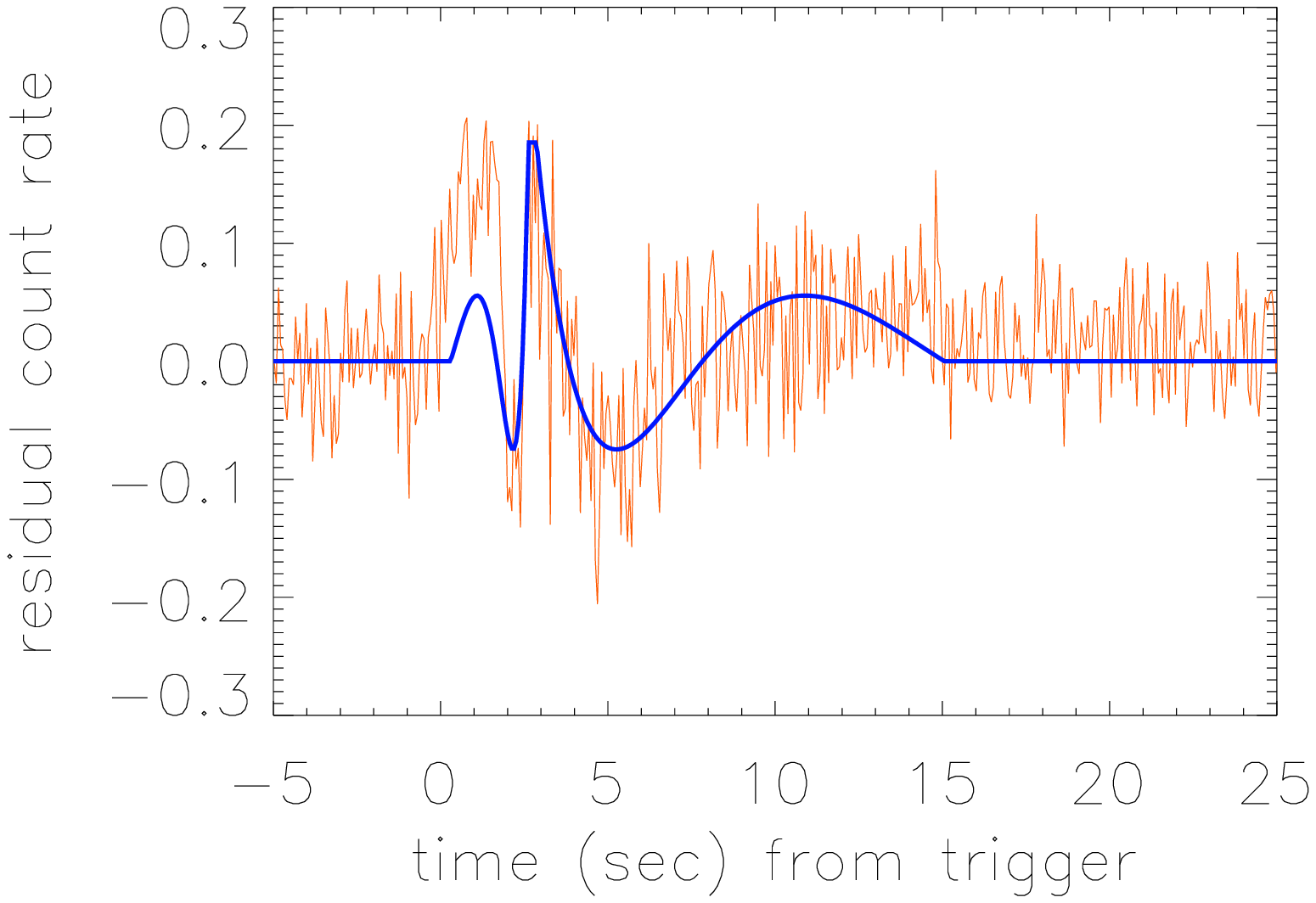}{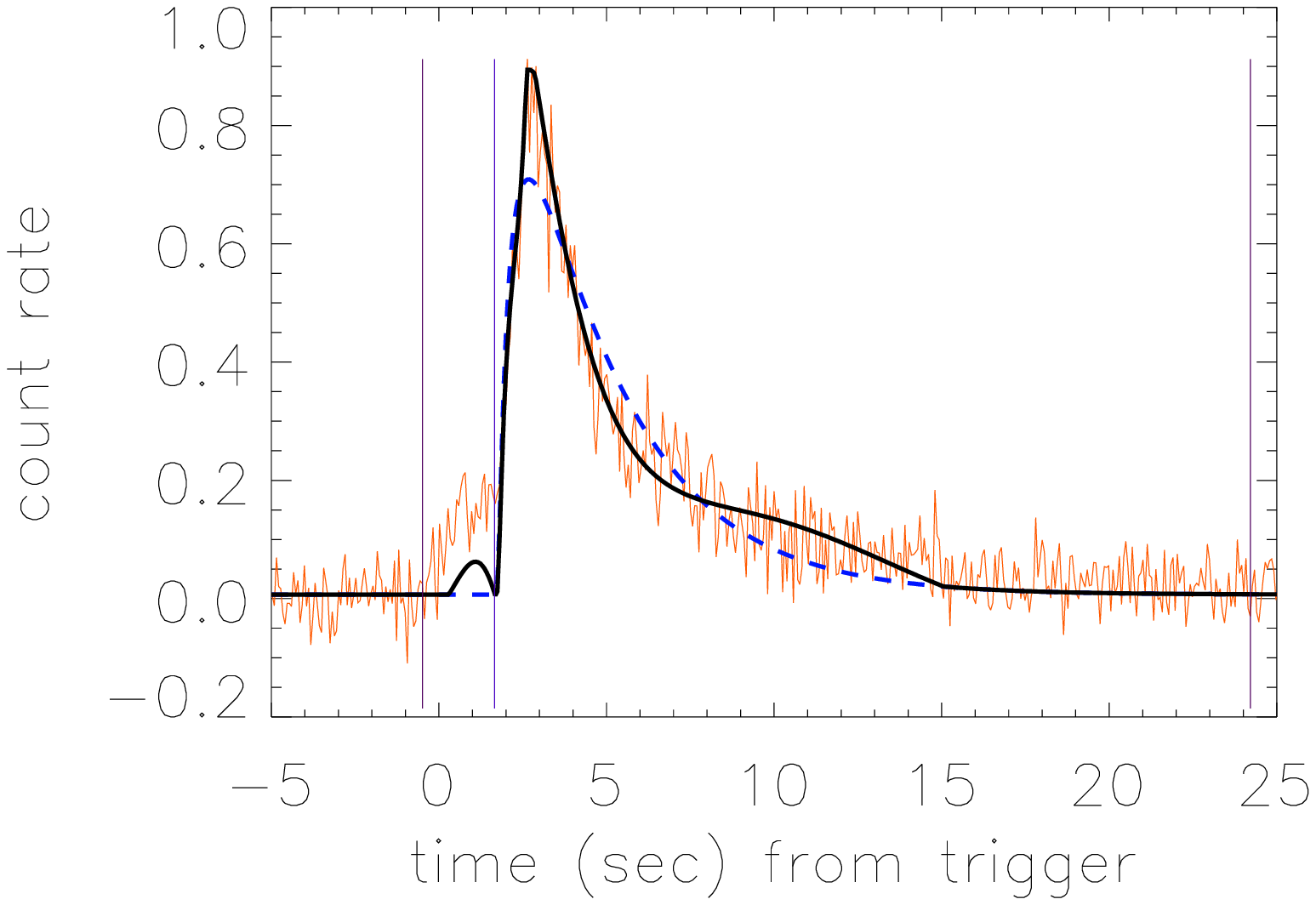}
\caption{GRB 061021: (left) fit to the residuals, (right) fit to the residuals plus pulse model.  \label{fig8}}
\end{figure}

\clearpage

\begin{figure}
\plottwo{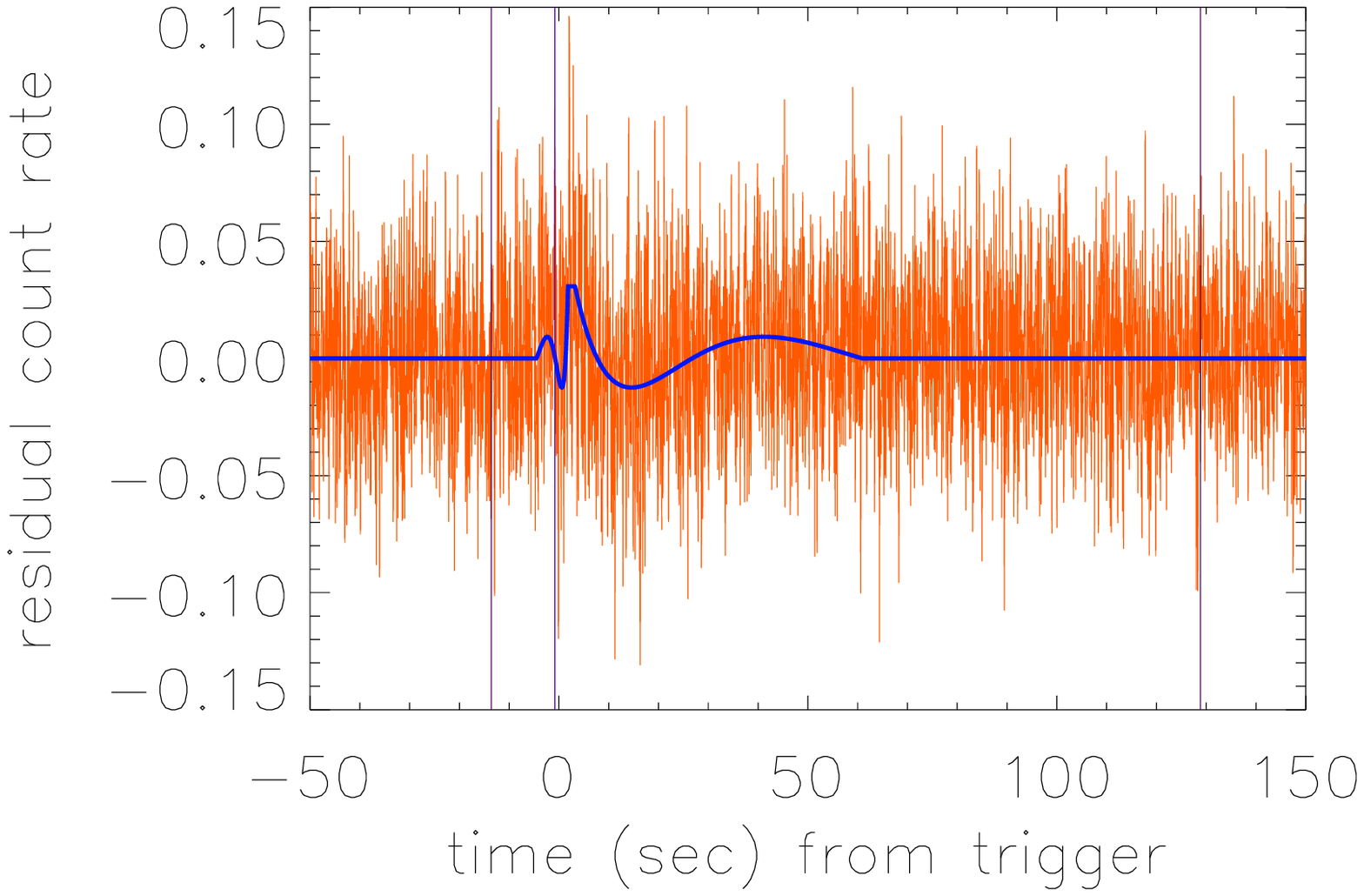}{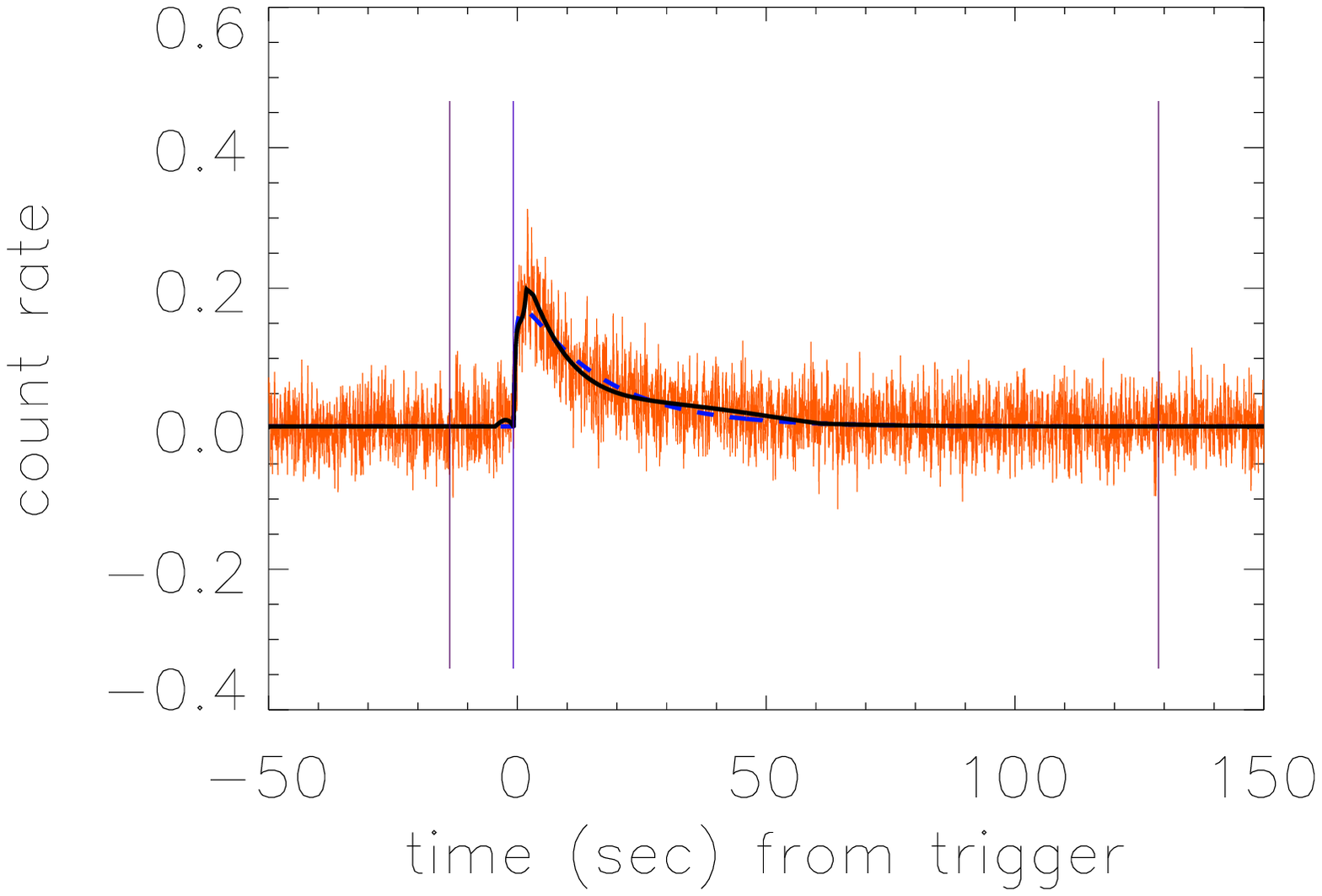}
\caption{GRB 070318: (left) fit to the residuals, (right) fit to the residuals plus pulse model.  \label{fig9}}
\end{figure}

%\clearpage

\begin{figure}
\plottwo{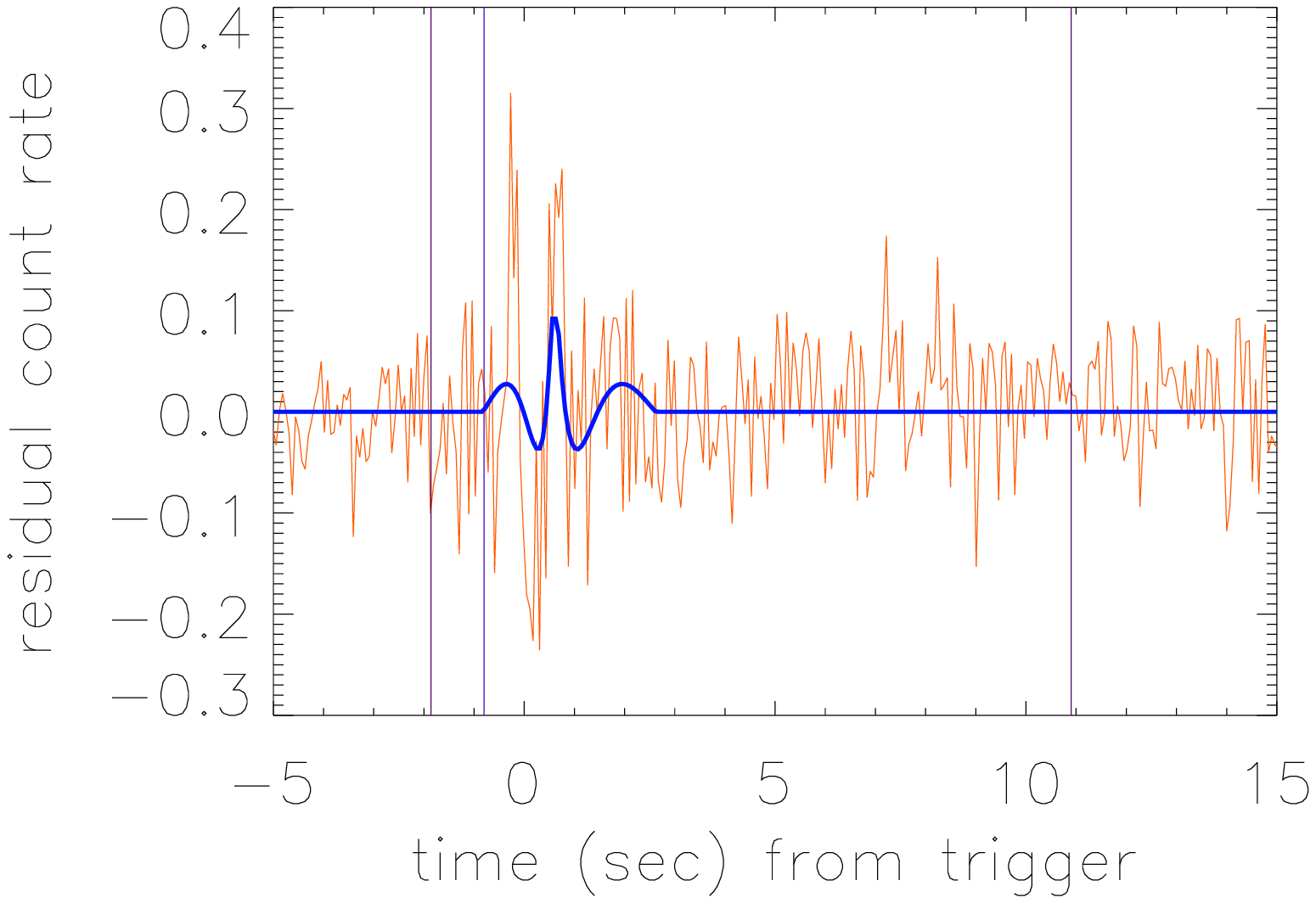}{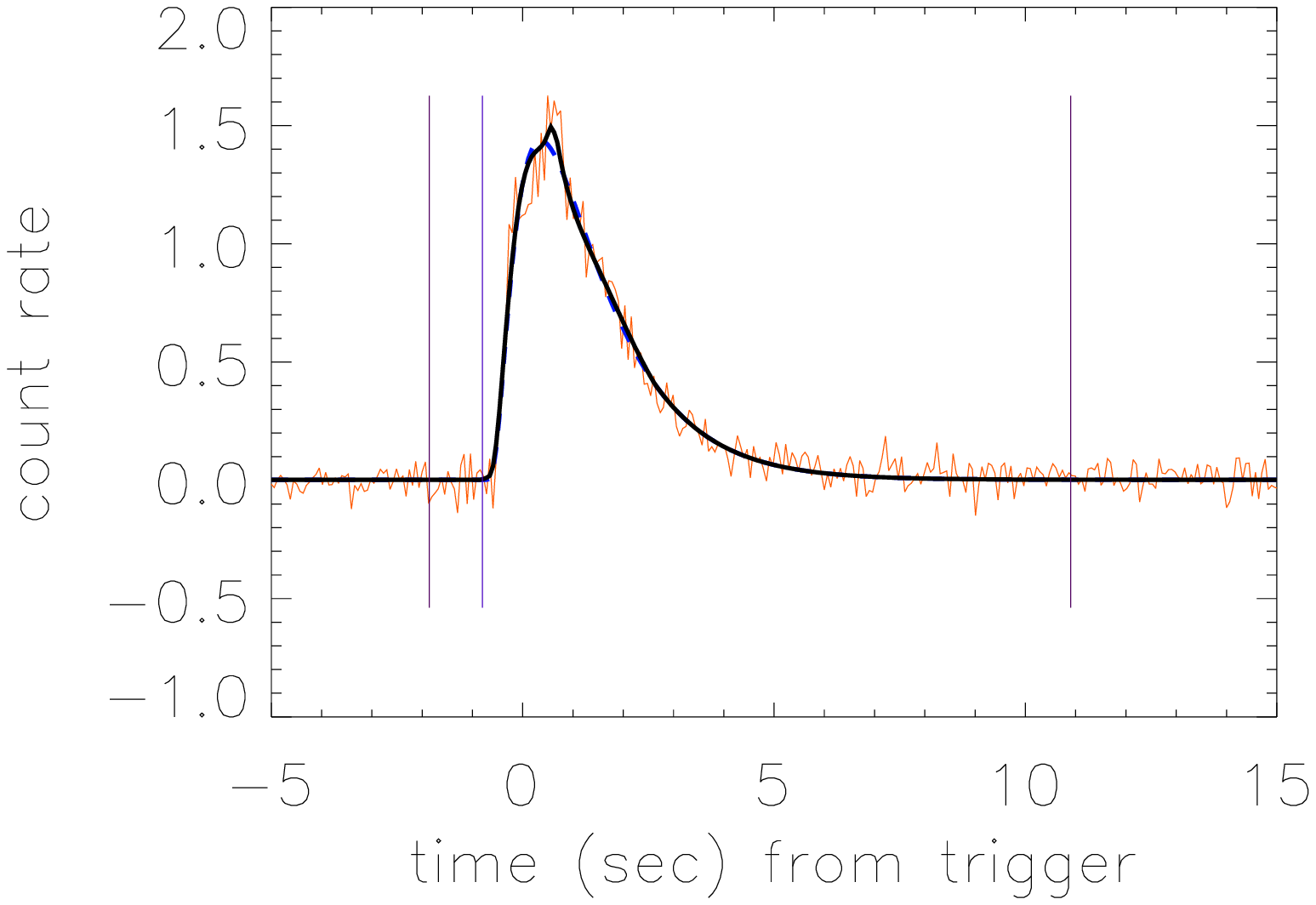}
\caption{GRB 071117: (left) fit to the residuals, (right) fit to the residuals plus pulse model.  \label{fig10}}
\end{figure}

%\clearpage

\begin{figure}
\plottwo{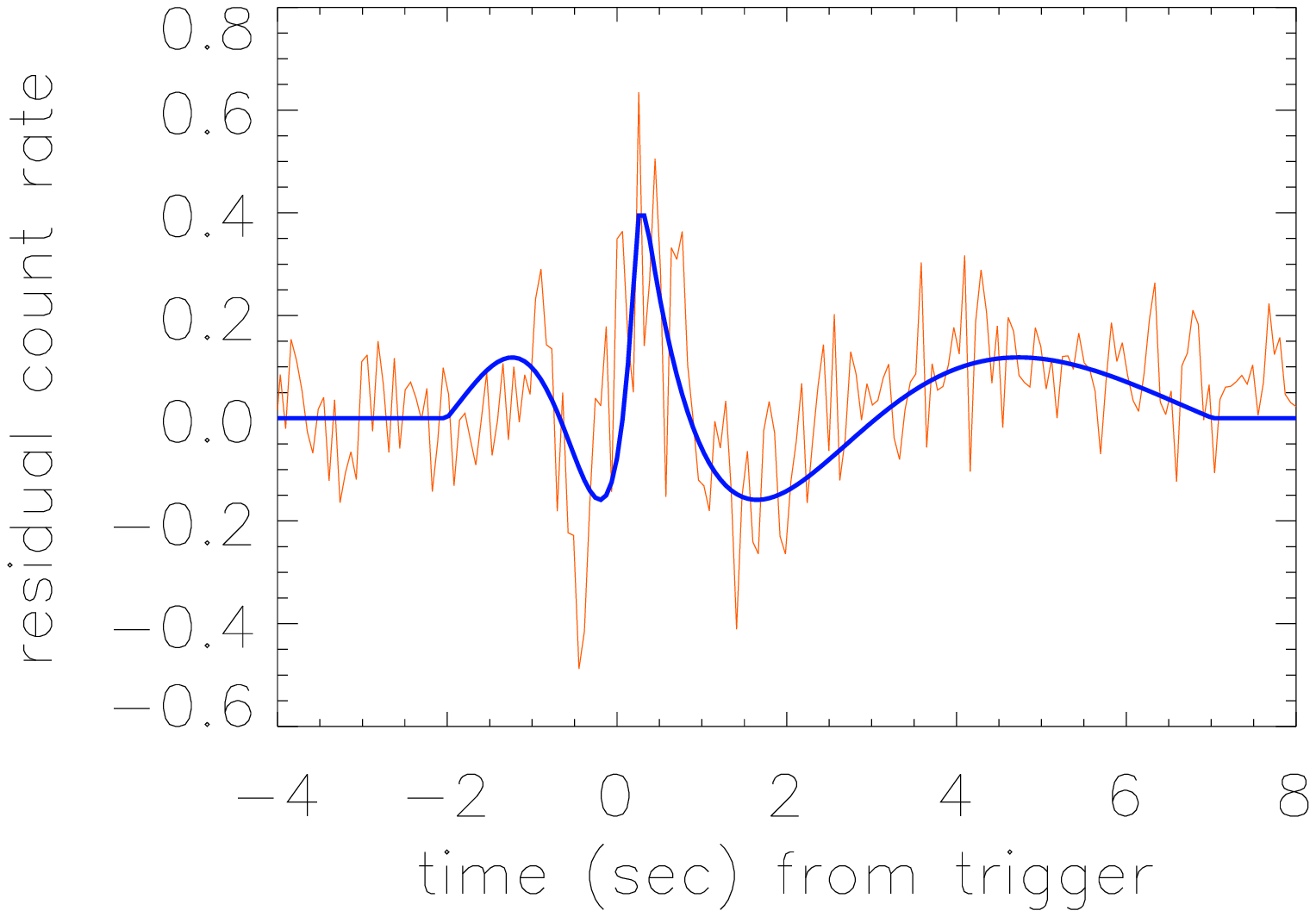}{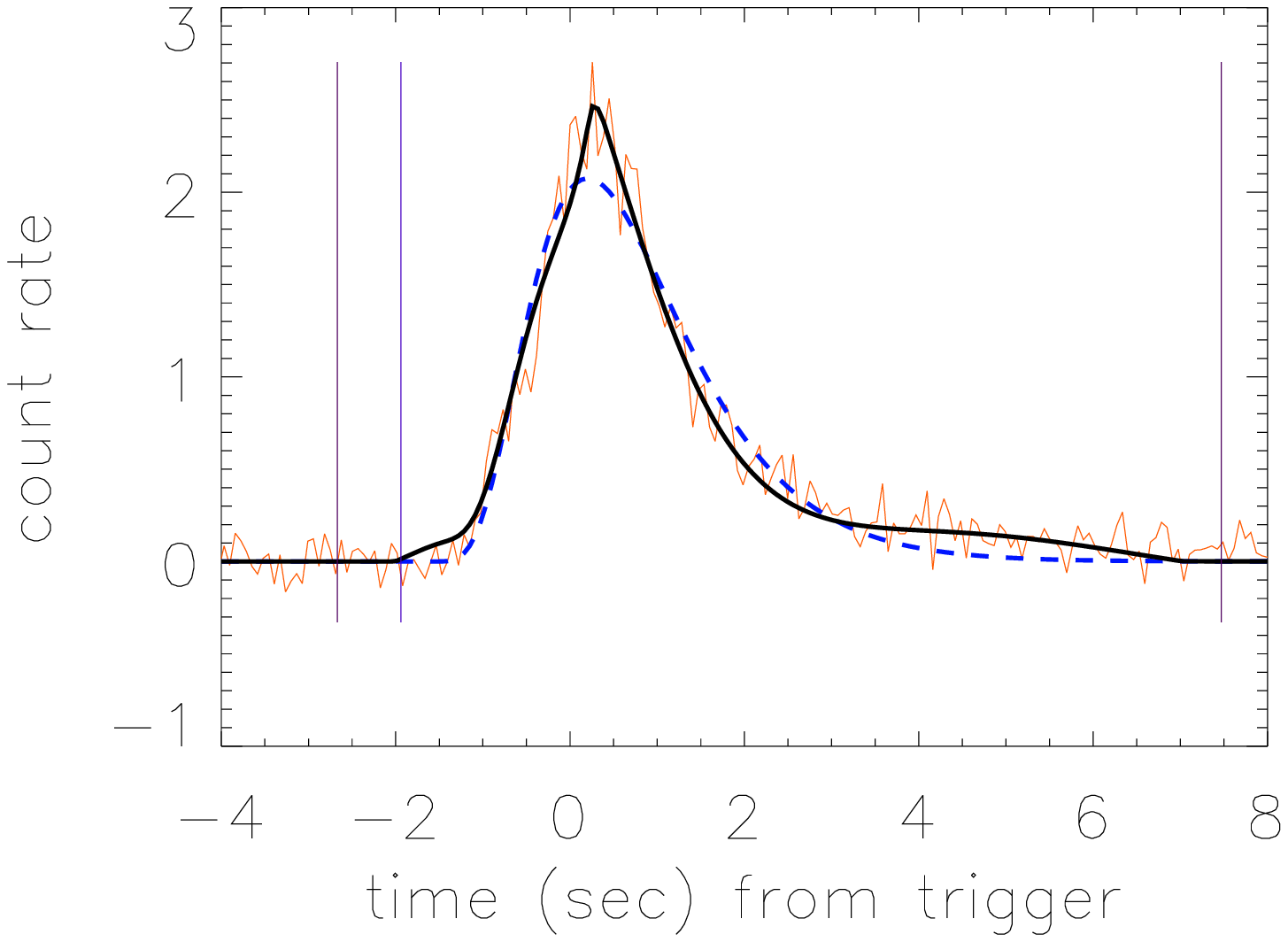}
\caption{GRB 80413B: (left) fit to the residuals, (right) fit to the residuals plus pulse model.  \label{fig11}}
\end{figure}

\clearpage

\begin{figure}
\plottwo{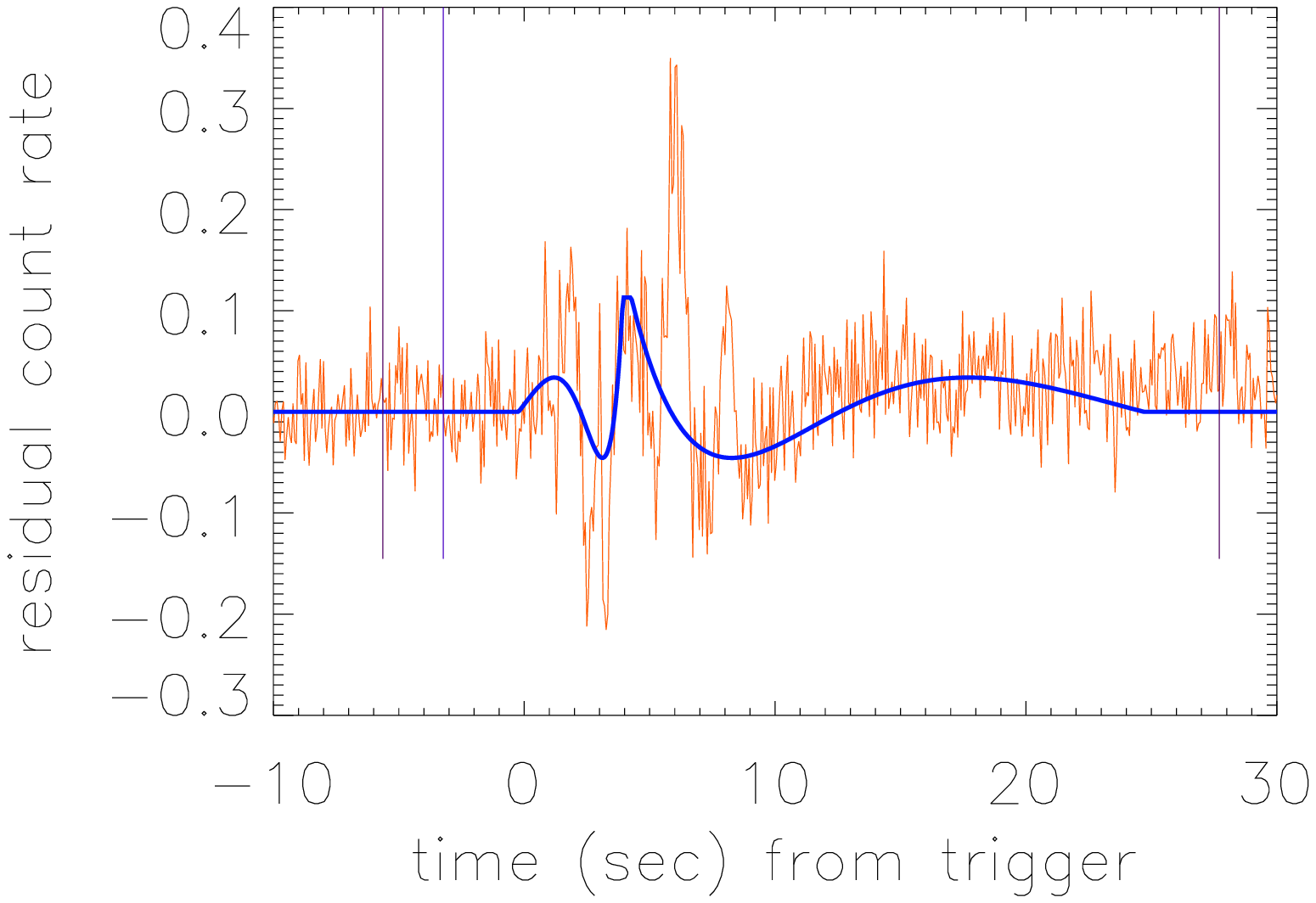}{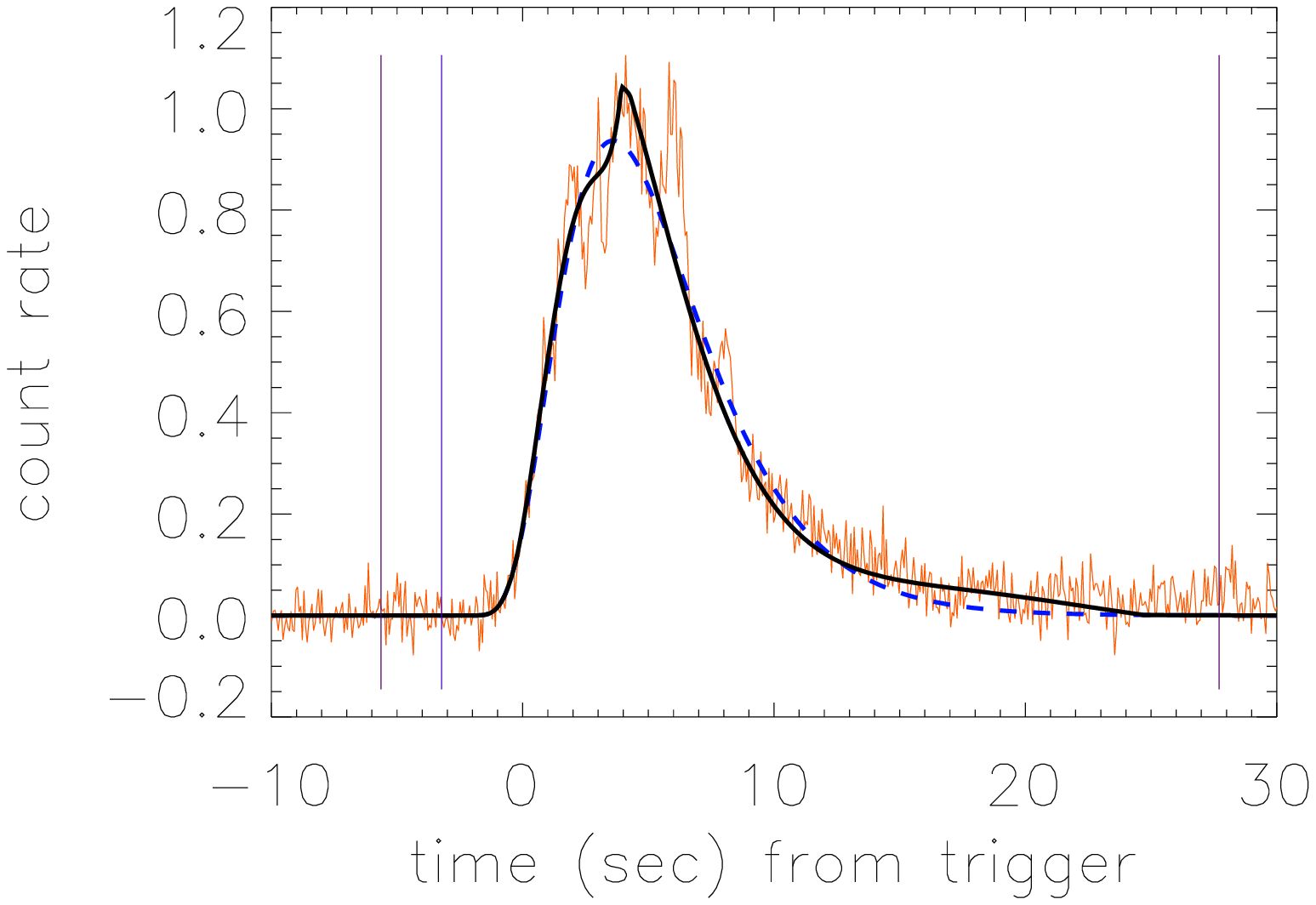}
\caption{GRB 081222: (left) fit to the residuals, (right) fit to the residuals plus pulse model.  \label{fig12}}
\end{figure}

%\clearpage

\begin{figure}
\plottwo{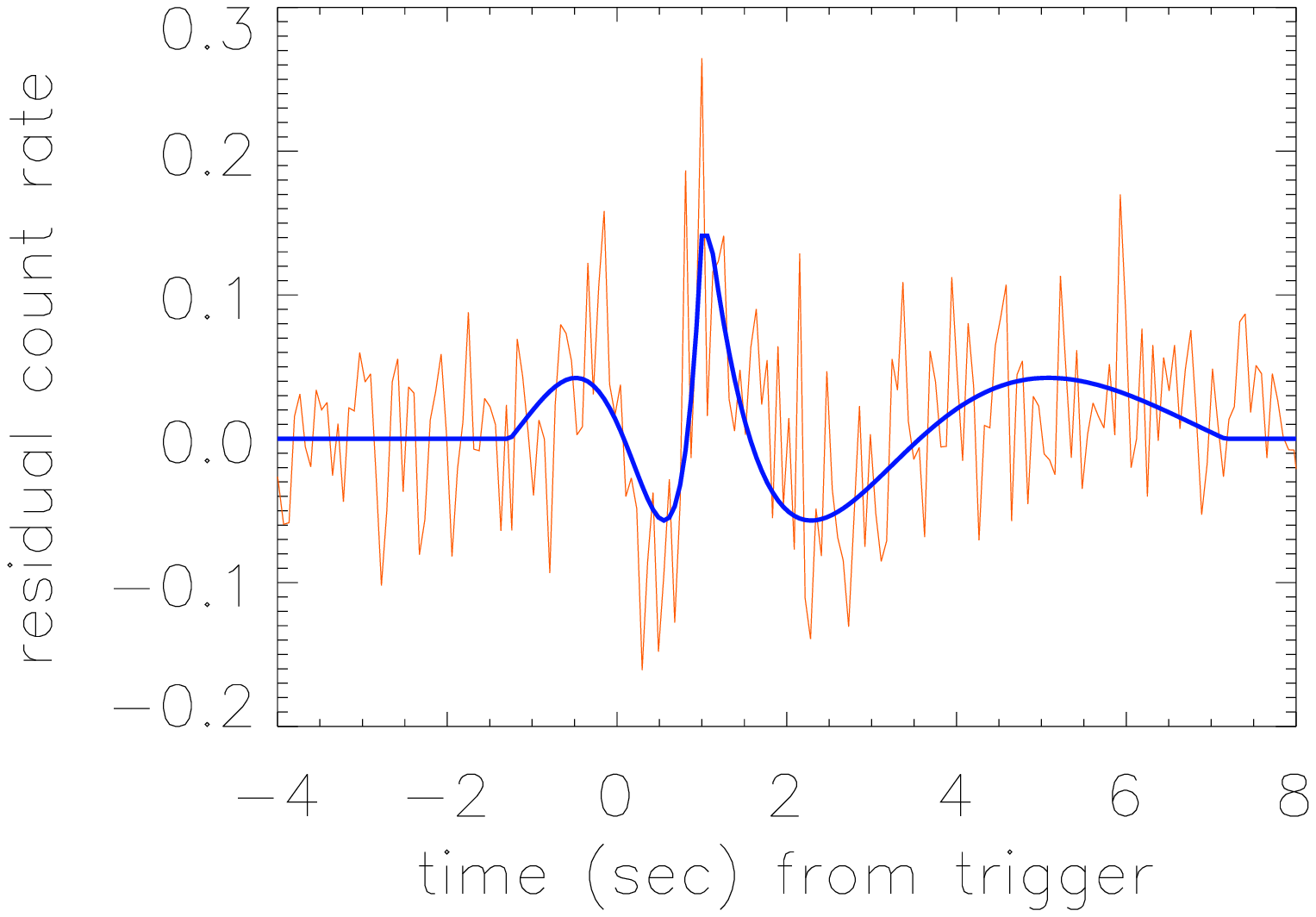}{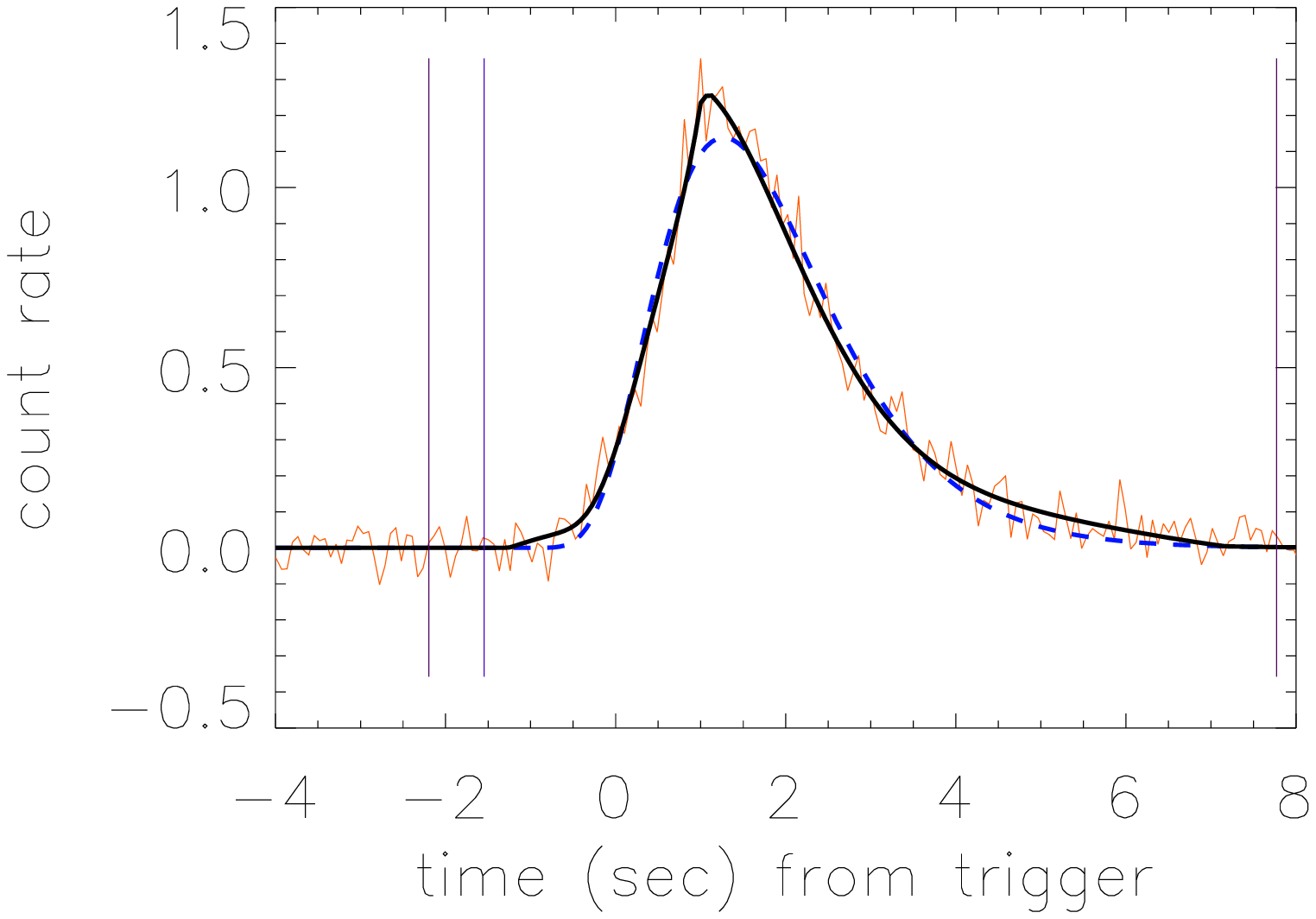}
\caption{GRB 091018: (left) fit to the residuals, (right) fit to the residuals plus pulse model.  \label{fig13}}
\end{figure}

\clearpage

\begin{figure}
\includegraphics{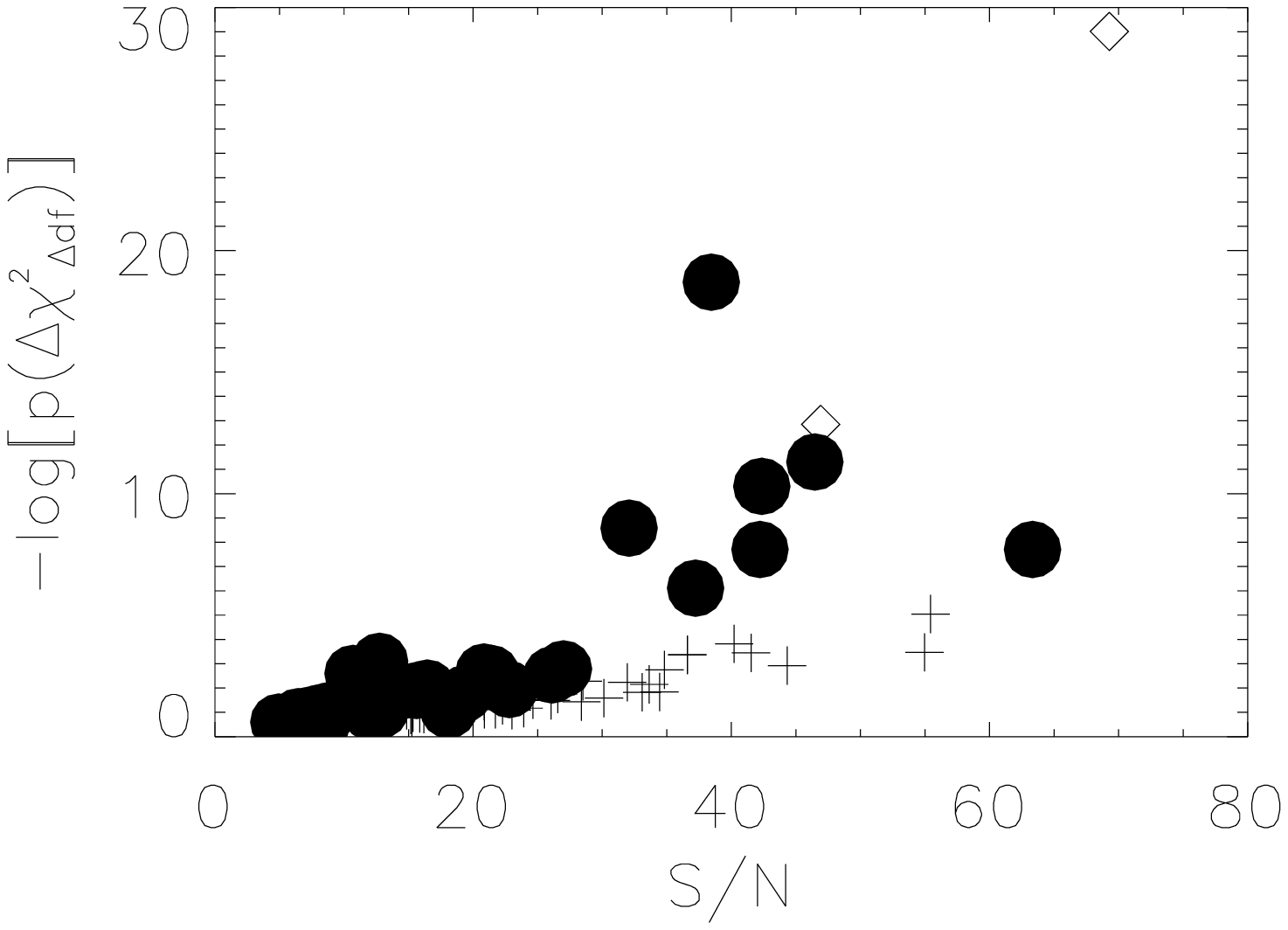}
 \caption{A measure of how much pulse fits have been improved by combining the \cite{hak14} residual fit to the \cite{nor05} fit ($-\rm{log}[p(\Delta \chi_{\Delta df}^2)]$) is shown as a function of the signal-to-noise ratio ($S/N$). A large $-\rm{log}[p(\Delta \chi_{\Delta df}^2)]$ value indicates improvement consistent with a better model fit, while a small one suggests only slight improvement. BATSE fits (crosses) and Fermi pulses (open diamonds) improve systematically with increasing signal-to-noise, suggesting that complex pulse shapes are normal at high signal-to-noise but get harder to recognize at lower signal-to-noise. Swift pulses (filled circles) have similar $S/N$ ratios to BATSE and Fermi pulses. \label{fig14}}
\end{figure}

\clearpage

\begin{figure}
  \includegraphics{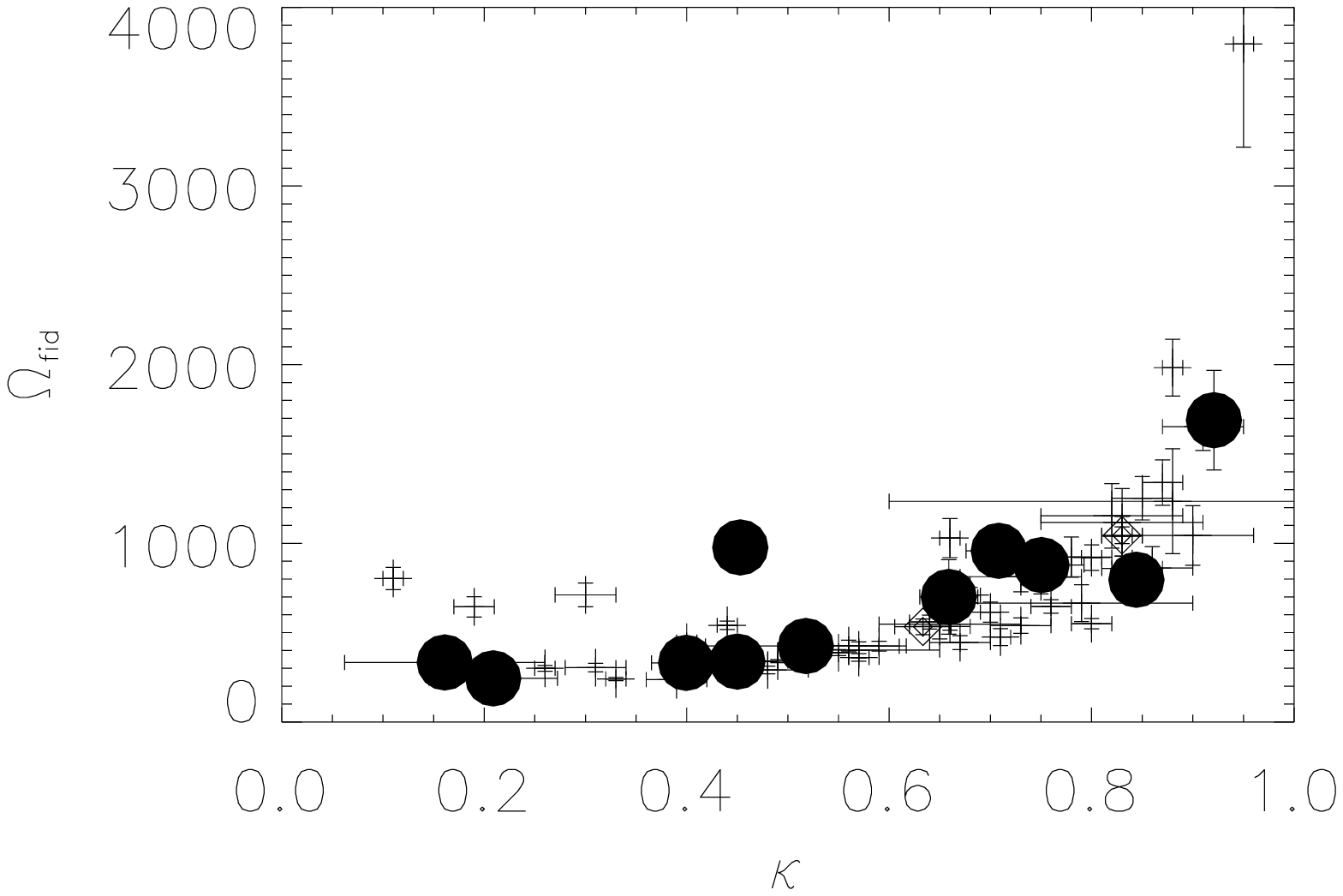}
  \caption{$\Omega_{\rm fid}$ vs. $\kappa$ for the pulses in this analysis. Swift pulses are indicated by filled circles, BATSE pulses by crosses, and Fermi pulses by diamonds. \label{fig15}}
\end{figure}

\clearpage

\begin{figure}
  \includegraphics{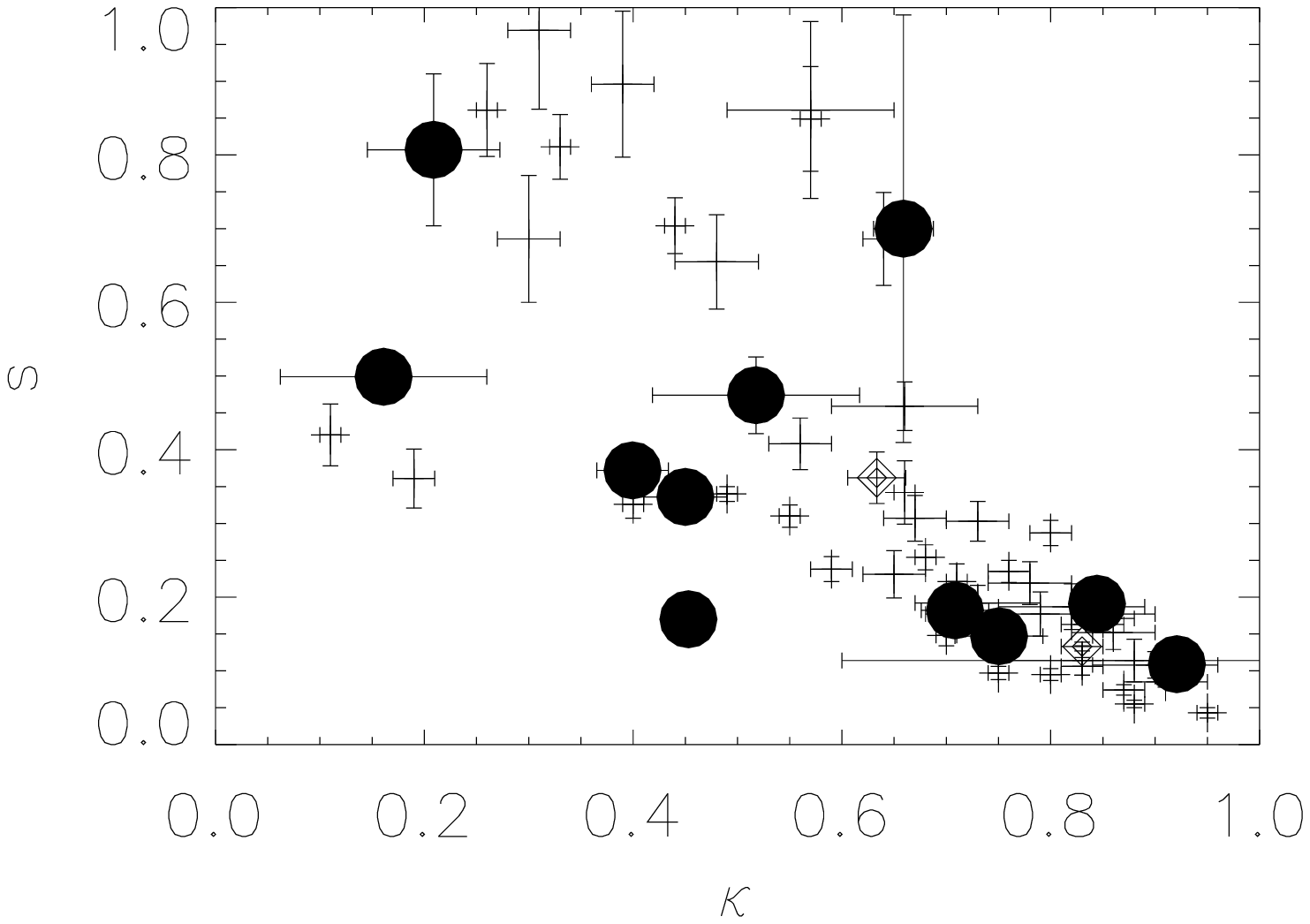}
  \caption{$s$ vs. $\kappa$ for the pulses in this analysis. Swift pulses are indicated by filled circles, BATSE pulses by crosses, and Fermi pulses by diamonds. \label{fig16}}
\end{figure}

\clearpage

\begin{figure}
  \includegraphics{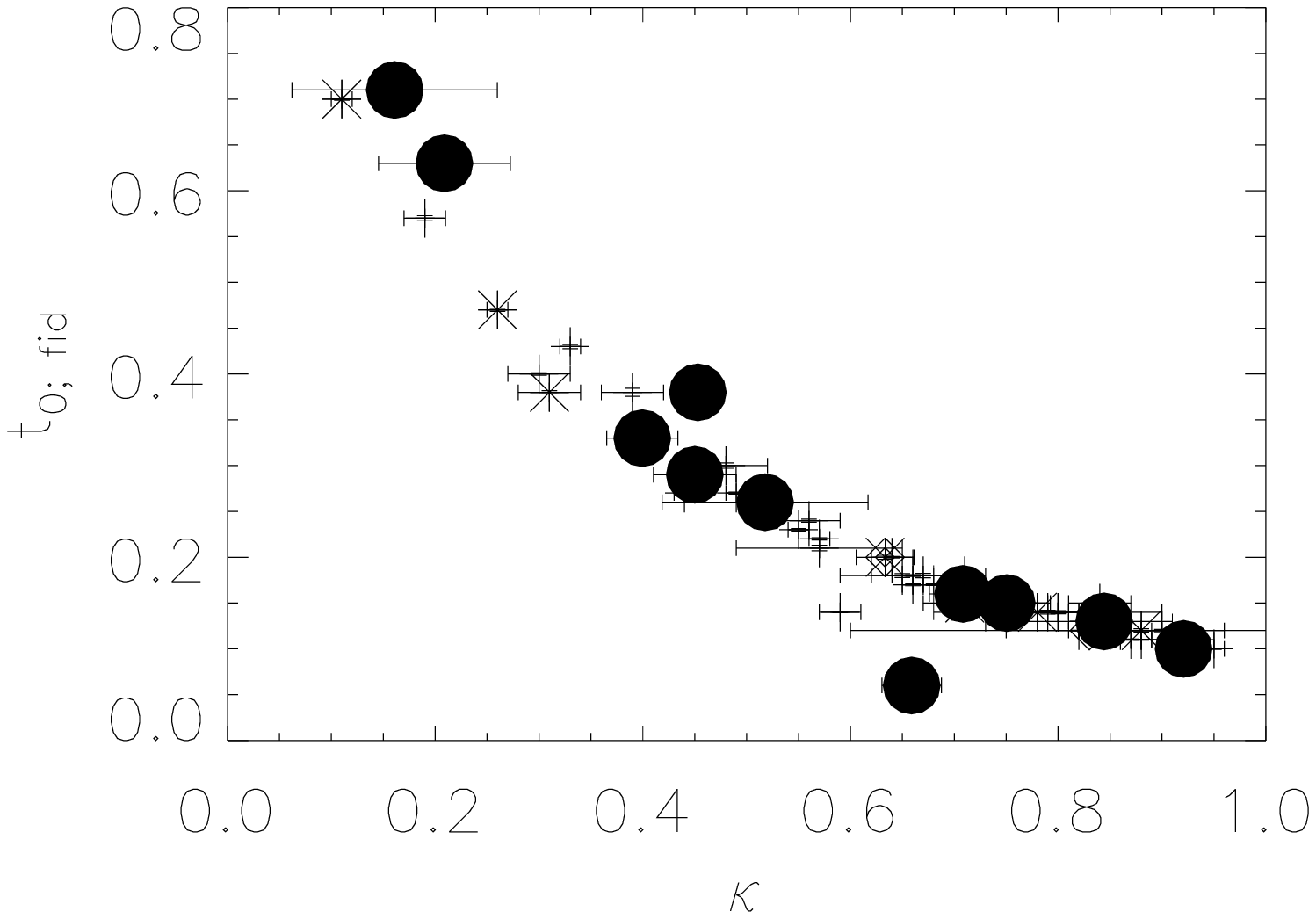}
  \caption{$t_0$ vs. $\kappa$ for the pulses in this analysis. Swift pulses are indicated by filled circles, BATSE pulses by crosses, and Fermi pulses by diamonds. \label{fig17}}
\end{figure}

\clearpage

\begin{figure}
  \includegraphics{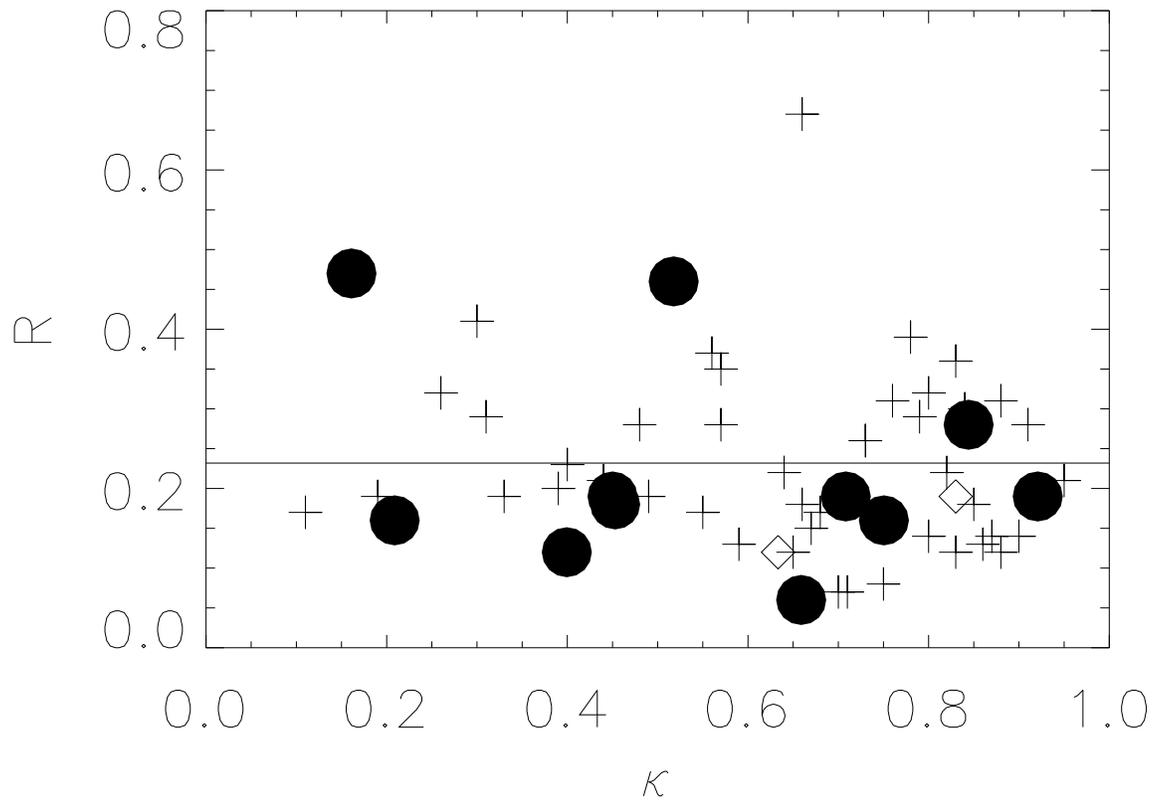}
  \caption{$R$ vs. $\kappa$ for the pulses in this analysis; there is no obvious correlation. Swift pulses are indicated by filled circles, BATSE pulses by crosses, and Fermi pulses by diamonds. {\bf The horizontal line shows the mean value $<R> \approx 0.23.$} \label{fig18}}
\end{figure}

\clearpage

\begin{figure}
\plottwo{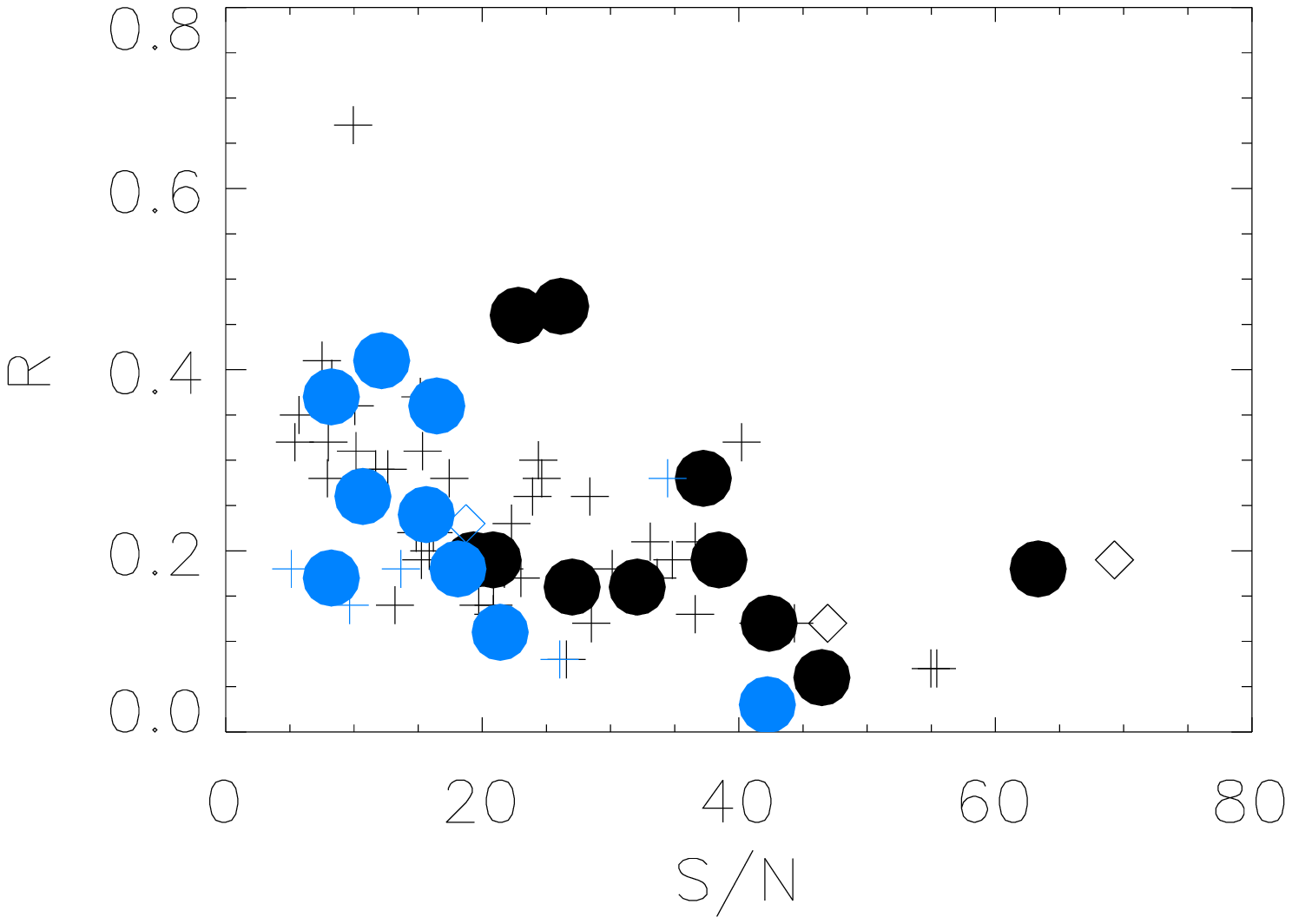}{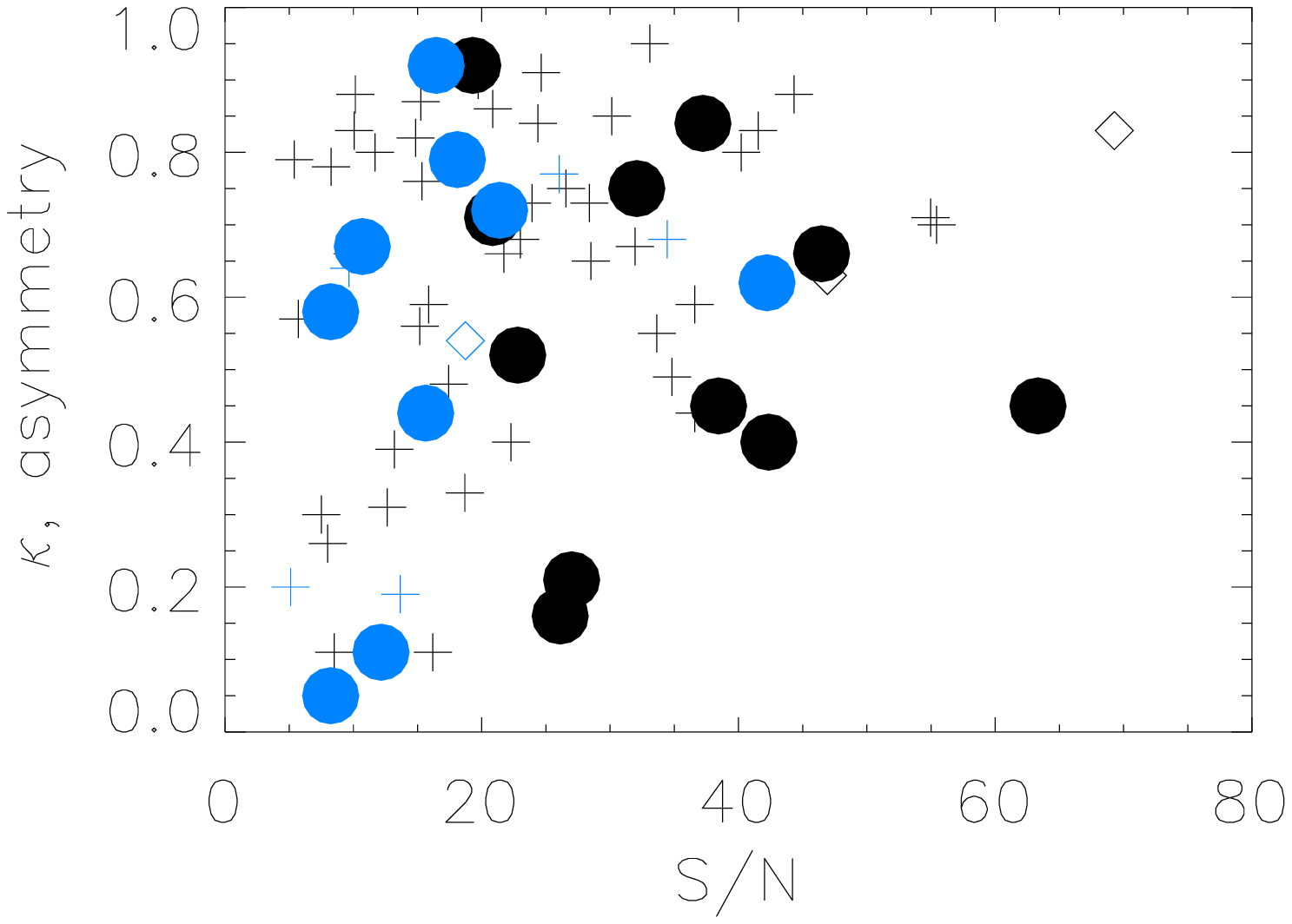}
\caption{Signal-to-noise ($S/N$) properties of GRB pulses that have (black; $p \le 0.001$) and that have not been (blue; $p > 0.001$) improved by the residual fits. Symbols are the same as those described in Figure 14. Pulse fits near the threshold (low $S/N$) only faintly exhibit the residual structure, suggesting that the structure is present for all isolated pulses and easier to detect for bright ones. Furthermore, improved residual fits of Swift pulses are obtained at higher $S/N$ ratios than for BATSE and Fermi pulses, suggesting that the residual features are more pronounced at higher energies. Left panel: Although faint improved pulses (black) tend to have larger relative amplitudes ($R=a/A$) than bright pulses, inclusion of unimproved pulses (blue) tends to offset this correlation somewhat. This again suggests that pulses with small residual amplitudes are present in the isolated pulse distribution but that pulses with large residual amplitudes are easier to identify. Right panel: symmetric pulses ($\kappa < 0.4$) are less common among bright pulses than among faint pulses.  \label{fig19}}
\end{figure}

\clearpage

\begin{figure}
\includegraphics{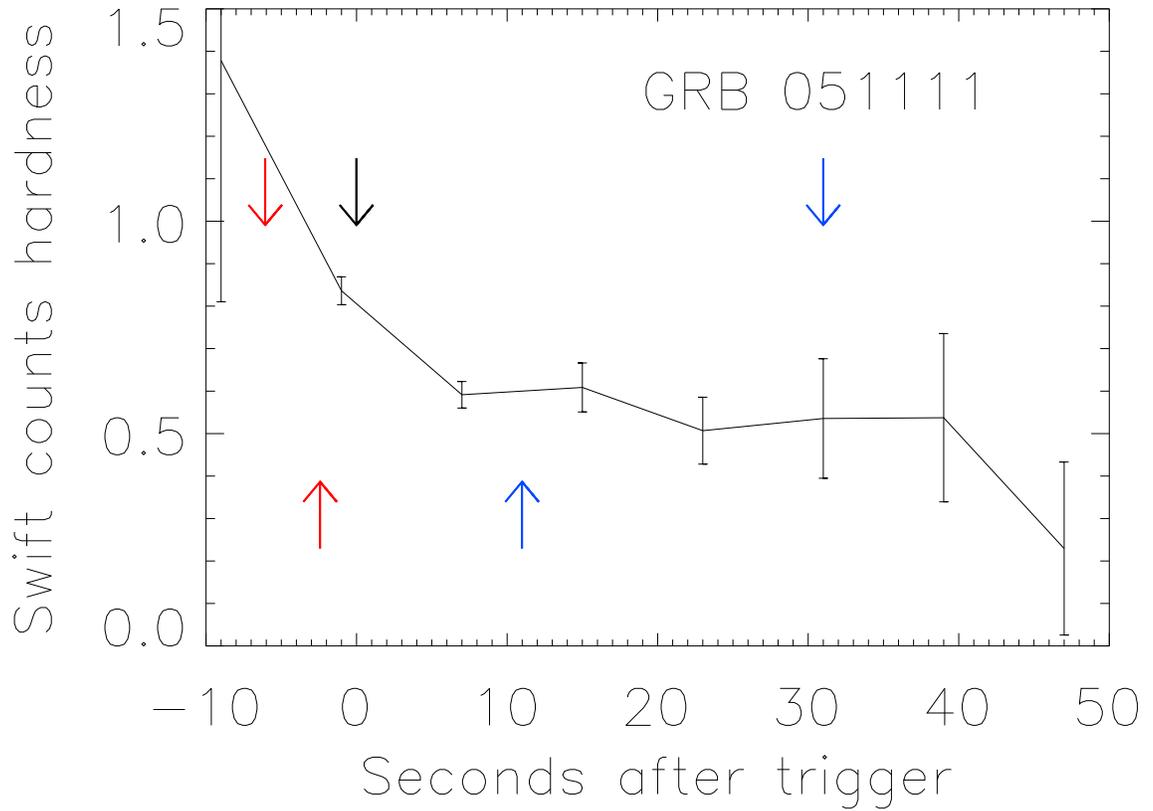}
\caption{Counts hardness evolution for GRB 051111.  In this and subsequent figures, each hardness plot is marked with the five maxima and minima in the residual light curve: the precursor peak (downward facing red arrow), the dip between the precursor peak and the central peak (upward facing red arrow), the central peak (downward facing {\bf black} arrow), the dip between the central peak and the decay peak (upward facing blue arrow), and the decay peak (downward facing blue arrow).  \label{fig20}}
\end{figure}

\clearpage

\begin{figure}
\plottwo{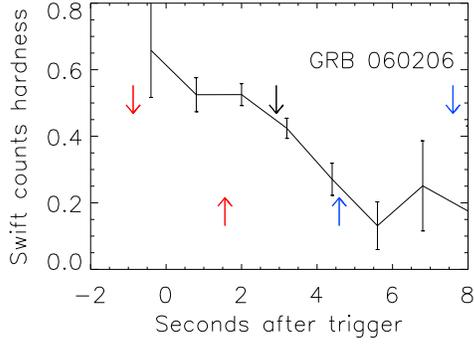}{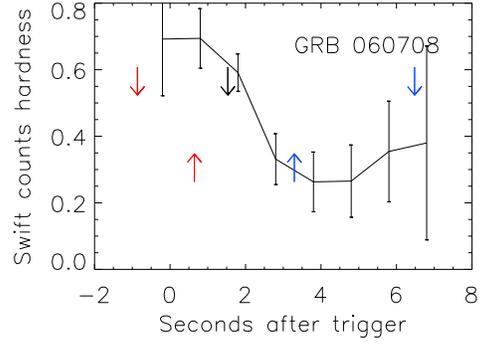}
\caption{Counts hardness evolution for GRB060206 (left) and GRB 060708 (right). \label{fig21}}
\end{figure}

%\clearpage

\begin{figure}
\plottwo{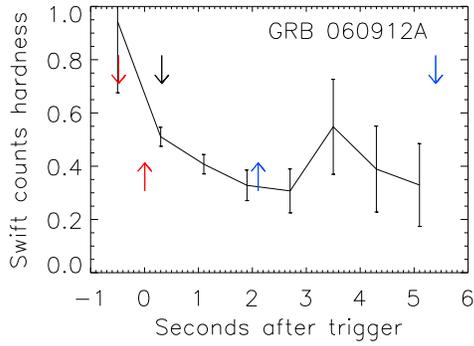}{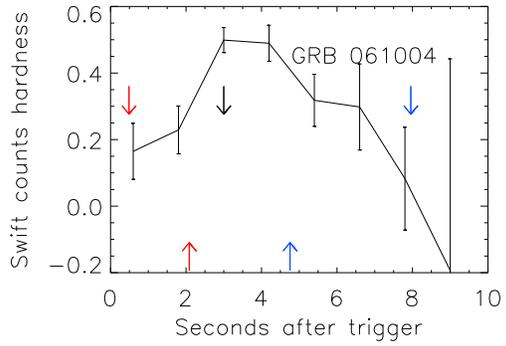}
\caption{Counts hardness evolution for GRB 060912A (left) and GRB 061004 (right).  \label{fig22}}
\end{figure}

%\clearpage

\begin{figure}
\plottwo{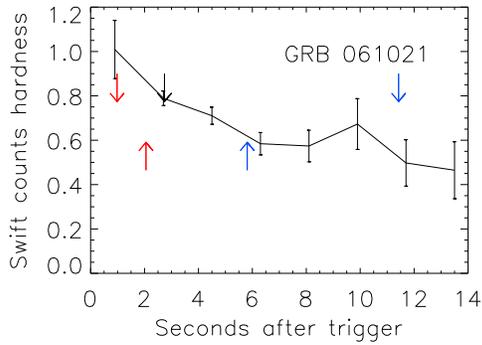}{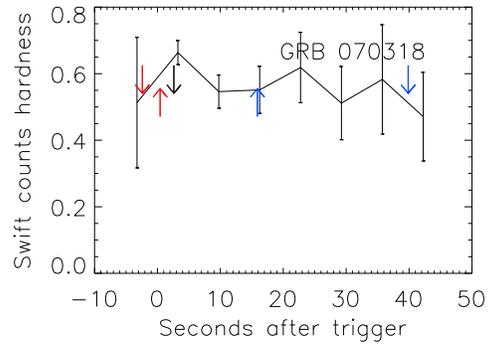}
\caption{Counts hardness evolution for GRB 061021 (left) and GRB 070318 (right).  \label{fig23}}
\end{figure}

\clearpage

\begin{figure}
\plottwo{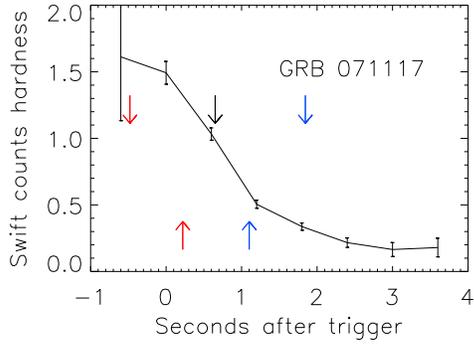}{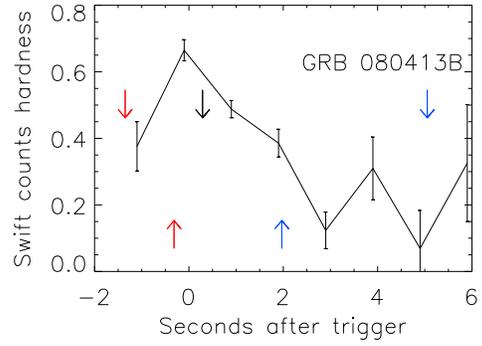}
\caption{Counts hardness evolution for GRB 071117 (left) and GRB 080413B (right).  \label{fig24}}
\end{figure}

%\clearpage

\begin{figure}
\plottwo{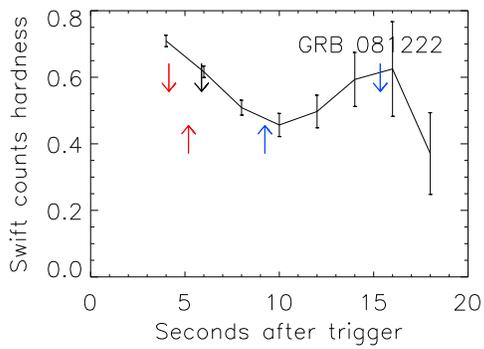}{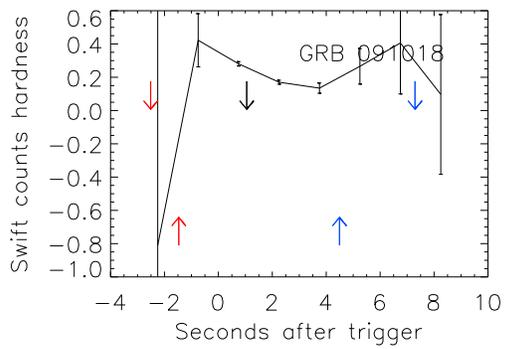}
\caption{Counts hardness evolution for GRB 081222 (left) and GRB 091018 (right).  \label{fig25}}
\end{figure}

\clearpage

\begin{figure}
  \includegraphics{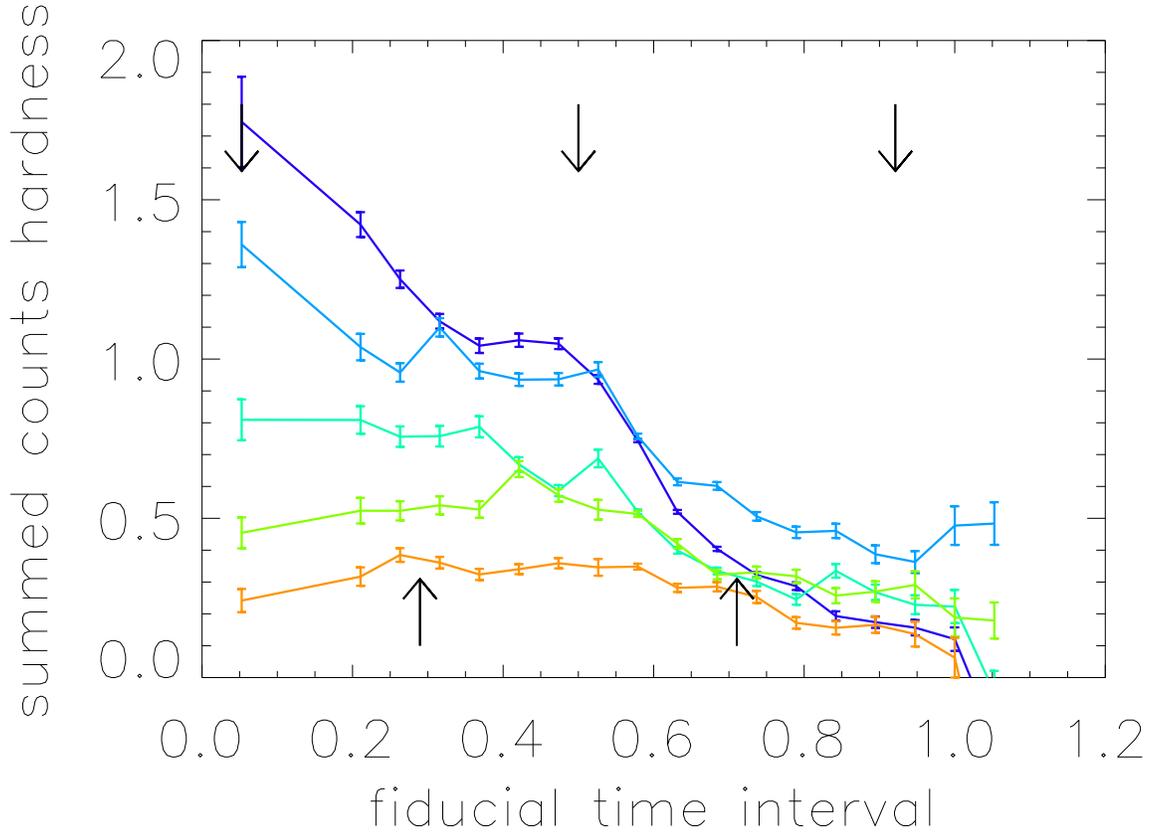}
  \caption{Hardness evolution for BATSE pulses grouped by maximum hardness. Individual pulses are partitioned using residual structures (similar to those shown in Figures 20 to 25, with the three downward-facing arrows indicating the fiducial times of the precursor peak, the central peak, and the decay peak, respectively, and the two upward-facing arrows indicating the fiducial times of the intervening valleys separating these peaks). The counts in each partition are summed for all pulses having similar maximum hardnesses. Hardnesses are measured for each group by dividing the summed high energy counts (100 keV to 1 MeV) in each partition by the summed low energy counts (25 keV to 100 keV). Seven pulses have HR$_{\rm max} \ge 1.6$ (dark blue), eight have $1.1 \le $HR$_{\rm max} < 1.6$ (light blue), nine have $0.8 < $HR$_{\rm max} < 1.1$ (dark green), ten have $0.6 \le $HR$_{\rm max} < 0.8$ (light green), and ten have HR$_{\rm max} \le 0.6$ (orange). The effect of the precursor peak is clearly important in defining whether a pulse undergoes hard-to-soft or intensity matching evolution. \label{fig26}}
\end{figure}

\clearpage

\begin{figure}
  \includegraphics{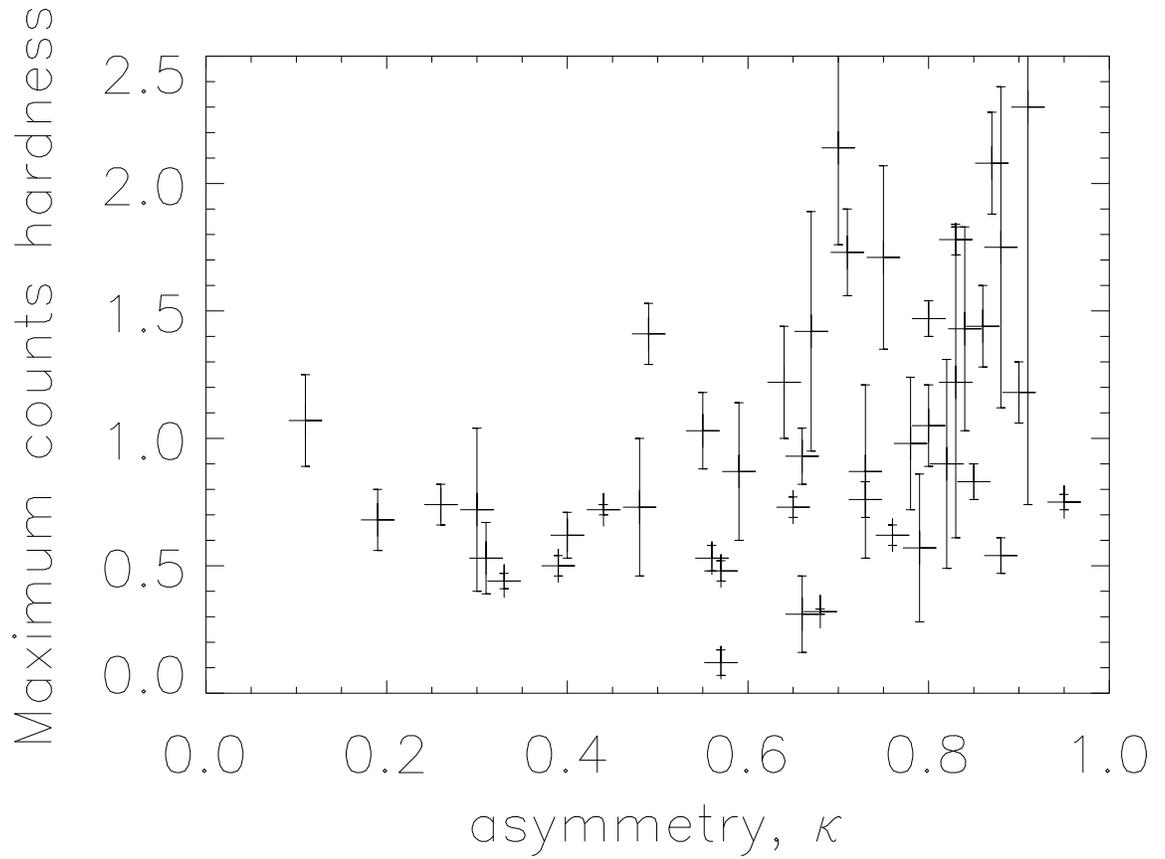}
  \caption{Peak hardness vs. asymmetry for the BATSE GRB pulses included in this analysis. A Spearman Rank Order test yields a p-value of only $p=6.1 \times 10^{-4}$ that these parameters are uncorrelated. \label{fig27}}
\end{figure}

\clearpage

\begin{figure}
  \includegraphics{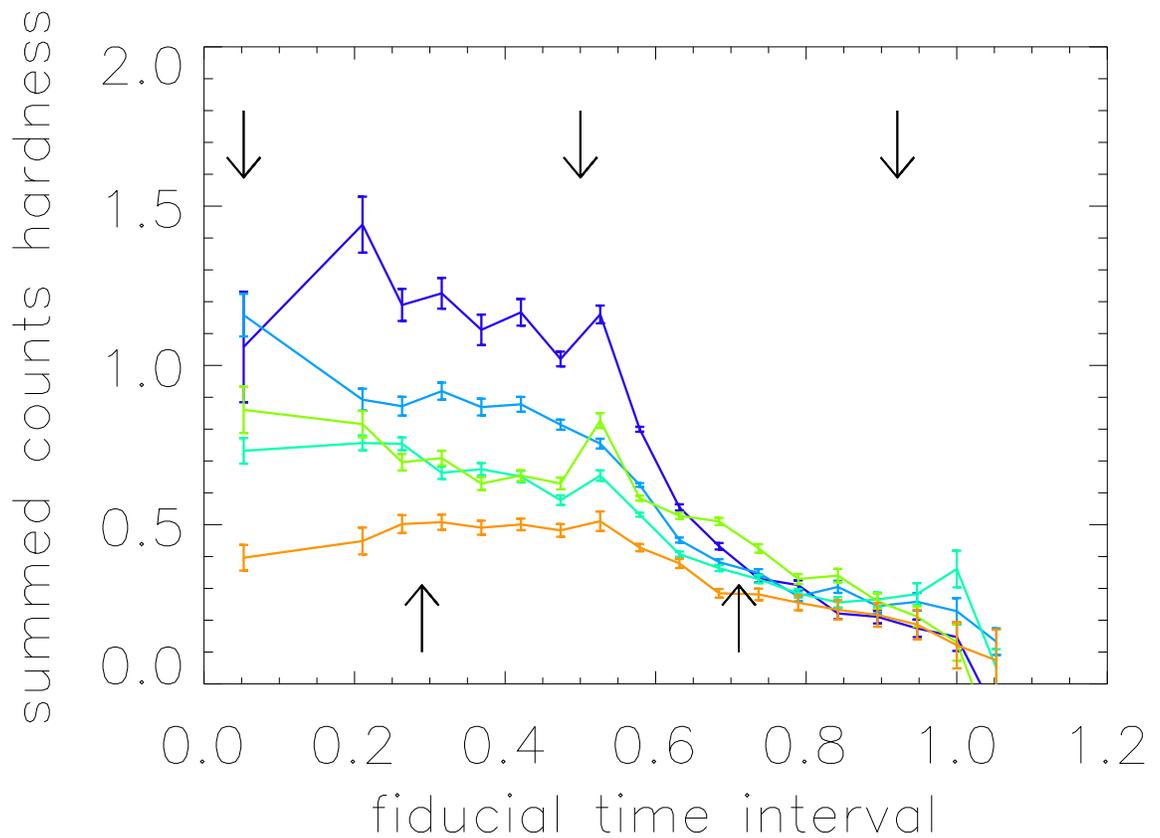}
  \caption{Hardness evolution for BATSE pulses grouped by asymmetry. Individual pulses are partitioned using residual structures (arrows are as described in Figure 26), then the counts in each partition are summed for all pulses having similar maximum hardnesses. Hardnesses are obtained for these binned groups as described for Figure 26. Eight pulses have $\kappa \ge 0.85$ (dark blue), ten have $0.75 \le \kappa < 0.85$ (light blue), nine have $0.65 \le \kappa < 0.75$ (dark green), eight have $0.45 \le \kappa < 0.65$ (light green), and nine have $ \kappa < 0.45$ (orange).\label{fig28}}
\end{figure}

\clearpage

\begin{figure}
  \includegraphics{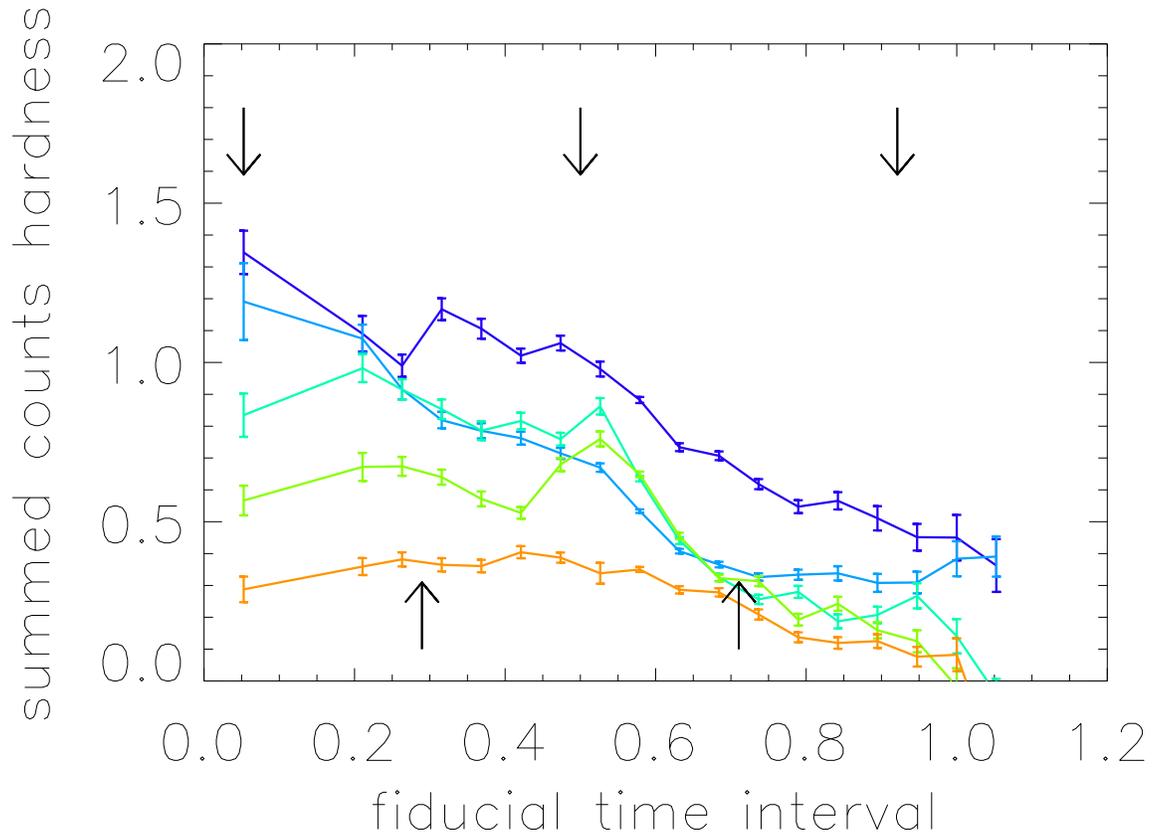}
  \caption{Hardness evolution for BATSE pulses grouped by fluence hardness ratio HR$_{43/21}$. Individual pulses are partitioned using residual structures (arrows are as described in Figure 26), then the counts in each partition are summed for all pulses having similar fluence hardnesses. Hardnesses are obtained for these binned groups as described for Figure 26. Eight pulses have HR$_{43/21} \ge 5.0$ (dark blue), nine have $2.75 \le $HR$_{43/21} < 5.0$ (light blue), eight have $2.0 \le $HR$_{43/21} < 2.75$ (dark green), eight have $1.0 \le $HR$_{43/21} < 2.0$ (light green), and eleven have $0.0 \le $HR$_{43/21} < 1.0$ (orange). \label{fig29}}
\end{figure}


\begin{thebibliography}{}

%\bibitem[Arimoto et al.(2010)]{ari10} Arimoto, M., Kawai, N., Asano, K., et al.\ 2010, \pasj, 62, 487 
%\bibitem[Band(2003)]{ban03} Band, D.~L.\ 2003, \apj, 588, 945 
%\bibitem[Basak \& Rao(2012)]{bas12} Basak, R., \& Rao, A.~R.\ 2012, \apj, 745, 76 
%\bibitem[Bhat et al.(2012)]{bha12} Bhat, P.~N., Briggs, M.~S., Connaughton, V., et al.\ 2012, \apj, 744, 141 
%\bibitem[Bo{\c c}i et al.(2010)]{boc10} Bo{\c c}i, S., Hafizi, M., \& Mochkovitch, R.\ 2010, \aap, 519, A76 
\bibitem[Bo{\v s}njak \& Daigne(2014)]{bol14} Bo{\v s}njak, {\v Z}., \& Daigne, F.\ 2014, \aap, 568, A45 
\bibitem[Broadbent(2014)]{bro14} Broadbent, M.~E.\ 2014, Ph.D.~Thesis. 
\bibitem[Burgess et al.(2011)]{bur11} Burgess, J.~M., Preece, R.~D., Baring, M.~G., et al.\ 2011, \apj, 741, 24 
\bibitem[Chincarini et al.(2010)]{chi10} Chincarini, G., et al.\ 2010, \mnras, 406, 2113 
%\bibitem[Connaughton(2002)]{con02} Connaughton, V.\ 2002, \apj, 567, 1028 
\bibitem[Crider et al.(1999)]{cri99} Crider, A., Liang, E.~P., Preece, R.~D., Briggs, M.~S., Pendleton, G.~N., Paciesas, W.~S., Band, D.~L., \& Matteson, J.~L.\ 1999, \aaps, 138, 401 
%\bibitem[Daigne \& Mochkovitch(2002)]{dai02} Daigne, F., \& Mochkovitch, R.\ 2002, \mnras, 336, 1271
\bibitem[Dermer(2004)]{der04} Dermer, C.~D.\ 2004, \apj, 614, 284 
\bibitem[Fenimore et al.(1996)]{fen96} Fenimore, E.~E., Madras, C.~D., \& Nayakshin, S.\ 1996, \apj, 473, 998 
%\bibitem[Fenimore \& Ramirez-Ruiz(2000)]{frr00} Fenimore, E.~E., \& Ramirez-Ruiz, E.\ 2000, arXiv:astro-ph/0004176 
%\bibitem[Fermi Large Area Telescope Team et al.(2012)]{FLAT12} Fermi Large Area Telescope Team, Ackermann, M., Ajello, M., et al.\ 2012, \apj, 754, 121 
\bibitem[Ford et al.(1995)]{for95} Ford, L.~A., et al.\ 1995, \apj, 439, 307 
%\bibitem[Gehrels \& M{\'e}sz{\'a}ros(2012)]{geh12} Gehrels, N., \& M{\'e}sz{\'a}ros, P.\ 2012, Science, 337, 932 
%\bibitem[Golenetskii et al.(1983)]{gol83} Golenetskii, S.~V., Mazets, E.~P., Aptekar, R.~L., \& Ilinskii, V.~N.\ 1983, \nat, 306, 451
%\bibitem[Gonz{\'a}lez et al.(2003)]{gon03} Gonz{\'a}lez, M.~M., Dingus, B.~L., Kaneko, Y., et al.\ 2003, \nat, 424, 749 
%\bibitem[Goodman(1986)]{goo86} Goodman, J.\ 1986, \apjl, 308, L47 
%\bibitem[Hakkila et al.(2003)]{hak03} Hakkila, J., Giblin, T.~W., Roiger, R.~J., Haglin, D.~J., Paciesas, W.~S., \& Meegan, C.~A.\ 2003, \apj, 582, 320 
%\bibitem[Hakkila et al.(2008)]{hak08} Hakkila, J., et al. \ 2008a, \apjl, 677, L81
%\bibitem[Hakkila \& Cumbee(2009)]{hak09} Hakkila, J., \& Cumbee, R.~S.\ 2009, in AIP Proc. 1133 (ed. Meegan, Gehrels, \& Kouveliotou), 379
\bibitem[Hakkila \& Preece(2011)]{hak11} Hakkila, J., \& Preece, R.~D.\ 2011, \apj, 740, 104
\bibitem[Hakkila \& Preece(2014)]{hak14} Hakkila, J., \& Preece, R.~D.\ 2014, \apj, 783, 88 
%\bibitem[Horv{\'a}th(1998)]{hor98} Horv{\'a}th, I.\ 1998, \apj, 508, 757 
%\bibitem[Horv{\'a}th et al.(2006)]{hor06} Horv{\'a}th, I., Bal{\'a}zs, L.~G., Bagoly, Z., Ryde, F., \& M{\'e}sz{\'a}ros, A.\ 2006, \aap, 447, 23 
%\bibitem[Jia \& Qin(2005)]{jia05} Jia, L.-W., \& Qin, Y.-P.\ 2005, \apjl, 631, L25
\bibitem[Kaneko et al.(2006)]{kan06} Kaneko, Y., Preece, R.~D., Briggs, M.~S., Paciesas, W.~S., Meegan, C.~A., \& Band, D.~L.\ 2006, \apjs, 166, 298 
\bibitem[Kino et al.(2004)]{kin04} Kino, M., Mizuta, A., \& Yamada, S.\ 2004, \apj, 611, 1021 
%\bibitem[Krimm et al.(2013)]{kri13} Krimm, H.~A., Holland, S.~T., Corbet, R.~H.~D., et al.\ 2013, \apjs, 209, 14 
%\bibitem[Kocevski et al.(2003)]{koc03} Kocevski, D., Ryde, F., \& Liang, E.\ 2003, \apj, 596, 389 
%\bibitem[Lee et al.(2000a)]{lee00a} Lee, A., Bloom, E.~D., \& Petrosian, V.\ 2000, \apjs, 131, 1 
%\bibitem[Lee et al.(2000b)]{lee00b} Lee, A., Bloom, E.~D., \& Petrosian, V.\ 2000, \apjs, 131, 21 
\bibitem[Li et al.(2012)]{li12} Li, L., Liang, E.-W., Tang, Q.-W., et al.\ 2012, \apj, 758, 27
\bibitem[Liang \& Kargatis(1996)]{lia96} Liang, E., \& Kargatis, V.\ 1996, \nat, 381, 49  
\bibitem[Lu et al.(2010)]{lu10} Lu, R.-J., Hou, S.-J., \& Liang, E.-W.\ 2010, \apj, 720, 1146 
%\bibitem[MacLachlan et al.(2012)]{mac12} MacLachlan, G.~A., Shenoy, A., Sonbas, E., et al.\ 2012, \mnras, 425, L32
\bibitem[Margutti et al.(2010)]{mar10} Margutti, R., Guidorzi, C., Chincarini, G., Bernardini, M.~G., Genet, F., Mao, J., \& Pasotti, F.\ 2010, \mnras, 406, 2149 
\bibitem[Markwardt(2009)]{mar09} Markwardt, C.~B.\ 2009, Astronomical Data Analysis Software and Systems XVIII, 411, 251 
\bibitem[Morris et al.(2014)]{mor14} Morris, D.~C., Neff, J.~E., \& Hakkila, J.~E.\ 2014, American Astronomical Society Meeting Abstracts \#223, 223, \#148.42 
%\bibitem[Medvedev et al.(2009)]{med09} Medvedev, M.~V., Pothapragada, S.~S., \& Reynolds, S.~J.\ 2009, \apjl, 702, L91
%\bibitem[Mukherjee et al.(1998)]{muk98} Mukherjee, S., Feigelson, E.~D., Jogesh Babu, G., et al.\ 1998, \apj, 508, 314 
%\bibitem[Norris et al.(1986)]{nor86} Norris, J.~P., Share, G.~H., Messina, D.~C., et al.\ 1986, \apj, 301, 213 
%\bibitem[Norris et al.(1996)]{nor96} Norris, J.~P., Nemiroff, R.~J., Bonnell, J.~T., et al.\ 1996, \apj, 459, 393 
%\bibitem[Norris et al.(2000)]{nor00} Norris, J.~P., Marani, G.~F., \& Bonnell, J.~T.\ 2000, \apj, 534, 248 
%\bibitem[Norris(2002)]{nor02} Norris, J.~P.\ 2002, \apj, 579, 386 
\bibitem[Norris et al.(2005)]{nor05} Norris, J.~P., Bonnell, J.~T., Kazanas, D., Scargle, J.~D., Hakkila, J., \& Giblin, T.~W.\ 2005, \apj, 627, 324 
%\bibitem[Norris \& Bonnell(2006)]{nor06} Norris, J.~P., \& Bonnell, J.~T.\ 2006, \apj, 643, 266
%\bibitem[Norris et al.(2011)]{nor11} Norris, J.~P., Gehrels, N., \& Scargle, J.~D.\ 2011, \apj, 735, 23 
\bibitem[Peng et al.(2009)]{pen09} Peng, Z.~Y., Ma, L., Lu, R.~J., Fang, L.~M., Bao, Y.~Y., \& Yin, Y.\ 2009, \na, 14, 311
\bibitem[Peng et al.(2012)]{pen12} Peng, Z.~Y., Zhao, X.~H., Yin, Y., Bao, Y.~Y., \& Ma, L.\ 2012, \apj, 752, 132  
\bibitem[Preece et al.(2015)]{pre15} Preece, R.~D., Goldstein, A., Bhat, N., Stanbro, M., Hakkila, J. \& Blalock, D. \ 2015 \apj, submitted
\bibitem[Qin(2002)]{qin02} Qin, Y.-P.\ 2002, \aap, 396, 705 
%\bibitem[Preece et al.(2013)]{pre13} Preece, R., Burgess, J.~M., von Kienlin, A., et al.\ 2013, arXiv:1311.5581 
%\bibitem{Raftery(1995){raf95} Raftery, A.~E. \1995, Sociological Methodology, 25, 111
\bibitem[Ramirez-Ruiz \& Fenimore(2000)]{rrf00} Ramirez-Ruiz, E., \& Fenimore, E.~E.\ 2000, \apj, 539, 712
\bibitem[Rees \& Meszaros(1994)]{ree94} Rees, M.~J., \& Meszaros, P.\ 1994, \apjl, 430, L93 
%\bibitem[Reichart et al.(2001)]{rei01} Reichart, D.~E., Lamb, D.~Q., Fenimore, E.~E., et al.\ 2001, \apj, 552, 57 
%\bibitem[Ryde(2005)]{ryd05} Ryde, F.\ 2005, \aap, 429, 869
\bibitem[Sakamoto et al.(2005)]{sak05} Sakamoto, T., Lamb, D.~Q., Kawai, N., et al.\ 2005, \apj, 629, 311 
\bibitem[Scargle(1998)]{sca98} Scargle, J.~D.\ 1998, \apj, 504, 405
\bibitem[Steiger et al.(1985)]{ste85} Steiger, J.~H., Shapiro, A., \& Browne, M.~B.\ 1985, Psychometrika, 50, 253.
%\bibitem[Schmidt \& Lipson(2009)]{sch09} Schmidt M., \& Lipson H.\ 2009, Science, 324, 81
%\bibitem[Stern \& Svensson(1996)]{ste96} Stern, B.~E., \& Svensson, R.\ 1996, \apjl, 469, L109 
%\bibitem[Ukwatta et al.(2010)]{ukw10} Ukwatta, T.~N., Stamatikos, M., Dhuga, K.~S., et al.\ 2010, \apj, 711, 1073
\bibitem[Willingale et al.(2010)]{wil10} Willingale, R., Genet, F., Granot, J., \& O'Brien, P.~T.\ 2010, \mnras, 403, 1296 
%\bibitem[Wolpert et al.(2011)]{wol11} Wolpert, R.~L., Clyde, M.~A., \& Tu, C.\ 2011, Ann. Statist., 39, 1916
%\bibitem[Zhang \& Qin(2005)]{zha05} Zhang, Z.-B., \& Qin, Y.-P.\ 2005, \mnras, 363, 1290 
%\bibitem[Zhang \& Yan(2011)]{zha11} Zhang, B., \& Yan, H.\ 2011, \apj, 726, 90

\end{thebibliography}
\end{document}